\documentclass[journal]{new-aiaa}
\usepackage[utf8]{inputenc}

\usepackage{graphicx}
\usepackage{amsmath}
\usepackage[version=4]{mhchem}
\usepackage{siunitx}
\usepackage{longtable,tabularx}
\usepackage{etoolbox}
\usepackage[usenames,dvipsnames]{color} 
\usepackage{ifthen}
\usepackage{geometry}
\usepackage{ulem}
\usepackage{newtxtext,newtxmath}
\DeclareSymbolFont{CMletters}{OML}{cmm}{m}{it}
\DeclareMathSymbol{\nu}{\mathord}{CMletters}{23}
\DeclareMathSymbol{v}{\mathord}{CMletters}{`v}
\newboolean{showcomments}
\setboolean{showcomments}{true}

\newcommand{\dennice}[1]{  \ifthenelse{\boolean{showcomments}}
{\textcolor{red}{(Dennice says:  #1)}}{}}

\newcommand{\rev}[1]{{\color{black}{#1}}}

\newcommand{\red}[1]{{\color{black}#1}}

\makeatletter
\def\blfootnote{\xdef\@thefnmark{}\@footnotetext}
\makeatother

\title{Spatial input--output analysis of large-scale structures in actuated turbulent boundary layers}

\author{Chang Liu$^1$} 
\author{Igal Gluzman$^2$}
\affil{Johns Hopkins University, Baltimore, MD, 21218}
\author{Mitchell Lozier$^3$}
\author{Samaresh Midya$^3$}
\author{Stanislav Gordeyev$^4$}
\author{Flint O. Thomas$^5$}
\affil{University of Notre Dame, Notre Dame, IN, 46556}
\author{Dennice F. Gayme$^6$}
\affil{Johns Hopkins University, Baltimore, MD, 21218}

\def\mline{\vrule width4pt height2.5pt depth -2pt}
\def\dashed{\mline\hskip3.5pt\mline\thinspace}
\def\bdot{\raise.2em\hbox to .15em{.}}
\def\dashed{\mline\hskip3.5pt\mline\thinspace}
\def\dashdot{\mline\ \bdot\,\  \mline\thinspace}
\def\dashdotdot{\mline\ \bdot\,\ \bdot\,\ \mline\thinspace}

\begin{document}
\footnotetext[1]{Graduate student, Department of Mechanical Engineering. Currently Postdoctoral scholar, Department of Physics, University of California, Berkeley, CA. Corresponding email: cliu124@alumni.jh.edu}
\footnotetext[2]{Postdoctoral Fellow, Department of Mechanical Engineering. Currently Postdoctoral Research Associate, Department of Aerospace and Mechanical Engineering,  University of Notre Dame, Notre Dame, IN.}
\footnotetext[3]{Graduate student, Department of Aerospace and Mechanical Engineering, AIAA Student member.}
\footnotetext[4]{Associate Professor, Department of Aerospace and Mechanical Engineering, AIAA Associate Fellow.}
\footnotetext[5]{Professor, Department of Aerospace and Mechanical Engineering, AIAA Associate Fellow.}
\footnotetext[6]{Associate Professor, Department of Mechanical Engineering.}
\blfootnote{Part of this work has been presented in AIAA Aviation 2021 Forum, August 2-6, 2021, Virtual event. The AIAA Paper Number is 2021-2873 https://doi.org/10.2514/6.2021-2873}
\maketitle

\begin{abstract}

This paper develops a spatial input--output approach to investigate the dynamics of a turbulent boundary layer subject to a localized single frequency  excitation. This method uses one-way spatial integration to reformulate the problem in terms of spatial evolution equations. The technique is used to examine the effect of localized periodic actuation at a given temporal frequency, based on an experimental set-up in which an active large-scale is introduced into the outer layer of a turbulent boundary layer.  First, the large-scale structures associated with the phase-locked modal velocity field obtained from spatial input-output analysis are shown to closely match those computed based on hot-wire measurements. The approach is then used  to further investigate the response of the boundary layer to the synthetically generated large-scale. A quadrant trajectory analysis indicates that the spatial input-output response produces shear stress distributions consistent with those in canonical wall-bounded turbulent flows in terms of both the order and types of events observed. The expected correspondence between the dominance of different quadrant behavior and actuation frequency is also observed. These results highlight the promise of a spatial input-output framework for analyzing the formation and streamwise evolution of structures in actuated wall-bounded turbulent flows.

\end{abstract}

\section*{Nomenclature}

{\renewcommand\arraystretch{1.0}
\noindent\begin{longtable*}{@{}l @{\quad=\quad} l@{}}
$x$  & streamwise location \\
$y$ &   wall-normal location \\
$z$& spanwise location \\
$x_0$ & streamwise location of forcing\\
$x_m$ & streamwise location of measurement \\
$k_x$ & streamwise wavenumber\\
$k_z$ & spanwise wavenumber\\
$\lambda_x$ & streamwise wavelength\\
$\lambda_z$ & spanwise wavelength\\
$t$ & time\\
$\omega$ & temporal frequency\\
$\eta$ & temporal growth rate\\
$\text{i}$ & imaginary unit $\text{i}=\sqrt{-1}$\\
$u_{\text{tot}}$ & streamwise velocity \\
$v_{\text{tot}}$ & wall-normal velocity \\
$w_{\text{tot}}$ & spanwise velocity \\
$p_{\text{tot}}$ & pressure\\
$\bar{(\cdot)}$ & time-averaging operation\\
$T$ & final time for time-averaging\\
$\bar{u}$ & streamwise component of time-averaged turbulent mean velocity\\
$\bar{p}$ & time-averaged pressure\\
$u$ & streamwise velocity fluctuation from the model or instantaneous streamwise velocity from experiments\\
$v$ & wall-normal velocity fluctuation \\
$w$ & spanwise velocity fluctuation \\
$\tilde{u}$ & phase-locked modal streamwise velocity\\
$u'$ & residual fluctuating turbulent streamwise velocity\\
$\omega_z$ & spanwise vorticity fluctuation\\
$p$ & pressure fluctuation\\
\rev{$\bar{\tau}_w$} & time-averaged mean shear stress at the wall\\
$\rho$ & density\\
$u_\tau$ & friction velocity \\
$\nu$ & kinematic viscosity\\
$\delta_v$ & the inner unit length scale\\
$(\cdot)^+$ & variable normalized by the inner units: $\delta_v$ and $u_\tau$\\
$\delta_{99}$ & boundary layer thickness\\
$\delta_{99}^+$ & boundary layer thickness in the inner unit\\
$Re_\tau$ & friction Reynolds number\\
$\boldsymbol{e}_x$ & streamwise unit vector\\
$f_x$ & streamwise component of body force modeling plasma actuation \\
$y_p$ & wall-normal location of plasma actuator plate \\
$y_f$  & wall-normal location of the center of body force  \\
$\sigma_p$ & the standard deviation of the Gaussian function of body force\\
$F_0$ & the magnitude of body force\\
$\delta(\cdot)$ & Dirac delta function\\
$\phi_0$ & the initial phase of body force\\
$\phi$ & phase\\
$\boldsymbol{\Psi}$ & operator mapping the state $\boldsymbol{q}_S$ at $x=x_0$ to  $\boldsymbol{q}_S$ at $x=x_m$ under the same frequency-wavenumber pair $(\omega,k_z)$\\
$\psi$ & a nominal variable\\
$\check{(\cdot)}$ & Laplace transform in the time domain and Fourier transform in the spanwise domain\\
\rev{$\check{v}_x$} & \rev{$\frac{\partial \check{v}}{\partial x}$}\\
\rev{$\check{w}_x$} & \rev{$\frac{\partial \check{w}}{\partial x}$}\\
$\check{\boldsymbol{q}}_S$ & state variable of spatial input-output analysis\\
$\check{\boldsymbol{A}}_S$ & state evolution operator of spatial input-output analysis\\
$\check{\boldsymbol{B}}_S$ & input operator of spatial input-output analysis\\
$\check{\boldsymbol{C}}_S$ & output operator of spatial input-output analysis\\
$M$ & $\nu (\partial_y^2-k_z^2)-\eta+\text{i}\omega$ employed to simplify notation\\
$\boldsymbol{\Lambda}$ & a diagonal matrix containing eigenvalues of $\boldsymbol{\check{A}}_S$\\
$\boldsymbol{V}$ &  a matrix containing eigenvectors of $\check{\boldsymbol{A}}_S$\\
$\boldsymbol{0}$ & zero matrix\\
$\mathbb{R}e[\cdot]$ & the real part of the argument\\
$\mathbb{I}m[\cdot]$ & the imaginary part of the argument\\
$u_s$ & downstream evolution of phase-locked streamwise velocity\\
$v_s$ & downstream evolution of phase-locked wall-normal velocity\\
$\omega_{z,s}$ & downstream evolution of phase-locked spanwise vorticity\\
\end{longtable*}}


\newtoggle{introduction_note}
\toggletrue{introduction_note}
\newtoggle{thesis}
\togglefalse{thesis}
\newtoggle{AIAA_Journal}
\toggletrue{AIAA_Journal}
\newtoggle{exp_spatial_with56}
\toggletrue{exp_spatial_with56}

\section{Introduction}
\label{sec:spatial_introduction}

Large-scale structures in turbulent boundary layers (TBL) are known to contribute significantly to the turbulent kinetic energy and Reynolds stress production \citep{Balakumar2007,Guala2006}, which influence the near-wall small-scale structures \citep{mathis2009large,mathis2009comparison,Marusic2010} and local skin friction \citep{hwang2017influence}. This influence of the large-scale structures on the TBL dynamics has been shown to increase with Reynolds number \citep{Smits2011}. Large-scale structures can also be manipulated to change the properties of the boundary layer; e.g. to reduce drag in a high Reynolds number TBL \citep{abbassi2017skin}; see e.g., review \citep{corke2018active}. Therefore, understanding their dynamics and interactions with the overall TBL can provide insight into the underlying physics.

The dynamics of large-scale structures can be studied by analyzing the flow response to an external large-scale perturbation. Single harmonic perturbations provide a particularly attractive approach to tracking the linear response of the turbulent boundary layer at the same frequency through phase-locked analysis. Investigating these types of actuated flows dates back to \citet{hussain1970mechanics,hussain1972mechanics}, where a thin vibrating ribbon near the wall is used to introduce perturbations into turbulent channel flow. They analyzed the experimental results by introducing a triple decomposition of the instantaneous velocity into a temporal mean, phase-locked harmonic perturbations (organized waves), and the remaining turbulence. Periodic perturbations have also been experimentally introduced into a turbulent boundary layer through a dynamic (temporally oscillating) roughness, which provides a reference phase to isolate \rev{the} synthetic large-scale and small-scale flow structures \citep{jacobi2011new,jacobi2011dynamic,jacobi2013phase,mckeon2018dynamic}. The introduced periodic perturbation is shown to alter the phase relation between large and small scales, and the associated modulation coefficient in a quasi-deterministic manner \citep{duvvuri2015triadic}. Temporal periodic perturbations can be also introduced by a wall jet \citep{bhatt2020linear,artham2021inner} or a wall-mounted piezoelectric actuator \citep{tang2019local,tang2020effect,tang2020local}. \rev{\citet{ranade2019turbulence}
introduced the perturbation outside (above) the boundary layer instead of at the wall.}  Their results support the existence of a critical layer inside the wake region that is responsible for the amplified level of turbulence in that region. \rev{\citet{lozierexperimental,lozier2020streamwise,lozier2021turbulent}  similarly introduced large-scale perturbations in the outer region of the boundary layer} through a dielectric barrier discharge (DBD) plasma actuator.  \rev{They then performed a triple decomposition with a} phase-locked velocity to obtain synthetic large-scale structures and  investigate their interactions with the residual turbulence.




In the above experiments, the dominant temporal frequency of the perturbation determines the frequency for the velocity decomposition and acts as an input in models of the phenomena. However, \rev{there remains a lack of understanding regarding a suitable choice of the streamwise wavenumber to e.g., specify a convective velocity, or for use in input-output based techniques that decompose the flow into a superposition over these wavenumbers}. \rev{As such, t}here have been a number of methods used to determine the streamwise wavenumber of interest. \citet{jacobi2011dynamic}  compared the phase-locked velocity measured in a TBL perturbed by dynamic roughness with predictions from resolvent analysis \citep{McKeon2010}. They determined \rev{a streamwise wavenumber for modeling of synthetic large-scale structures} based on a least-squares fit over  several downstream measurements. \rev{This method led to good agreement with the experimental data for that particular case, however, in general, spatially localized perturbations are} known to break the shift-invariance in the streamwise direction assumed in the resolvent model. Therefore the streamwise variation may be more accurately characterized by a complex wavenumber that would also capture downstream growth or decay \citep{jacobi2011dynamic,huynh2020characterization}. In addition,  a single frequency perturbation has been shown to be associated with a broad band of streamwise wavenumbers \citep{huynh2020characterization}. \rev{For example, the single frequency that is introduced through the perturbation will result in different streamwise wavenumbers at different wall-normal heights depending on their local mean velocity \citep{jacobi2011dynamic,huynh2020characterization}. Therefore, limiting the analysis to a single streamwise wavenumber may restrict the range of behaviors that can be studied. }

In this work, \rev{we develop a
 spatial input--output analysis approach that does not require specification of a single streamwise wavenumber. Our approach uses the integration method of \citet{towne2015one} } to reformulate the problem in terms of well-posed and exact one-way spatial evolution equations that inherently represents the behavior across the streamwise spectra. \rev{This reformulation also results in a natural embedding of a  wall-normal dependent phase speed that enables specification of a local (wall-normal direction dependent) convective velocity. We apply the proposed approach to analyze the phase-locked velocity and evolution of large-scale structures in 
a low Reynolds number TBL, where a synthetic large-scale structure is introduced through a spanwise-uniform DBD plasma actuator based on the experimental set-up in \cite{lozierexperimental,lozier2020streamwise,lozier2021turbulent}.} We first demonstrate \rev{the ability to produce phase-locked} velocities with large-scale structures reminiscent of those obtained by experimental measurements employing hot-wire anemometry and a phase-locked analysis. \rev{Both the theory and experiments indicate that the actuated large-scale structures become more inclined towards the wall as they propagate downstream, which is indicative of the changes in phase speed with distance from the wall}.  Quadrant analysis \citep{wallace1972wall,Wallace2016}\rev{ indicates that the shear stress distribution of the spatially evolving flow field shows an ordering (spatial progression of quadrant behaviors) consistent with that observed in turbulent pipe flows \citep{nagano1995coherent}. }

 \rev{The last part of this work exploits the analytical structure of the approach to take some steps towards characterizing the effect of varying the height and actuation frequency, thereby addressing a gap in the literature that has thus far focused on periodic perturbations injected at the wall \citep{hussain1970mechanics,hussain1972mechanics,reynolds1972mechanics,jacobi2011dynamic,duvvuri2016nonlinear,duvvuri2017phase,bhatt2020linear,artham2021inner,huynh2020characterization,tang2019local,tang2020effect,tang2020local} and  perturbations introduced in the outer layer \citep{ranade2019turbulence,lozierexperimental,lozier2020streamwise,lozier2021turbulent}.  Our studies indicate that the actuation frequency influences the characteristic streamwise length scale and that the response to higher frequency actuation decays faster in the downstream direction.  The ordering of the shear stress patterns from the quadrant analysis is found to be independent of actuator height. The perturbation location  instead determines a phase speed for flow structures associated with the local mean velocity at that height. In contrast, changes in  actuation frequency affects the most commonly occurring shear stress patterns in a manner consistent with canonical wall-bounded turbulence. These results support the notion that  synthetic large-scale structures interact with the TBL in a manner consistent with naturally occurring large-scale structures. They also 
 highlight the promise of combining such an analysis with experimental studies to provide further insight into scale interactions in the TBL.}

The remainder of this paper is organized as follows. Section \ref{sec:spatial_spatial_IO} describes the spatial input--output analysis framework. In Section \ref{sec:spatial_exp}, we present the experimental setup of interest and compare results obtained from a spatial input--output analysis using a model of the experimental actuation with the experimentally obtained data. Section \ref{sec:spatial_downstream} employs spatial input-output analysis to analyze the downstream evolution of actuated large-scale structures, as well as the influence of actuation frequency and wall-normal height on the actuated large-scale structures. Section \ref{sec:spatial_spatial_conclusion} concludes this paper.

\section{Spatial input--output analysis of an actuated turbulent boundary layer}
\label{sec:spatial_spatial_IO}

We consider \rev{incompressible zero-pressure-gradient turbulent boundary layer}, where $x,y,z$ are the streamwise, wall-normal, and spanwise directions, respectively. In order to approximate boundary layer flow, we invoke the quasi-parallel assumption that the streamwise variation of mean velocity is negligible, which is quantitatively shown to be a reasonable assumption by spatio-temporal measurement \rev{in} \citep{huynh2020characterization}. We decompose the velocity field, $\boldsymbol{u}_{\text{tot}} = \begin{bmatrix} u_{\text{tot}} & v_{\text{tot}} & w_{\text{tot}}\end{bmatrix}^{\text{T}}$ and the pressure field, $p_{\text{tot}}$ into mean and fluctuating quantities $\boldsymbol{u}_{\text{tot}} = \bar{u}(y)\boldsymbol{e}_x + \boldsymbol{u}$ and $p_{\text{tot}} = \bar{p} + p$, where $^{\text{T}}$ denotes the transpose, $\boldsymbol{e}_x$ denoting the streamwise unit vector and, overbars represent time-averaged quantities, $\bar{\phi} = \underset{T\rightarrow\infty}{\text{lim}} \frac{1}{T}\int_{0}^{T} \phi(t)\,dt$.

\begin{figure}
    \centering
    \includegraphics[width=0.6\textwidth]{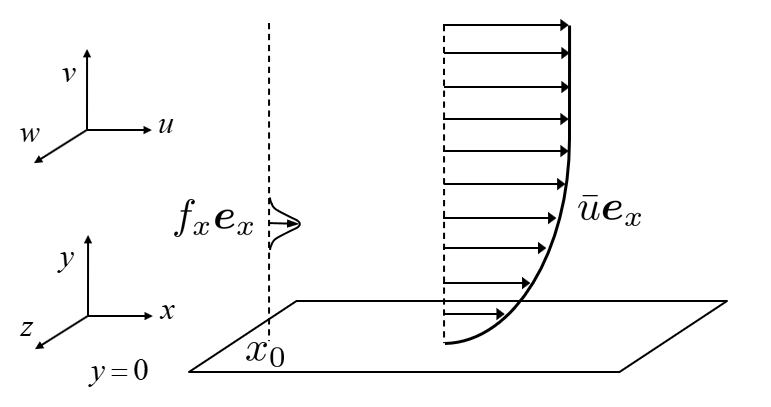}
    \caption{Illustration of  spatially localized actuation applied to a turbulent boundary layer.}
    \label{fig:spatial_figure1_illustration}
\end{figure}

We are interested in the flow response of turbulent boundary layer subject to a localized temporally harmonic oscillation at a fixed frequency applied using a spanwise uniform DBD plasma actuator. We model the effect of the plasma actuation as a streamwise body force $f_x \boldsymbol{e}_x$ and neglect any induced body forces in the wall-normal or spanwise directions. \rev{The use of this type of model is supported by the observation that plasma-induced body force typically shows a much smaller wall-normal forcing than streamwise forcing \citep{kotsonis2011measurement}.} This forcing is localized; i.e., applied at a particular wall-normal  and streamwise location as illustrated in figure \ref{fig:spatial_figure1_illustration}, the form of the forcing prescribed for this study is described in detail in \S\ \ref{subsec:spatial_model_phase_lock}.

The dynamics of the forced fluctuations $\boldsymbol{u}$ and $p$ linearized around the turbulent mean velocity are governed by:
\begin{subequations} \label{eq:spatial_NS_All}
\begin{align}
\partial_{t} \boldsymbol{u}  
+  \bar{u}\partial_x \boldsymbol{u} + v  \frac{d\bar{u} }{dy}\boldsymbol{e}_x +\frac{\boldsymbol{\nabla} p}{\rho} 
-\nu {\nabla}^2 \boldsymbol{u}
 &=f_x\boldsymbol{e}_x, \label{eq:spatial_NSDecompf1} \\
\boldsymbol{\nabla} \cdot \boldsymbol{u}&=0. \label{eq:spatial_NSDecompf2}
\end{align}
\end{subequations} Here, $\rho$ is the density of the fluid, $\nu$ is kinematic viscosity. The friction Reynolds number is defined as $Re_{\tau} :=\delta_{99} u_{\tau}/\nu$, where $\delta_{99}$ is the boundary layer thickness and the friction velocity is defined as \rev{$u_\tau \equiv \sqrt{\bar{\tau}_{\text{w}}/\rho }$}, and \rev{$\bar{\tau}_{\text{w}}$} is the time-averaged mean shear stress at the wall. We denote the velocity normalized by the friction velocity with a superscript $^+$; i.e., $u^+=u/u_\tau$. We also use superscript $^+$ to denote the length normalized by the inner unit length scale $\delta_v:=\nu/u_\tau$ and the time normalized by $\delta_v/u_\tau$; i.e., $y^+=y/\delta_v$ and $t^+=tu_\tau/\delta_v$. 

We next derive a spatial mapping operator to obtain the solution to spatial input-output by assuming solutions of the form: 
\begin{align}
\psi(x,y,z,t)=\check{\psi}(x,y;\omega,\eta, k_z)\text{e}^{\text{i}(k_z z-\omega t) }\text{e}^{\eta t},
\label{eq:spatial_FTdefn}
\end{align} 
where $k_z = 2\pi/\lambda_z$ is spanwise wavenumber and $\text{i}=\sqrt{-1}$ is the imaginary unit. $\eta$ and $\omega$ respectively denote temporal growth rate and frequency, where $\eta$ is introduced here for partitioning the upstream and downstream modes following \citet{towne2015one} based on \citet{briggs1964electron}'s criteria. These assumptions allow us to rewrite equation \eqref{eq:spatial_NS_All} as:
\begin{align}
        \frac{\partial }{\partial x}\boldsymbol{\check{q}}_S&=\boldsymbol{\check{A}}_S\boldsymbol{\check{q}}_S+\boldsymbol{\check{B}}_{S,x}{\check{f}}_x,\label{eq:spatial_spatial_ABC}
\end{align}
where $\boldsymbol{\check{q}}_S:=\begin{bmatrix}
    \check{u}&
    \check{v}&
    \check{v}_x&
    \check{w}&
    \check{w}_x&
    \check{p}/\rho
    \end{bmatrix}^{\text{T}}$, \rev{$\check{v}_x:=\frac{\partial \check{v}}{\partial x}$, $\check{w}_x:=\frac{\partial \check{w}}{\partial x}$,} and the operators $\boldsymbol{\check{A}}_S$ and $\boldsymbol{\check{B}}_{S,x}$ are respectively defined as:
    
\begin{align}
    \boldsymbol{\check{A}}_S(y;\omega,\eta,k_z)&:=\begin{bmatrix}
    0 & -\partial_y & 0 & -\text{i}k_z & 0 & 0\\
    0 & 0 & 1 & 0 & 0 & 0\\
    0 & - M/\nu & \bar{u}/\nu & 0 & 0 & \partial_y/\nu\\
    0 & 0 & 0 & 0 & 1 & 0\\
    0 & 0 & 0 & -M/\nu &\bar{u}/\nu & \text{i}k_z /\nu\\
    M & -\frac{d\bar{u}}{dy}+\bar{u}\partial_y & -\nu\partial_y  & \text{i}k_z \bar{u} & - \text{i}k_z\nu & 0
    \end{bmatrix},\;\boldsymbol{\check{B}}_{S,x}:=\begin{bmatrix}
    0 \\
     0\\
    0 \\
    0 \\
    0 \\
    1
    \end{bmatrix}
    \label{eq:spatial_A_S_B_S}
\end{align}
with 
\begin{align}
    M:=\nu(\partial_y^2-k_z^2)-\eta+\text{i}\omega.
\end{align} 
The operator similar to $\boldsymbol{\check{A}_S}$ in equation \eqref{eq:spatial_A_S_B_S} is previously defined in \citet[equations (7.110)- (7.111)]{schmid2012stability}. We impose boundary conditions of the form:
\begin{subequations}
\begin{align}
    \check{u}(y=0)=&\check{u}(y=\infty)=0,\\
    \check{v}(y=0)=&\check{v}(y=\infty)=0,\;\;\text{and}\\
    \check{w}(y=0)=&\check{w}(y=\infty)=0,
\end{align}
\end{subequations}
which correspond to no-slip at the wall and no fluctuation at the free-stream location.

In order to obtain the solution to equation \eqref{eq:spatial_spatial_ABC}, we need to first identify the upstream and downstream modes contained in $\boldsymbol{\check{A}}_S(y;\omega,\eta=0,k_z)$ to eliminate numerical instabilities associated with upstream decaying modes  growing in the downstream direction. Here, we implement the one-way spatial equation \citep{towne2015one} to explicitly identify upstream modes based on \citet{briggs1964electron}'s criteria; see e.g., \citep[Section 3.3]{briggs1964electron,huerre1990local,towne2015one,huerre2000open}. Following \citet{towne2015one}, we identify the eigenvalue associated with $\text{i}k_x(\omega,\eta=0,k_z)$ of $\boldsymbol{\check{A}}_S(y;\omega,\eta=0,k_z)$ by tracking the eigenvalues $\text{i}k_x(\omega,\eta,k_z)$ of $\boldsymbol{\check{A}}_S(y;\omega,\eta,k_z)$ as a function of $\eta$. This mode $k_x(\omega,\eta=0,k_z)$ is propagating downstream if
\begin{align}
    \underset{\eta\rightarrow+\infty}{\text{lim}}\mathbb{I}\text{m}[k_x(\omega, \eta, k_z)]=&+\infty,
    \label{eq:spatial_eig_downstream}
\end{align}
and propagating upstream if
\begin{align}
    \underset{\eta\rightarrow+\infty}{\text{lim}}\mathbb{I}\text{m}[k_x(\omega, \eta, k_z)]=&-\infty,
    \label{eq:spatial_eig_upstream}
\end{align}
where $\mathbb{I}\text{m}[\cdot]$ represents the imaginary part of the argument. We then perform an eigenvalue decomposition 
\begin{align}
    \boldsymbol{\check{A}}_S(y;\omega,\eta=0,k_z)
    =&\boldsymbol{V} \boldsymbol{\Lambda} \boldsymbol{V}^{-1},
\end{align}
where the diagonal elements of the diagonal matrix $\boldsymbol{\Lambda}$,  and matrix $\boldsymbol{V}$ respectively  contain the eigenvalues and eigenvectors of $\boldsymbol{\check{A}}_S(y;\omega,\eta=0,k_z)$. We then eliminate the upstream modes by defining an $x$ dependent matrix $\boldsymbol{D}$ as
\begin{align}
    \boldsymbol{D}_{ii}(x)=\begin{cases}
    e^{\boldsymbol{\Lambda}_{ii}x},\;\;&\text{if}\;\boldsymbol{\Lambda}_{ii}\; \text{is an eigenvalue corresponding to a downstream mode},\\
    0,\;\;&\text{if}\;\boldsymbol{\Lambda}_{ii}\;\text{is an eigenvalue corresponding to an upstream mode},
    \end{cases}
\end{align}
where the subscript $ii$ represents the $i^{\text{th}}$ diagonal element of the matrix $\boldsymbol{D}$ or $\boldsymbol{\Lambda}$. The operator mapping the state $\boldsymbol{q}_S(x_0,y;\omega,k_z)$ at $x=x_0$ to the state  $\boldsymbol{q}_S(x_m,y;\omega,k_z)$ at another downstream location $x=x_m$ under the same spatio-temporal wavenumber-frequency pair $(\omega,k_z)$; i.e., $\boldsymbol{q}_S(x_m,y;\omega,k_z)=\boldsymbol{\check{\Psi}}(x_m,x_0,y;\omega,k_z)\boldsymbol{q}_S(x_0,y;\omega,k_z)$ is then given by:
\begin{align}
    \boldsymbol{\check{\Psi}}(x_m,x_0,y;\omega,k_z):=&\boldsymbol{V}\boldsymbol{D}(x_m-x_0)\boldsymbol{V}^{-1}. 
    \label{eq:state_transition}
\end{align}

Using \eqref{eq:state_transition}, we can obtain the state response $\boldsymbol{\check{q}}_S(x_m,y;\omega, k_z)$ at given frequency-wavenumber pair $(\omega,k_z)$ and certain downstream location $x_m$ due to an input forcing function $\check{f}_x(x,y; \omega, k_z)$ with $\boldsymbol{\check{q}}_S(x_0,y;\omega, k_z)=0$ as 
\begin{align}
    \boldsymbol{\check{q}}_S(x_m,y;\omega, k_z)=\int_{x_0}^{x_m}\boldsymbol{\check{\Psi}}(x_m,x,y;\omega,k_z)\boldsymbol{\check{B}}_{S,x}\check{f}_x(x,y;\omega,k_z) d x.
\end{align}
With the additional definition of an output operator $\boldsymbol{\check{C}}$, we can  obtain the response of an output variable $\boldsymbol{\check{\phi}}(x_m, y;\omega, k_z)$ for a given $(\omega,k_z)$ pair and downstream location $x_m$:
\begin{align}
    \boldsymbol{\check{\phi}}(x_m, y;\omega, k_z)=\boldsymbol{\check{C}}\boldsymbol{\check{q}}_S(x_m,y;\omega, k_z).
\end{align}
\rev{The formulation here is general for a wide range of $k_z$ that can be determined based on actuator geometry, while we focus on the case $k_z=0$ representing spanwise uniform actuation corresponding to the experimental setup of interest in this paper.}

\subsection{Numerical method}
\label{subsec:spatial_numerical}
We compute the spatial mapping matrix associated with the operator in \eqref{eq:state_transition} by first discretizing the operators in equation \eqref{eq:spatial_A_S_B_S} using the Chebyshev differential matrices generated by the MATLAB routines of \citet{Weideman2000}. The mean profile $\bar{u}$ employed in this work is the asymptotic consistent turbulent boundary layer profile obtained from \citet{monkewitz2007self} as detailed in Appendix \ref{sec:spatial_appendix_TBL_mean}. The numerical implementation of the spatial framework is validated against the results of the spatial eigenvalue problem in \citet[figure 7.18]{schmid2012stability}. We implement algebraic stretching following \citet[equations (A.53)-(A.54)]{schmid2012stability}, and this stretched grid is validated against eigenvalue results for the Blasius boundary layer in \citet[Table A.4]{schmid2012stability}. We use $N_y=82$ grid points in the range $y^+\in[0,1690]$ with half of the grid points in the range of $y^+\in[0,345]$. \rev{This resolution was deemed sufficient by verifying that the relative difference between the results reported and those obtained when the number of  grid points is increased to $N_y=122$ is less than 1\%.} We identify upstream and downstream modes in equations \eqref{eq:spatial_eig_downstream}-\eqref{eq:spatial_eig_upstream} through the \texttt{eigenshuffle} \citep{Errico2020eigenshuffle} function, which tracks the variation of each eigenvalue numerically based on its continuity with varying parameter $\eta$. This numerical method is selected because analytical tracking is typically challenging; see e.g., \citet{alves2019identifying}. For results in this work, we use 60 logarithmically spaced values in the range $\eta^+\in[10^{-3},10]$ to approximate $\eta\rightarrow \infty$ in equations \eqref{eq:spatial_eig_downstream}-\eqref{eq:spatial_eig_upstream}. We verified that this is sufficient by checking that the results do not change if we increase this to 90 logarithmically spaced values in the range $\eta^+\in [10^{-4},10^2]$. \rev{These two grids on $\eta^+$ also give the same eigenvalue of the operator $\boldsymbol{\check{A}}_S(y;\omega,\eta=0,k_z)$ associated with the largest real part for the set of downstream eigenmodes.}

\section{Comparison with experimental results}
\label{sec:spatial_exp}

In this section, we compare the large-scale structures from the model with those obtained from experimental measurements. We describe the experimental setup in subsection \ref{subsec:spatial_exp_setup}. Then in subsection  \ref{subsec:spatial_model_phase_lock}, we describe the model calibration and provide a comparison between the phase locked velocity obtained through the analytical approach and the experimental results.

\subsection{Experimental setup and phase-locked decomposition}
\label{subsec:spatial_exp_setup}
\begin{figure}
    \centering
    \includegraphics[width=0.6\textwidth]{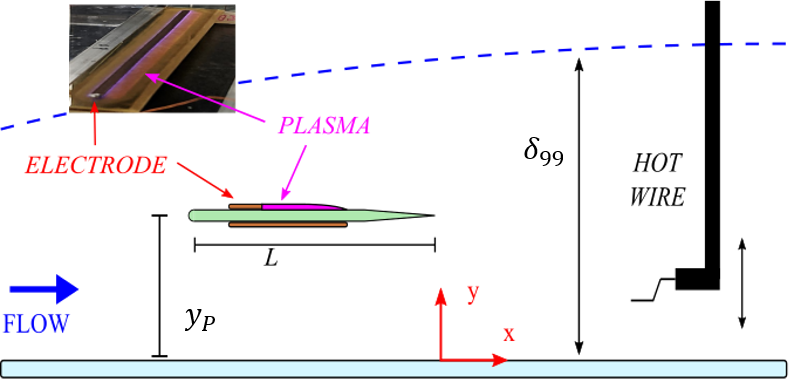}
    \caption{Schematic of the experimental set-up in which an active large-scale structures actuator (ALSSA) comprised of a dielectric barrier discharge plasma  actuator is mounted on a plate at a prescribed vertical height $y_p$.}
    \label{fig:spatial_actuated_TBL}
\end{figure}

The experiments are performed in one of the low-turbulence, subsonic, in-draft wind tunnels located at the Hessert Laboratory for Aerospace Research at the University of Notre Dame. The wind tunnel has an inlet contraction ratio of 6:1 and a series of 12 turbulence-management screens in front of the inlet to achieve tunnel free stream turbulence levels of less than 0.1\% (0.06\% for frequencies above 10 Hz). Experiments are performed in a test section with a $0.610$~m square cross-section and is \rev{$1.83$~m} long.  A schematic of the full experimental set-up is shown in Figure \ref{fig:spatial_actuated_TBL}.

For this study, a two-meter long boundary layer development  plate with a distributed roughness element attached to the leading edge is installed in the central height of the tunnel test section. \rev{The leading edge of the boundary layer development plate is aligned with the test section inlet, while its trailing edge goes into the diffuser. The inlet section of the diffuser matches the test section so that $\approx$17~cm of boundary layer development plate is inside the diffuser; this is not expected to influence the development of the turbulent boundary layer within the test section.} A constant temperature anemometer (CTA) with a single boundary layer hot-wire probe (Dantec 55P15) with diameter $5$~$\mu m$ and length $l= 1.25$~mm is used to collect time-series measurements of the streamwise velocity component. A computer-controlled traversing stage is inserted through the top wall of the tunnel along the midpoint of the tunnel span to allow the hot-wire probe to traverse the test section and make measurements at different wall-normal $(y)$ locations. 
The streamwise position of the hot-wire probe traverse system is adjustable and the following four streamwise locations are selected for this study:  $x=51$~mm, $102$~mm, $170$~mm, and $272$~mm, which correspond to $1.5\delta_{99}$, $3\delta_{99}$, $5\delta_{99}$, and $8\delta_{99}$, respectively, based on the experimentally measured boundary layer thickness, $\delta_{99}$, near the actuator trailing edge. \rev{The wall-normal position of the hot-wire probe varies between $y/\delta_{99}=0.0069$ and $y/\delta_{99}=0.9724$ for a total of 21 sampling points.}
\rev{The data was sampled at $f_s=30$ kHz which corresponds to $\Delta t^+=(1/f_s)u_\tau^2/\nu=0.2$ for a total period of 90 seconds, or about 15,000 $\delta_{99}/U_\infty$ in each test. With this sampling frequency and sampling time, there should be no loss of turbulence information as described in \citep{hutchins2009hot}.}

A plasma-based active large-scale structure actuator (ALSSA) device is used  to modify the dynamics of the outer layer of the boundary layer with periodic plasma-induced force.
This device is attached to the top side of the boundary layer development plate at a fixed streamwise location of $140$~cm from the leading edge of the boundary layer development plate, as shown in Figure \ref{fig:spatial_actuated_TBL}.  The plasma actuator is supported above the boundary layer development plate by vertical, symmetrical NACA0010 airfoils on both sides. These airfoils are $4$~mm thick, have a $50$~mm wide chord.  The plates are made at height intervals, $y_p$, at $10$~mm  $(0.3\delta_{99})$ so that  the synthetic large-scale structures can be introduced into the TBL at different heights. The plasma actuator is $W = 25$~cm ($\approx $8 $\delta_{99}$) wide in the spanwise direction and $L = 32$~mm ($\approx$1 $\delta_{99}$) long in the streamwise direction. The actuator plate is made of a $2$~mm thick sheet of Ultem dielectric polymer. An upper surface electrode of $0.05$~mm thick copper foil tape is located 15~mm from the plate leading edge and is 4~mm in length and 22~cm in width. On the lower surface, a second copper foil electrode is located 15~mm from the leading edge in line with the top electrode and is 12~mm in length and 22~cm in width. The corners of the electrodes are rounded, and they are mounted in alignment to eliminate extraneous regions of plasma generation and regions of highly concentrated plasma. The leading edge of the actuator plate is rounded, and the last 10~mm of the trailing edge is linearly tapered to reduce the separation region behind the trailing edge of the plate. The alternating current dielectric barrier discharge (AC-DBD) plasma formed on the actuator is produced using a high voltage AC source consisting of a function generator, power amplifiers, and a transformer \citep{thomas2009optimization}. The electrodes placed on the top and bottom of the actuator are connected to the high voltage AC source which provides a 40~kV peak-to-peak sinusoidal waveform excitation to the electrodes at a frequency of 4~kHz. The peak-to-peak voltage is maintained within -5\% of the expected excitation voltage during experiments. At the 4~kHz carrier frequency, the plasma actuator operates in a quasi-steady mode, essentially creating a spanwise-uniform steady jet in the streamwise direction. To introduce periodic forcing with frequency $f_p$, the sinusoidal waveform is modulated by a square wave with a fifty percent duty cycle. \rev{Previous analysis demonstrates that this form of square wave forcing does not produce a significant TBL response at frequencies besides $f_p$, see e.g. \citet[figure 3]{lozier2020streamwise}.}

\begin{table}
    \centering
    \begin{tabular}{ccccccc}
    \hline
        $\delta_{99}$ & $U_{\infty}$ & $u_{\tau}$ & $C_f$ & $H$ & $Re_{\theta}$ & $Re_{\tau}$  \\
        \hline
        $34.8$~mm & $7$~m/s & $0.298$~m/s & $0.0036$ & 1.33 & $1857$ & $690$ \\
        \hline
    \end{tabular}
    \caption{Turbulent boundary layer parameters from the experiment, measured at  $x=5\delta_{99}$.}
    \label{tab:experimental_parameter_TBL}
\end{table}

\begin{figure}
    \centering
    (a) \hspace{0.5\textwidth} (b)
    
    \includegraphics[width=0.49\textwidth]{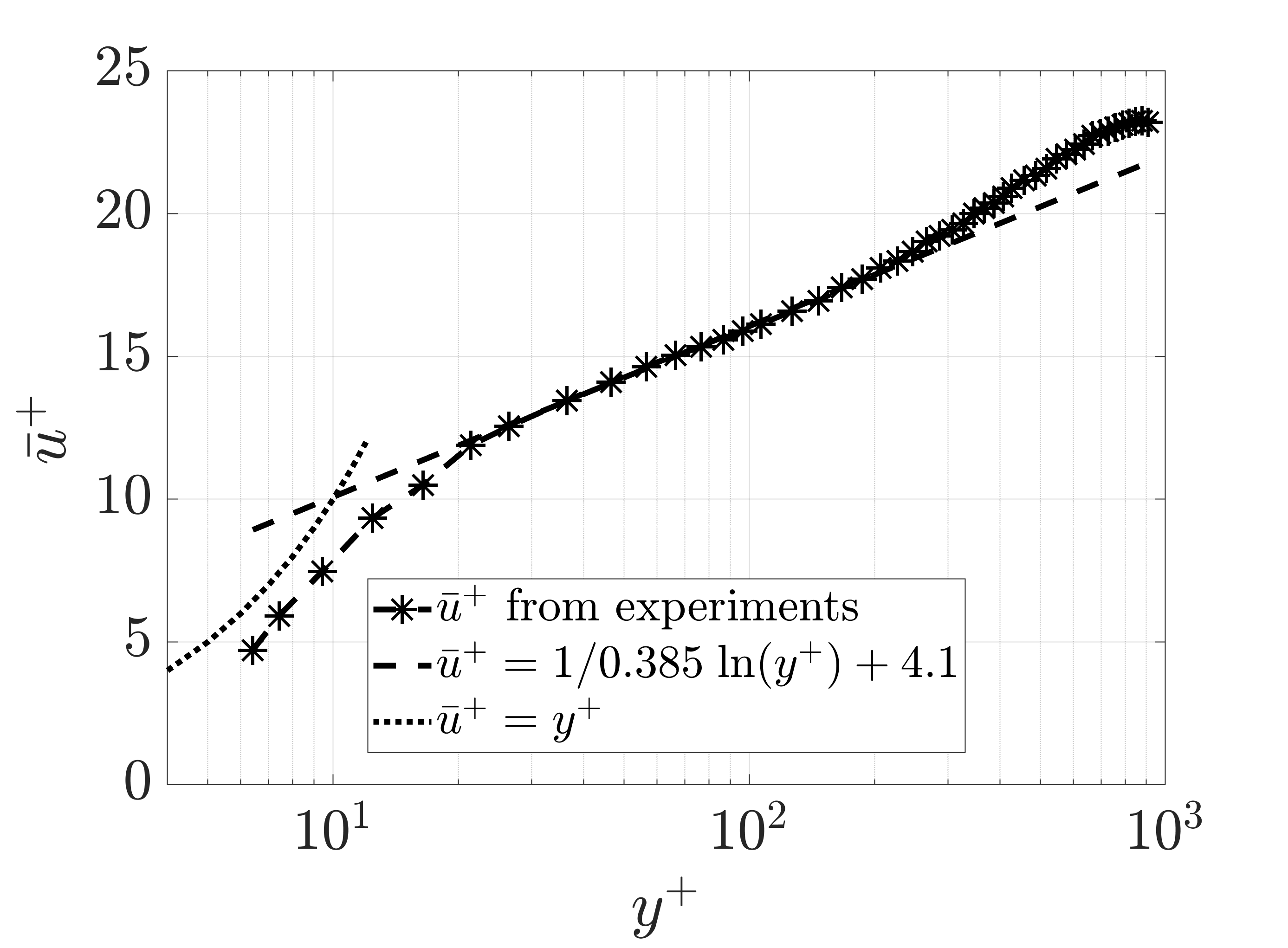}
    \includegraphics[width=0.49\textwidth]{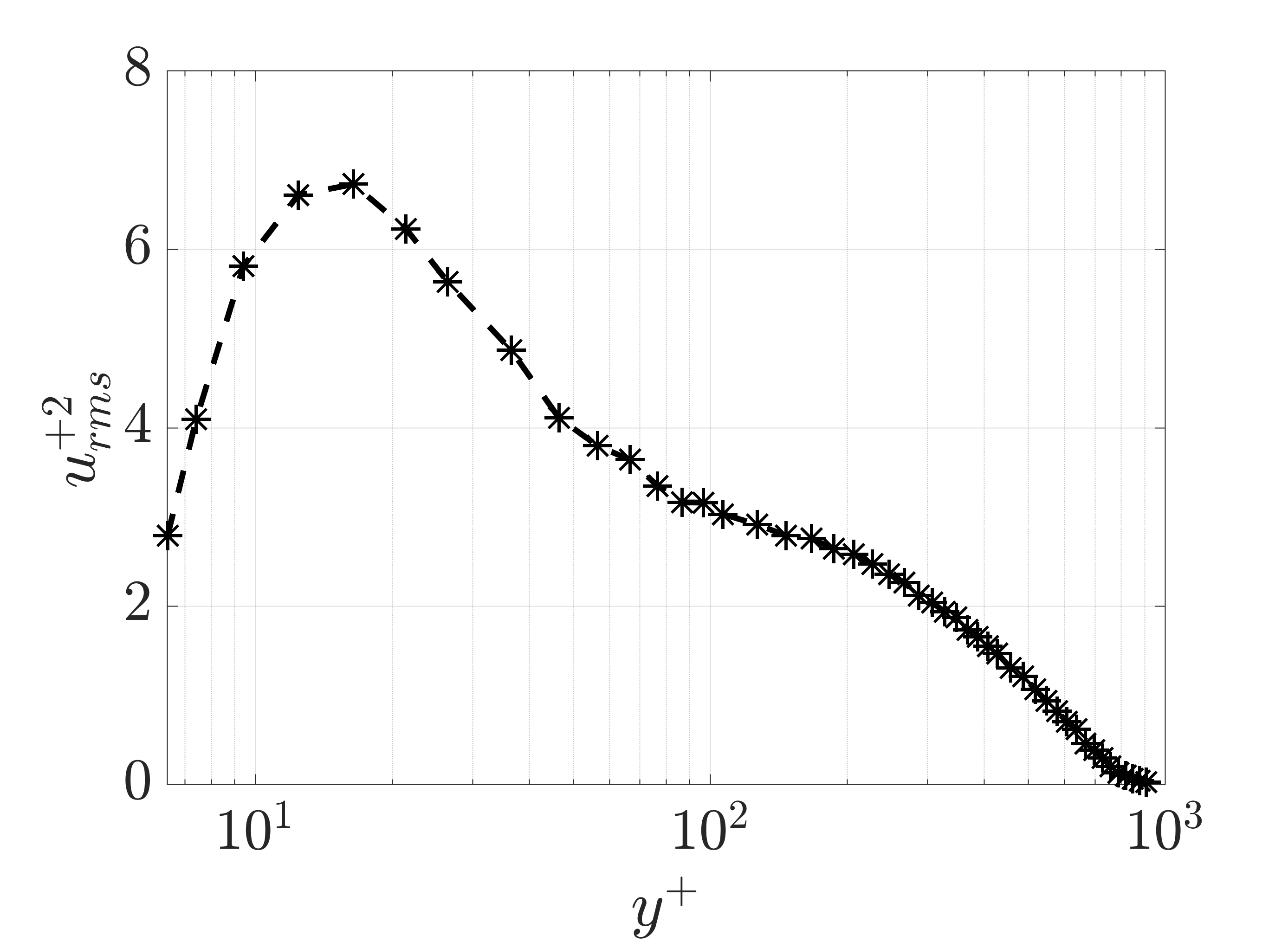}
    \caption{Experimentally measured (a) mean velocity $\bar{u}^+$ and (b) mean square of the streamwise velocity fluctuation $u^{+2}_{rms}$ for the canonical turbulent boundary layer at $x_m/\delta_{99}=8$. }
    \label{fig:canonical_TBL_X8}
\end{figure}
 
The measured velocity time series are then processed by a narrow bandstop filter around 4~kHz to eliminate electronic noise associated with the high voltage AC source supplying the actuator.  Since the actuator introduced periodic forcing into the flow, it is convenient to phase-lock the results to the actuation frequency. To do so, a triple phase-locked Reynolds decomposition of the velocity is considered, as shown in equation \eqref{eq:spatial_triple_decomposition} where $u$ is the instantaneous velocity, $\bar{u}$ is the time mean component of velocity, $\tilde{u}$ is a phase-locked modal velocity component, $u'$ is a residual fluctuating turbulent component, $\phi$ is the phase, defined by the relationship in equation \eqref{eq:spatial_triple_decomposition_phase}, and $n$ is the number of realizations as described below
\begin{subequations}
\label{eq:phase_lock_exp}
\begin{align}
    u(y,t)=&\bar{u}(y)+\tilde{u}(y,\phi)+u'(y,\phi,n),\label{eq:spatial_triple_decomposition}\\
    \phi=&\left(\frac{t_n}{T_p}-n \right)2\pi.
    \label{eq:spatial_triple_decomposition_phase}
\end{align}
\end{subequations}
Here, $t_n$ is a time in the $n^{\text{th}}$ realization, which is related to the phase angle, $\phi$, by the period of the forcing repetition cycle, $T_p=1/f_p$. The output of the function generator is used to ensure the data is phase-locked with the repetition cycle of the plasma. These $n$ realizations are then ensemble-averaged to find the modal component of velocity, $\tilde{u}(y,\phi)$, as a function of the phase angle.

 \rev{A set of representative characteristic parameters of the canonical turbulent boundary layer are measured at the downstream location of $x_m=5\delta_{99}$ using the hot-wire probe, and the associated data is summarized in Table \ref{tab:experimental_parameter_TBL} for reference.} The skin friction velocity $u_\tau$ is found using the Clauser method and the friction Reynolds number is $Re_\tau=690$. \rev{We use this $u_\tau$ to normalize all of the experimental results. All of the experiments employ a} wind tunnel free stream velocity is $7$~m/s that is measured to be within $\pm1\%$ of the expected free stream velocity before each test. \rev{The mean velocity and mean square of the streamwise velocity fluctuations of canonical TBL at $x_m/\delta_{99}=8$ shown in Figure \ref{fig:canonical_TBL_X8} demonstrate that the boundary layer is fully developed at this location. Additional  statistics for this experimental setup are reported in \citep{lozier2022PIV}. }

\subsection{Model calibration and comparison results}
\label{subsec:spatial_model_phase_lock}

In this subsection, we will describe the forcing model and calibrate the parameters of the forcing function $\check{f}_x$ in equation \eqref{eq:spatial_spatial_ABC} to closely match the effect of the plasma actuation on the flow field.  We will then compare the computed results to those from the experimental measurements to demonstrate the efficacy of the spatial input-output analysis described in section~\ref{sec:spatial_spatial_IO} in reproducing the phase-locked velocity.

Based on the actuator geometry described in \S\ \ref{subsec:spatial_exp_setup}, we model the effect of actuation on the flow by assuming the streamwise body force $\check{f}_x$ is in the form of a Gaussian function over the wall-normal direction, a Dirac delta function over the streamwise direction, and a uniform function in the spanwise direction:
\begin{align}
    \check{f}_x(x,y;\omega,k_z)=F_0\text{e}^{-\frac{(y-y_f)^2}{2\sigma_p^2}}\delta(x-x_0)\text{e}^{\text{i}\phi_0},
    \label{eq:spatial_body_force}
\end{align}
where $F_0$ represents the magnitude of this body force and $\phi_0$ represents the initial phase induced by the plasma actuator. We select the initial phase of the body force model as $\phi_0=1.15\pi$ and the magnitude as $F_0^+=38.2$ based on experimental measurements of phase-locked velocity at $x_m=1.5\delta_{99}$. The values of the parameters $\phi_0$ and $F_0^+$ do not influence the shape of phase-locked velocity due to linearity. In analogy to the vibrating ribbon problem \citep{ashpis1990vibrating} in the study of transitional boundary layers or the signaling problem \citep[Section 3]{huerre2000open,huerre1985absolute}, the streamwise variation of this body force in \eqref{eq:spatial_body_force} is modeled as a Dirac delta $\delta(x-x_0)$ function over the streamwise direction that is localized at the streamwise position $x_0$. Here, we impose $x_0=0$. The Gaussian function in the wall-normal direction is motivated by \citep{jovanovic2001spatio,vadarevu2019coherent}, where this function is also employed to model localized forcing. The parameters $y_f$ and $\sigma_p$ in the Gaussian function are, respectively, the center of the peak and standard deviation determining the wall-normal shape of plasma-induced body force. We set $\sigma_p^+=60$ and the body force center to be $y_f=0.13\delta_{99}+y_p$, i.e., $0.13\delta_{99}$ higher than the actuator plate height. This height correction and the standard deviation $\sigma_p^+$ of the forcing function are selected in order to match the ALSSA device induced peak phase-locked velocity at $x_m/\delta_{99}=1.5$ from the experiments. The calibrated values $F_0^+$, $\phi_0$, $y_f$, and $\sigma_p^+$ are then kept constant. The spanwise wavenumber in equation \eqref{eq:spatial_A_S_B_S} is set to  $k_z=0$ because the plasma actuation in the experiment is spanwise-uniform, and the fact that the experimental measurements of flow response do not show significant spanwise variation. We set the frequency to $\omega^+=2\pi f_p^+$ to match that of the plasma actuation. We specify the Reynolds number $Re_\tau=690$ to match the experimental conditions described in table \ref{tab:experimental_parameter_TBL} in both the determination of the mean velocity profile and the computations. \rev{
The actuator plate in the experimental setup will also introduce a wake leading to a velocity deficit in the mean flow, but this wake will decay as the flow continues downstream \citep{lozier2021turbulent}. The clear peak in the premultiplied streamwise energy spectrum with plasma actuation suggests a much stronger effect of plasma actuation than the actuator plate \citep[figure 4]{lozier2022PIV}. We therefore make the simplification of neglecting the effect of the actuator plate and employ the canonical TBL mean velocity profile here. The use of a canonical TBL mean velocity profile does not require experimental measurements to specify the mean profile and provides greater flexibility to explore other flow regimes or actuation schemes.  Incorporating a spatially developing mean profile requires modifications to the formulation or recalculation at each location of interest and both increase computational time. These computations along with a study of the trade-off between additional accuracy and computational time are beyond the current scope.}

The corresponding solution of equation \eqref{eq:spatial_spatial_ABC} at downstream measurement position $x_m$ with respect to the streamwise localized forcing $\check{f}_x$ can be computed using the spatial mapping operator in \eqref{eq:state_transition}:
\begin{align}
\label{eq:spatial_frequency_response}
    \check{\boldsymbol{q}}_S(x_m,y;\omega,k_z)=\boldsymbol{\check{\Psi}}(x_m,x_0,y;\omega,k_z)    \boldsymbol{\check{B}}_{S,x} F_0 \text{e}^{-\frac{(y-y_f)^2}{2\sigma_p^2}}\text{e}^{\text{i}\phi_0}. 
\end{align}
In order to compare with the hot-wire measurements from the experimental set up described in \ref{subsec:spatial_exp_setup}, we select the streamwise velocity as the output, i.e.
\begin{subequations}
\begin{align}
     \check{u}=&\boldsymbol{\check{C}}_{S,u}\boldsymbol{\check{q}}_S,\\ 
     \boldsymbol{\check{C}}_{S,u}:=&\begin{bmatrix}
    1 & 0 & 0 & 0 & 0 & 0 
    \end{bmatrix}.
\end{align}
\end{subequations}
We then obtain the phase-locked velocity defined  \eqref{eq:phase_lock_exp} at a certain downstream measurement location $x_m$ by multiplying $\text{e}^{-\text{i}\phi}$ to shift the phase:
\begin{align}
    \tilde{u}(x_m,y;\omega,k_z,\phi)=\mathbb{R}\text{e}[\check{u}(x_m,y;\omega,k_z)\text{e}^{-\text{i}\phi}],
    \label{eq:spatial_phase}
\end{align}
where $\mathbb{R}\text{e}[\cdot]$ represents the real part of the argument. Note that the minus sign of $\text{e}^{-\text{i}\phi}$ in equation \eqref{eq:spatial_phase} is based on the fact that an increased phase corresponding to later time moments is consistent with the phase-locked analysis in equation \eqref{eq:phase_lock_exp}   and the Ansatz in equation \eqref{eq:spatial_FTdefn}.

\begin{figure}
    \centering
    
    (a)$x_m/\delta_{99}=1.5$ \hspace{0.11\textwidth} (b)$x_m/\delta_{99}=3$ \hspace{0.11\textwidth} (c)$x_m/\delta_{99}=5$  \hspace{0.11\textwidth} (d)
    $x_m/\delta_{99}=8$
    
    \includegraphics[width=0.24\textwidth]{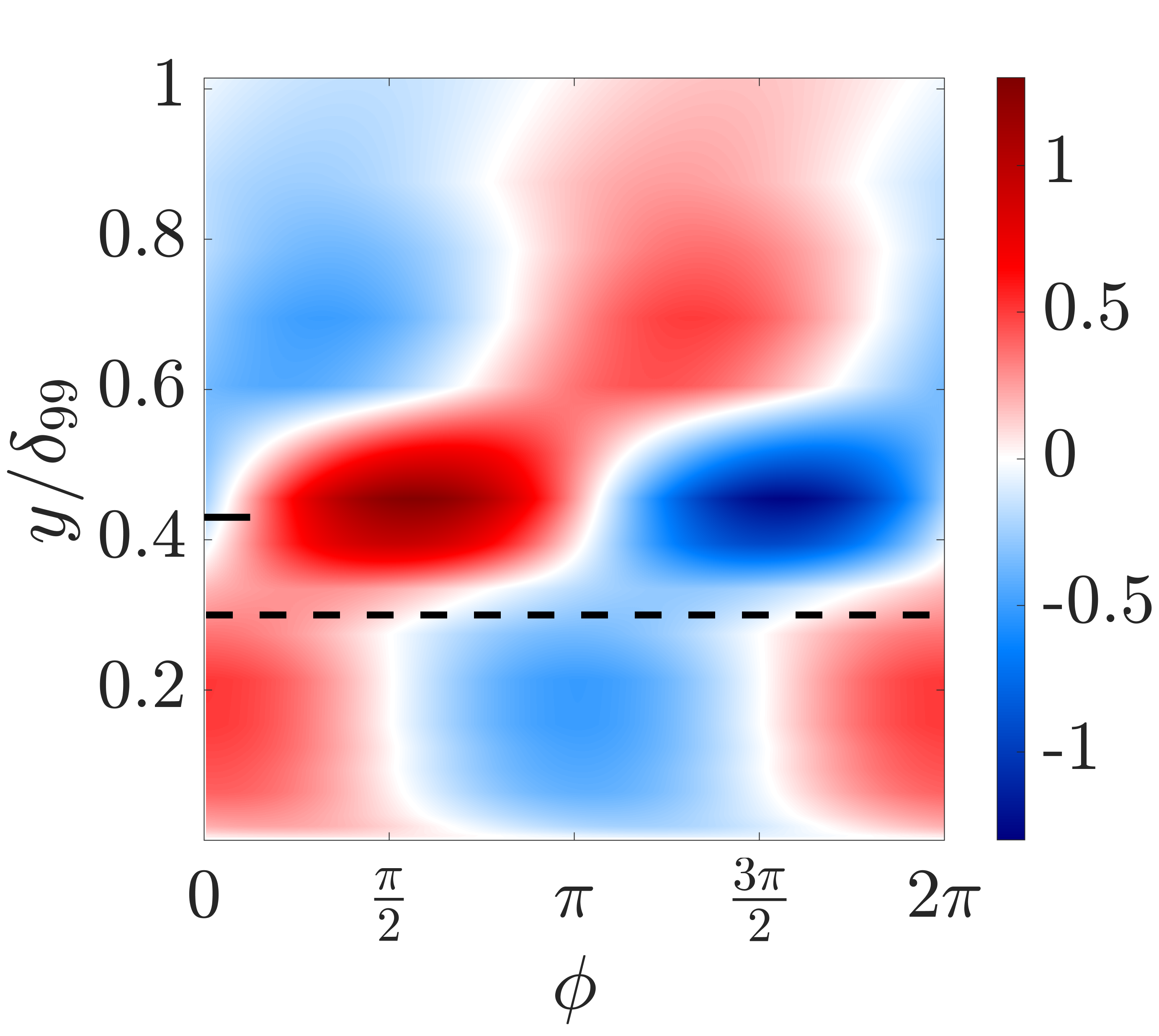}
    \includegraphics[width=0.24\textwidth]{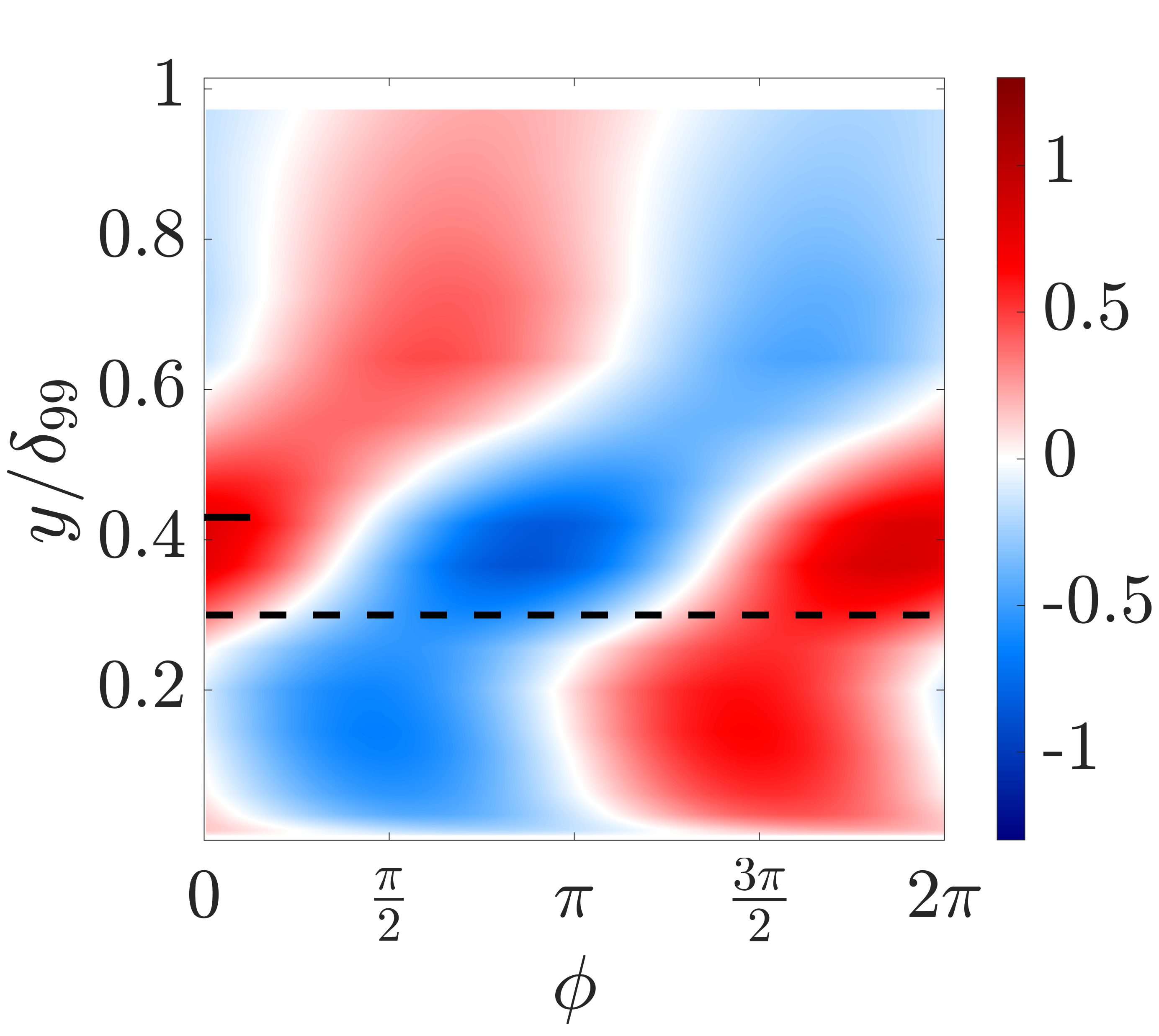}
    \includegraphics[width=0.24\textwidth]{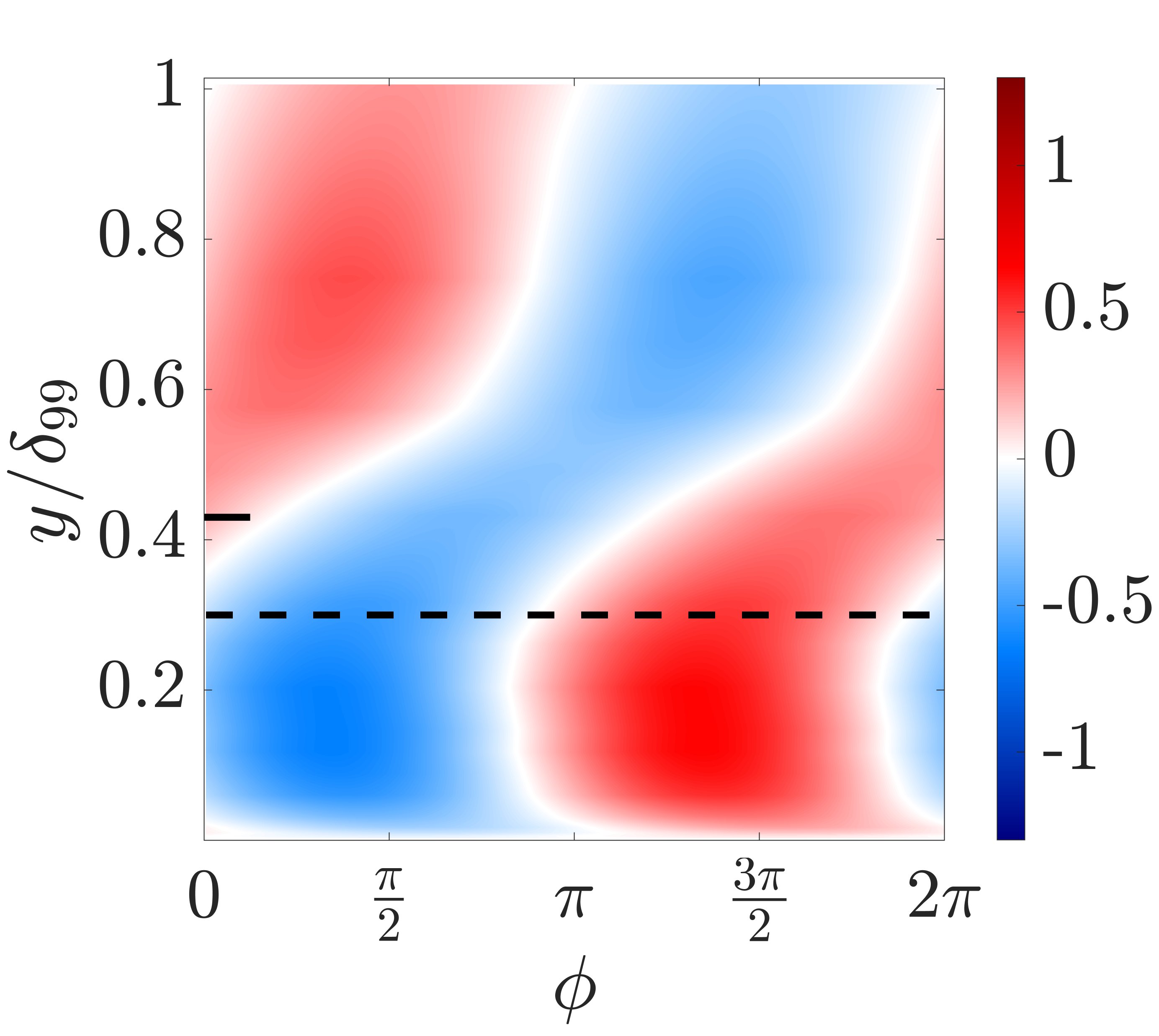}
    \includegraphics[width=0.24\textwidth]{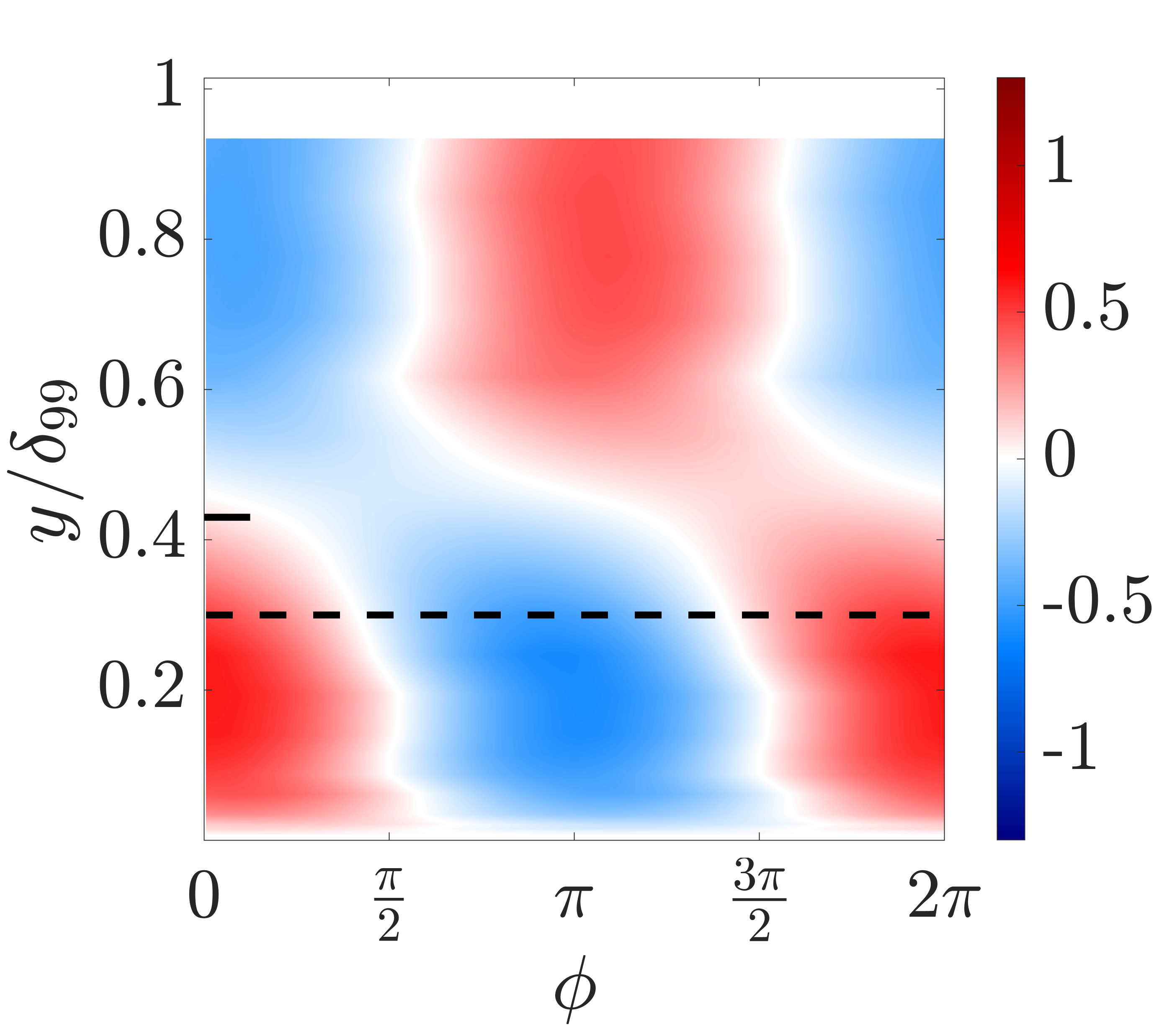}
    
        (e) $x_m/\delta_{99}=1.5$ \hspace{0.11\textwidth} (f)$x_m/\delta_{99}=3$ \hspace{0.11\textwidth} (g) $x_m/\delta_{99}=5$ \hspace{0.11\textwidth} (h) $x_m/\delta_{99}=8$
        
      \includegraphics[width=0.24\textwidth]{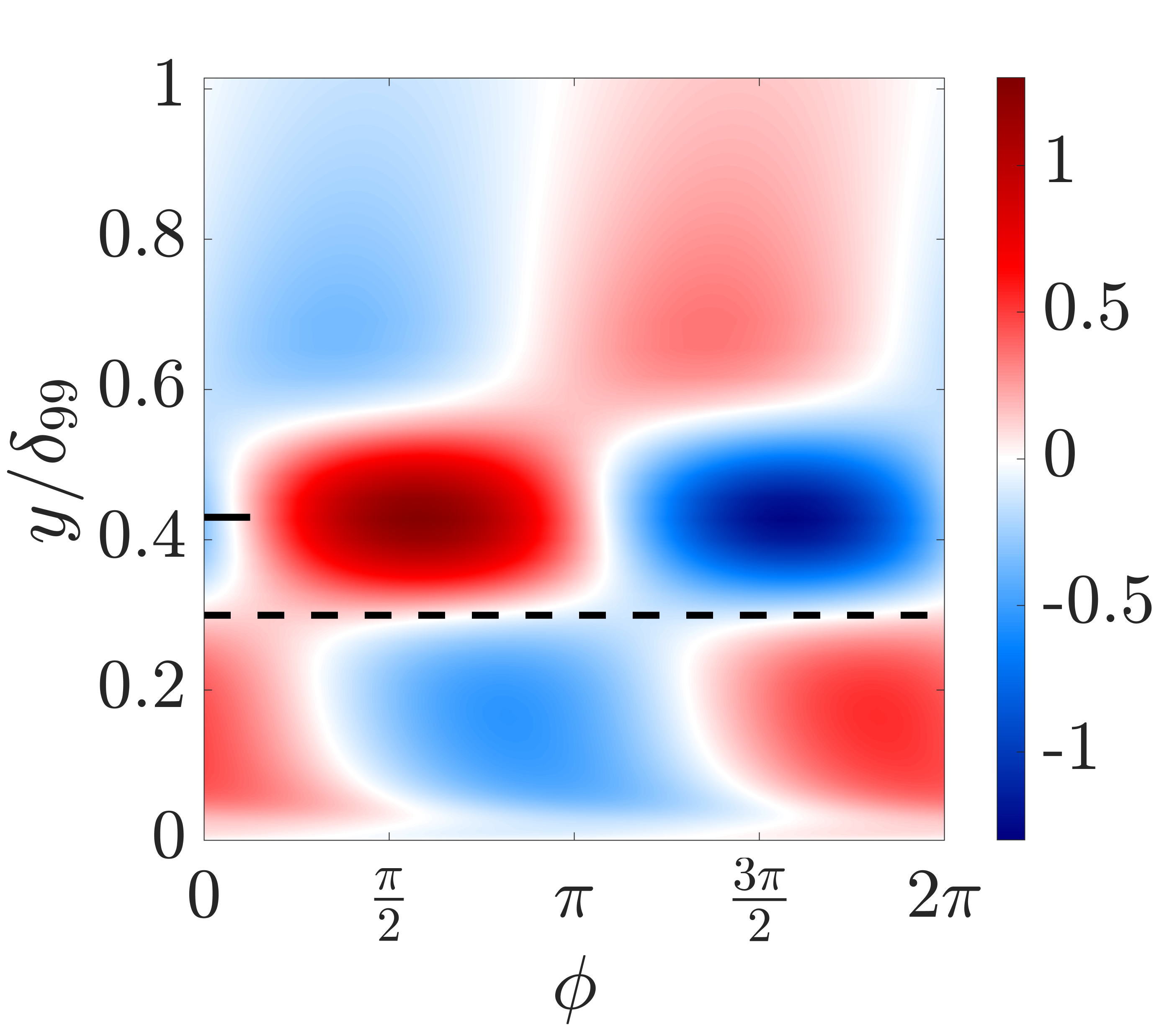}
    \includegraphics[width=0.24\textwidth]{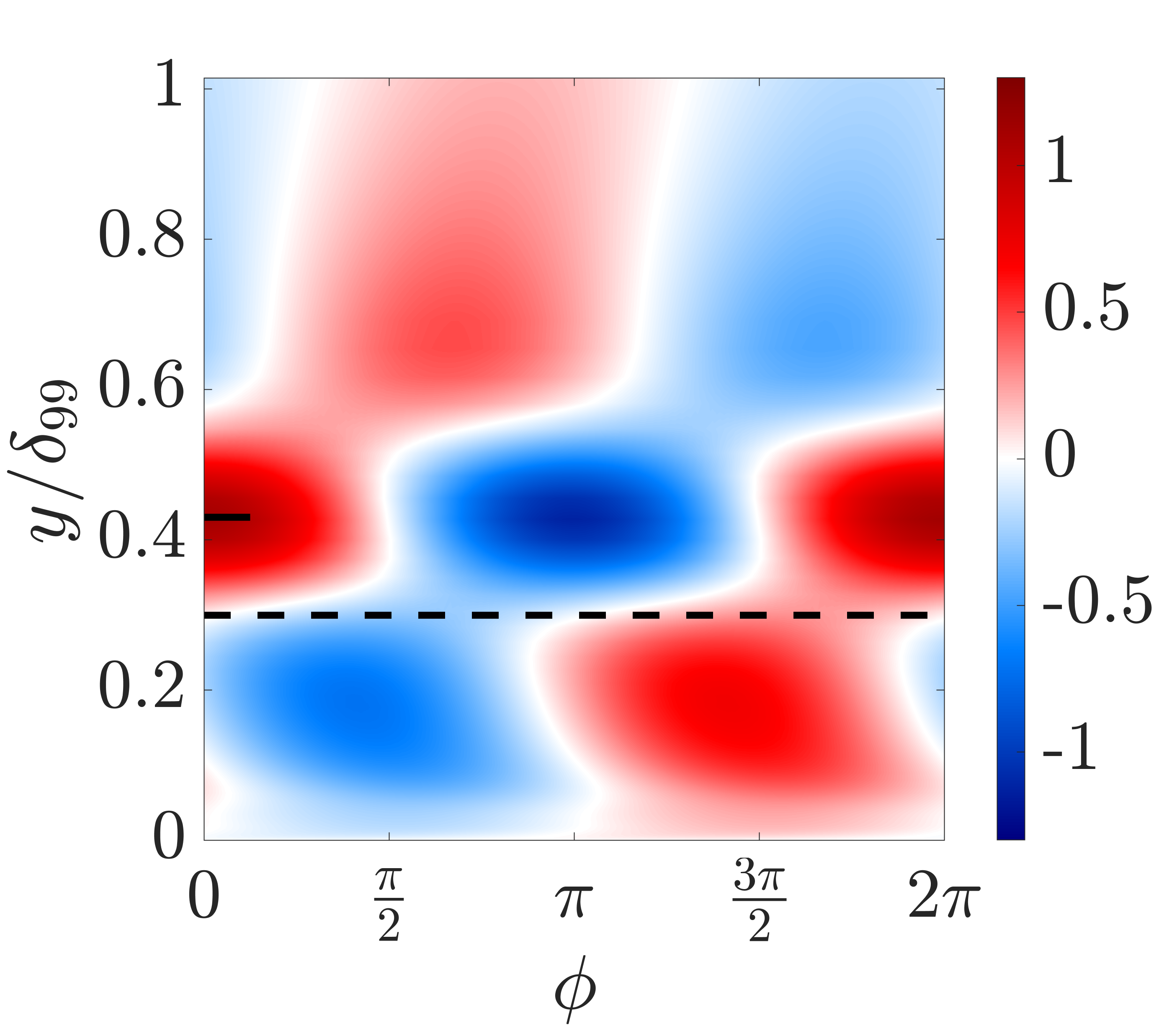}
    \includegraphics[width=0.24\textwidth]{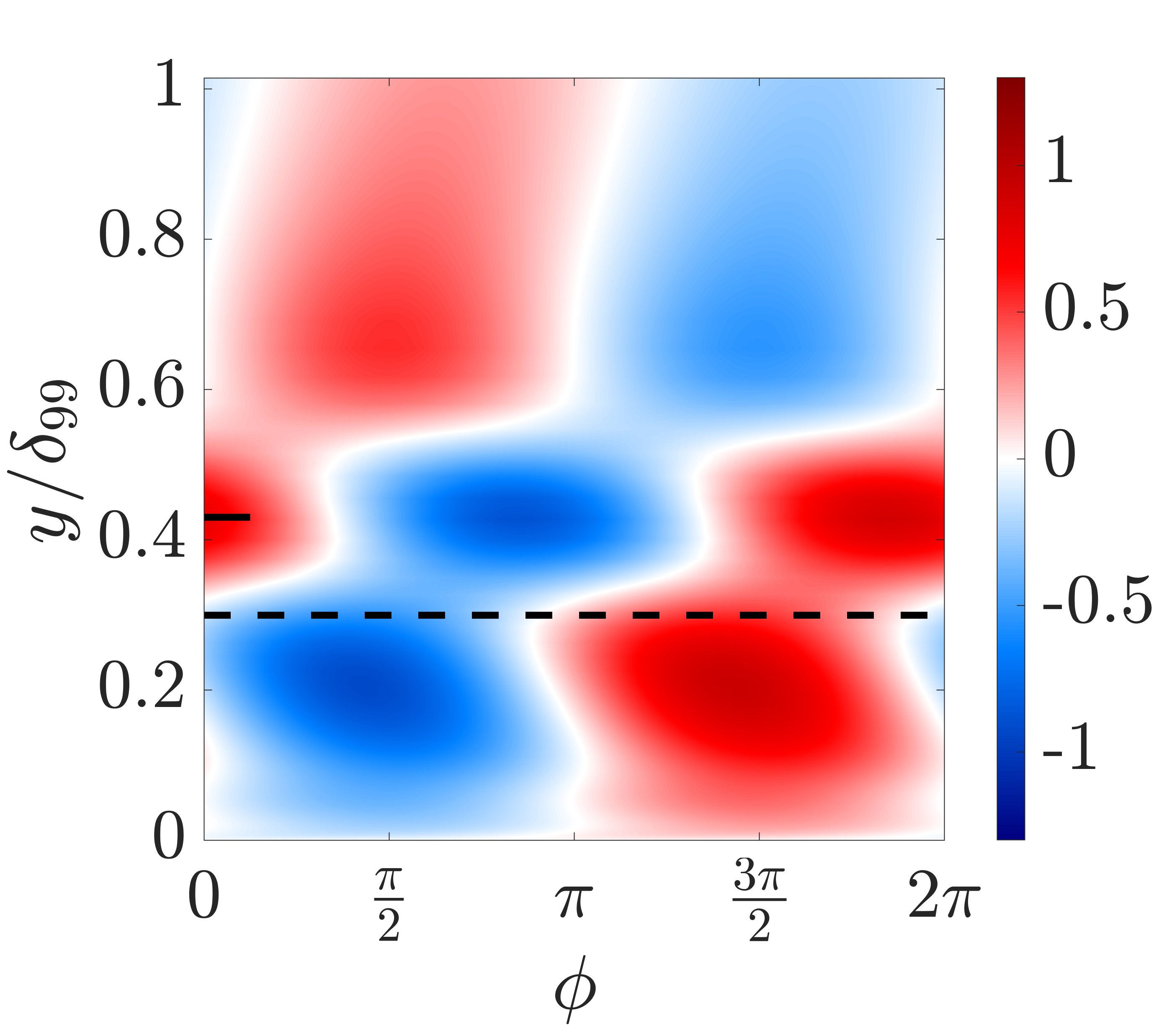}
    \includegraphics[width=0.24\textwidth]{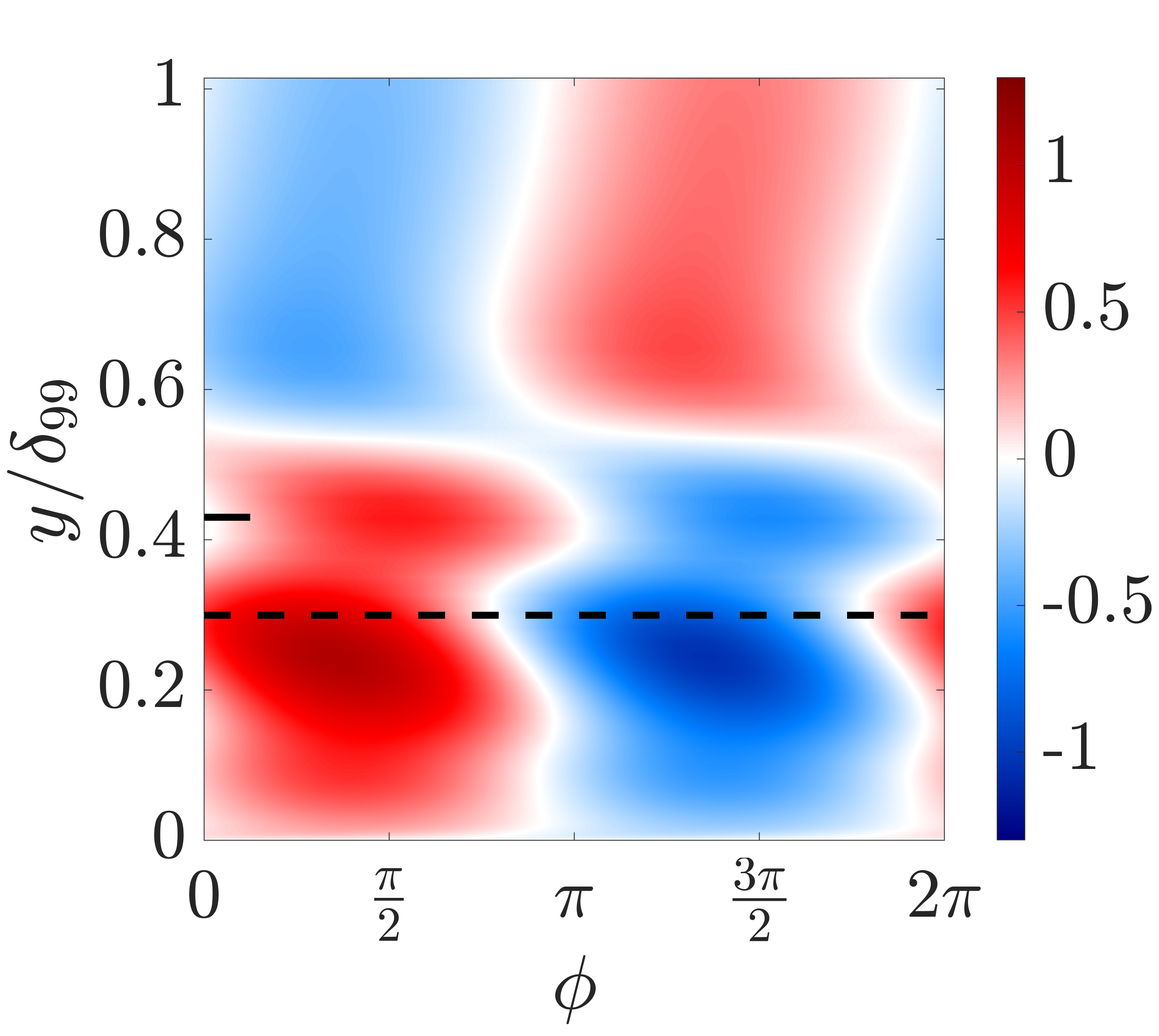}

    \caption{A comparison of phase-locked modal velocity $\tilde{u}^+$ computed from experiments (panels (a)-(d)) using equation \eqref{eq:phase_lock_exp} and that computed from spatial input-output analysis by equation   \eqref{eq:spatial_phase} (panels (e)-(h)) obtained for $y_p/\delta_{99}=0.3$ and $f_p=80$~Hz. Here, ($\dashed$) indicates the plate height of $y_p/\delta_{99}=0.3$ employed in experiments; and ($\mline\mline$) represents the body force center $y_f=y_p+0.13\delta_{99}$ employed in the body force term described in equation \eqref{eq:spatial_body_force}.
 }
    \label{fig:spatial_external_forcing_comparison_exp_modal}
\end{figure}

We compare the phase-locked velocity obtained from the proposed spatial input--output analysis against results from experimental measurements associated with an actuation frequency $f_p=80$~Hz ($0.3983{U}_\infty/\delta_{99}$ and $f_p^+=0.0135$)  and an actuator plate height $y_p/\delta_{99}=0.3$. This actuator plate height $y_p/\delta_{99}=0.3$ corresponds to the top boundary of the log-law layer \citep{pope2000turbulent}. Figure \ref{fig:spatial_external_forcing_comparison_exp_modal} compares the phase-locked velocity obtained from experimental measurements (top panels) and the model (bottom panels) at the four different downstream measurement locations $x_m/\delta_{99}=1.5$, $x_m/\delta_{99}=3$, $x_m/\delta_{99}=5$, and $x_m/\delta_{99}=8$. In all panels, the long black dashed line ($\dashed$, black) corresponds to the height of the actuator plate $y_p$ and the short black solid line ($\mline\mline$, black) is the height of the body force center $y_f$ in equation \eqref{eq:spatial_body_force}. Here, we can see that the model provides good qualitative agreement with experimental measurement. At the downstream location $x_m/\delta_{99}=1.5$, the phase-locked velocity is isolated into three distinct regions across the boundary layer thickness. We refer to the region below the plate \rev{$y/\delta_{99}\in [0,0.3]$} as the bottom region, the region \rev{$y/\delta_{99}\in (0.3,0.56)$} with $y_f$ in the middle as the central region, and the region \rev{$y/\delta_{99}\in [0.56,1)$} as the top region. As expected, the central region is most strongly influenced by the actuation. The figures indicate that there is a clear phase shift between these regions in both the experimental and model results at all measurement locations. The behavior in the central and top regions is reminiscent of the results from previous studies that showed two similar regions above an actuator (mounted at the wall) \citep{jacobi2011dynamic,duvvuri2016nonlinear,duvvuri2017phase,bhatt2020linear,huynh2020characterization}.  The bottom region observed here is not visible in these previous studies \citep{jacobi2011dynamic,duvvuri2016nonlinear,duvvuri2017phase,bhatt2020linear,huynh2020characterization} as their actuation is introduced by wall-mounted actuation. The phase-locked velocities in figure \ref{fig:spatial_external_forcing_comparison_exp_modal}(a)-(c) have a larger phase $\phi$ at a larger $y$ that is opposite to what is shown in figure \ref{fig:spatial_external_forcing_comparison_exp_modal}(d), which suggests that the direction of phase-locked velocity is also changing along the streamwise directions. The results from the model in figure \ref{fig:spatial_external_forcing_comparison_exp_modal}(e)-(h) also capture this phenomenon qualitatively. This phenomenon is likely due to different phase speeds at different wall-normal heights and suggests the importance of a model that allows a range of streamwise wavenumbers. This notion will be examined further in the next section. \rev{One of the differences in the spatial input-output results and experimental data is the smoother variation between the top, center, and bottom regions (figure \ref{fig:spatial_external_forcing_comparison_exp_modal}(a)-(d)), versus sharper interface that appears to occur at a single phase and wall normal height in figures \ref{fig:spatial_external_forcing_comparison_exp_modal}(e)-(h). This difference may be due to the choice of a single temporal frequency $f_p$ and single spanwise wavenumber $k_z=0$. An interesting direction for future work is the evaluation of the influence of additional temporal frequencies and spanwise wavenumbers, which can be introduced by triadic nonlinear interactions, see e.g., \citep{mckeon2013experimental,mckeon2017engine}.  These types of nonlinear effects have also been partially captured in traditional input-output approaches through the addition of an eddy viscosity model, see e.g., \citep{Cossu2009,hwang2010amplification,Hwang2010Linear}, and incorporating such a model into this framework provides another avenue for ongoing investigation.}

\begin{figure}
    \centering
    \includegraphics[width=0.7\textwidth]{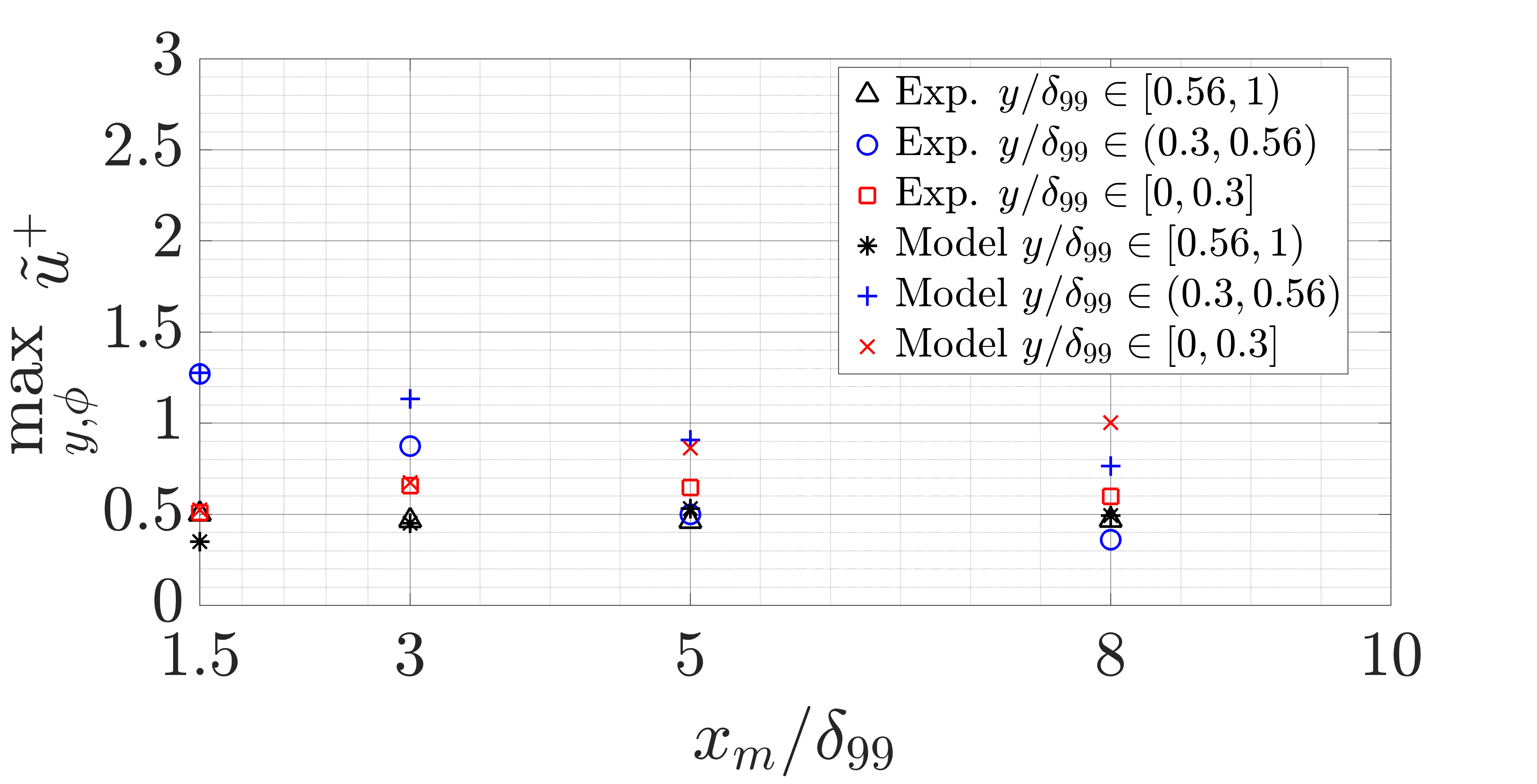}
    \caption{The experimental measured and model-predicted value of $\underset{y,\phi}{\text{max}}\;\tilde{u}^+$ for the top region $y/\delta_{99}\in [0.56,1)$, the center region $y/\delta_{99}\in (0.3,0.56)$, and the bottom region $y/\delta_{99}\in [0,0.3]$ at downstream locations $x_m/\delta_{99}=$1.5, 3, 5, and 8 corresponding to results in figure \ref{fig:spatial_external_forcing_comparison_exp_modal}. }
    \label{fig:phase_lock_amplitude}
\end{figure}

\rev{Figure \ref{fig:phase_lock_amplitude} compares the experimentally measured maximum velocity with the model predicted maximum amplitude of phase-locked velocity $\tilde{u}^+$ in the top, center, and bottom regions for downstream locations $x_m/\delta_{99}=$1.5, 3, 5, and 8 corresponding to the results in figure \ref{fig:spatial_external_forcing_comparison_exp_modal}. Here it is clear that the magnitude of phase-locked velocity $\tilde{u}$ in the central region $y/\delta_{99}\in (0.3,0.56)$ decays with downstream distance in both the experimental data and model results, although the decay rate of the analytical results is slower, particularly between $x_m/\delta_{99}=$1.5 and 3. This difference could be a result of the simplified actuator model.} This downstream spatio-temporal characteristic of the phase lock velocity as it decays is consistent with recent work  using particle image velocimetry (PIV) measurements to directly track the streamwise evolution of the velocity field \citep{huynh2020characterization}. \rev{In the top region $y/\delta_{99}\in [0.56,1)$, the data from the model shows a slightly lower maximum value close to the actuator, but grows to match the experiment at further downstream distances.} The experimental results show that the magnitude of phase-locked velocity $\tilde{u}$ in the bottom region \rev{($y/\delta_{99}\in [0,0.3]$)} at $x_m/\delta_{99}=3$ is larger than that at $x_m/\delta_{99}=1.5$, a trend that is also reflected in the model prediction. The larger phase-locked velocity amplitude below the center of perturbation suggests a spatial transient growth mechanism for the near-wall region. \rev{The trend near the wall far downstream of the actuator differs slightly at the two furthest measurement locations $x_m/\delta_{99}=5$ and 8 as the model predicts a slightly larger maximum while the experimental results remain relatively constant within the experimental error. The influence of nonlinear interactions of the small scales in the near-wall region is expected to be larger, which may account for the differences.}

\section{Downstream propagation of large-scale structures}
\label{sec:spatial_downstream}

We next examine the downstream evolution of the streamwise phase-locked velocity, we focus on the streamwise, wall-normal velocity components as the 
experimental set-up  leads to a flow that is dominated by the $(u,v)$ velocity in the $(x,y)$ plane and nearly uniform in the spanwise direction due to a spanwise-uniform actuation.  In order to obtain the wall-normal velocity we modify the output operator to obtain the wall-normal velocity:
\begin{subequations}
\begin{align}
     \check{v}=&\boldsymbol{\check{C}}_{S,v}\boldsymbol{\check{q}}_S,\\
     \boldsymbol{\check{C}}_{S,v}:=&\begin{bmatrix}
    0 & 1 & 0 & 0 & 0 & 0 
    \end{bmatrix}. 
\end{align}
\end{subequations}
Based on the experimental configuration the spanwise vorticity is of primary interest, and this quantity can be obtained as
\begin{subequations}
\begin{align}
     \check{\omega}_z=&\boldsymbol{\check{C}}_{S,\omega_z}\boldsymbol{\check{q}}_S,\\
     \boldsymbol{\check{C}}_{S,\omega_z}:=&\begin{bmatrix}
    -\partial_y & \text{i}k_x & 0 & 0 & 0 & 0 
    \end{bmatrix}.
\end{align}
\end{subequations}
We quantify the  downstream evolution of phase-locked velocity for these qualities at each downstream measurement location $x_m$ as
\begin{subequations}
    \label{eq:spatial_u_s_all}
\begin{align}
    u_s(x_m,y;\omega,k_z)=&\mathbb{R}\text{e}[\check{u}(x_m,y;\omega,k_z)], \label{eq:spatial_u_s}\\
    v_s(x_m,y;\omega,k_z)=&\mathbb{R}\text{e}[\check{v}(x_m,y;\omega,k_z)],\label{eq:spatial_v_s}\\
    \omega_{z,s}(x_m,y;\omega,k_z)=&\mathbb{R}\text{e}[\check{\omega}_z(x_m,y;\omega,k_z)].\label{eq:spatial_omega_s}
\end{align}
\end{subequations}

\begin{figure}
    \centering
    \includegraphics[width=0.5\textwidth]{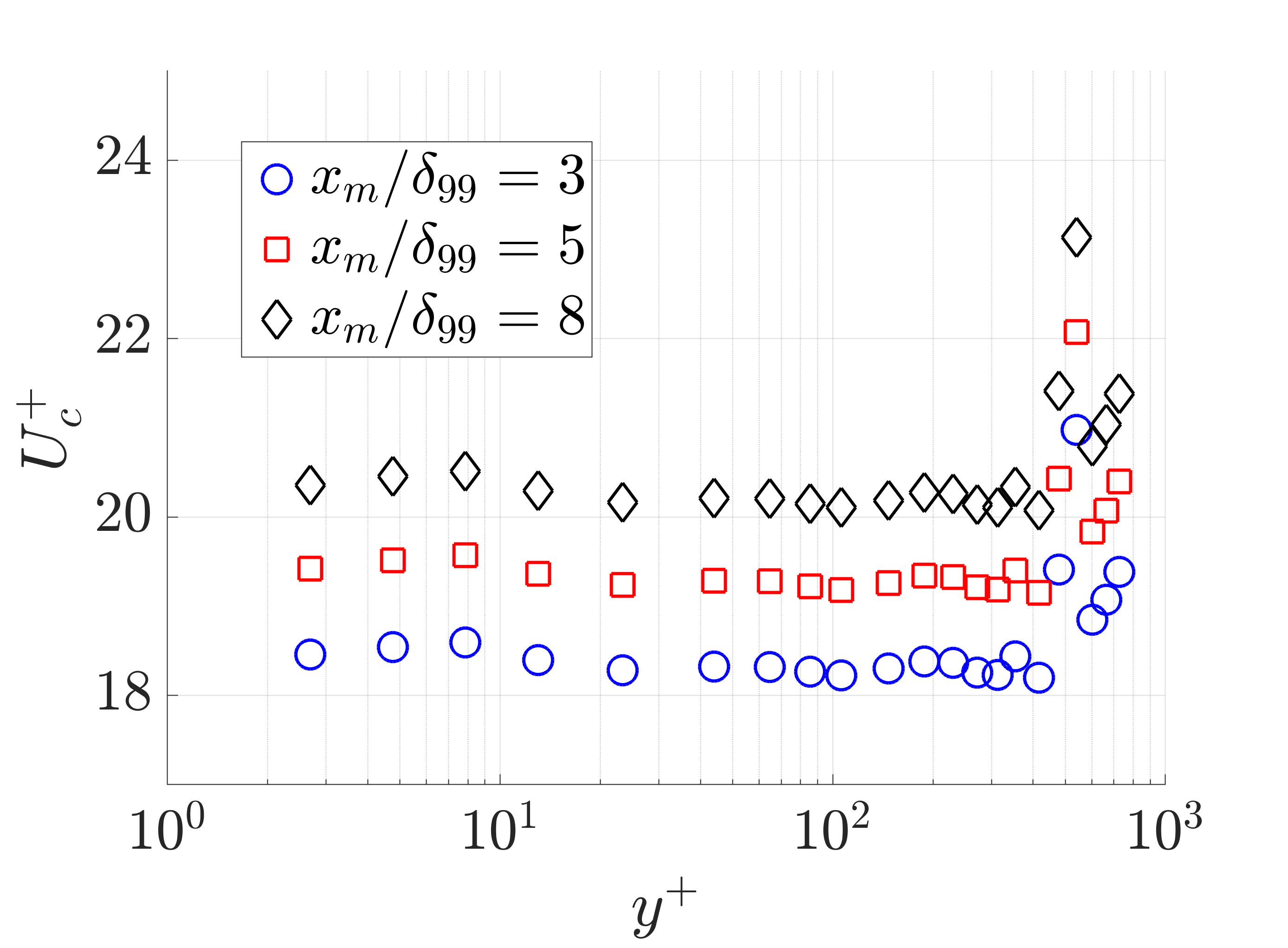}
    \caption{ \red{Phase speed} $U_c^+(x_m,y)$ \red{of phase-locked velocity} employed to construct pseudo-spatial evolution of phase-locked velocity.}
    \label{fig:convective_velocity_U_c}
\end{figure}

The experimental measurement of phase-locked velocity is used to construct a pseudo-spatial evolution of phase-locked streamwise velocity $u_s(x_s, y)$ at the pseudo-streamwise location 
\begin{align}
    x_s=x_{m}-\frac{\phi}{2\pi}\frac{1}{f_p}U_c.
    \label{eq:x_m_x_s}
\end{align}
\rev{Figure \ref{fig:convective_velocity_U_c} displays the \red{phase speed} $U_c^+(x_m,y)$ \red{of phase-locked velocity associated with the actuation frequency}, which is employed to construct the pseudo-spatial evolution of the phase-locked velocity using equation \eqref{eq:x_m_x_s}. At each wall-normal location $y$, this $U_c^+(x_m,y)$ is obtained by applying a power law fit between the time-delay of the zero-crossing of the streamwise phase-locked velocity and the downstream measurement location $x_m$. Then the slope of fitted results at each downstream measurement location $x_m$ and wall-normal location $y$ is employed to obtain $U_c^+(x_m,y)$; for a more detailed description see \citep{lozier2020streamwise,lozier2022PIV}.} \red{The phase speed of phase-locked velocity in figure  \ref{fig:convective_velocity_U_c} asymptotes to a constant value near the wall that is  much larger than turbulent mean velocity. This behavior is qualitatively similar to the convective velocity of large-scale structures observed in the analysis of canonical turbulent channel flow using DNS data \citep[figures 4-5]{DelAlamo2009} and input-output based methods \citep[figures 5-6]{liu2020input_convective}.}
\rev{We then extrapolate the hot-wire measurement at $x_m/\delta_{99}=3$, 5, and 8 upstream for approximately one period using the phase speed $U_c^+(x_m,y)$ in figure \ref{fig:convective_velocity_U_c}. Near the wall, this corresponds to extrapolate the measurement at $x_m/\delta_{99}=3$ to obtain $u_s(x_s,y)$ at $x_s/\delta_{99}\in [0.9, 3]$, at $x_m/\delta_{99}=5$ to obtain $u_s(x_s,y)$ at $x_s/\delta_{99}\in [2.9,5]$, and at $x_m/\delta_{99}=8$ to obtain $u_s(x_s,y)$ at $x_s/\delta_{99}\in [5.7,8]$. When there is any overlap, we perform a linear interpolation using the values at the boundary of overlap region for a smooth transition between locations. }The wall-normal modal velocity $v_s(x_s, y)$ is computed using  $u_s(x_s, y)$ and the two-dimensional $(x,y)$-plane continuity equation.

\begin{figure}[h!]
    \centering
      (a) $u_s^+$ \hspace{0.35\textwidth} (b) $v_s^+$
      
    \includegraphics[width=0.49\textwidth]{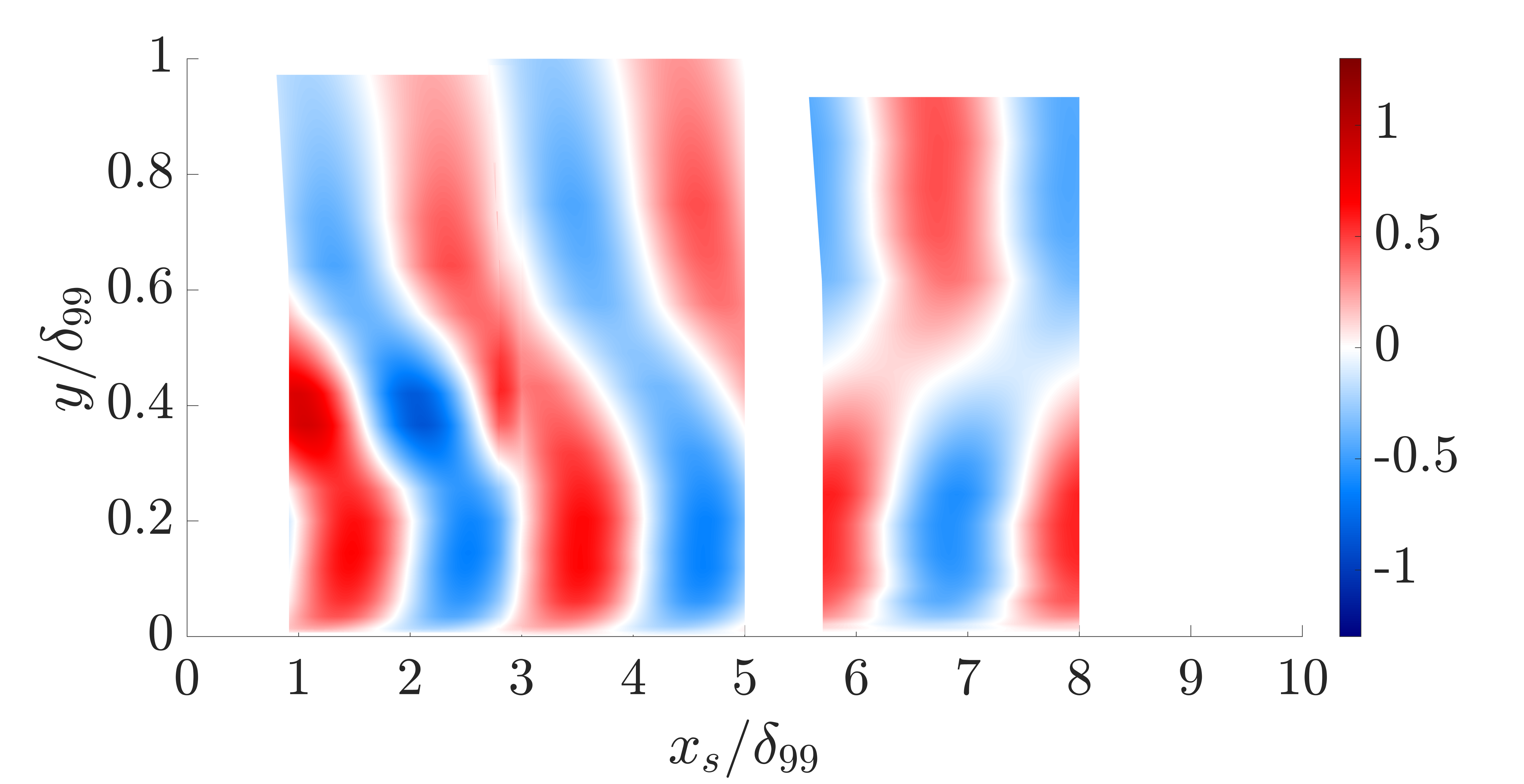}
    \includegraphics[width=0.49\textwidth]{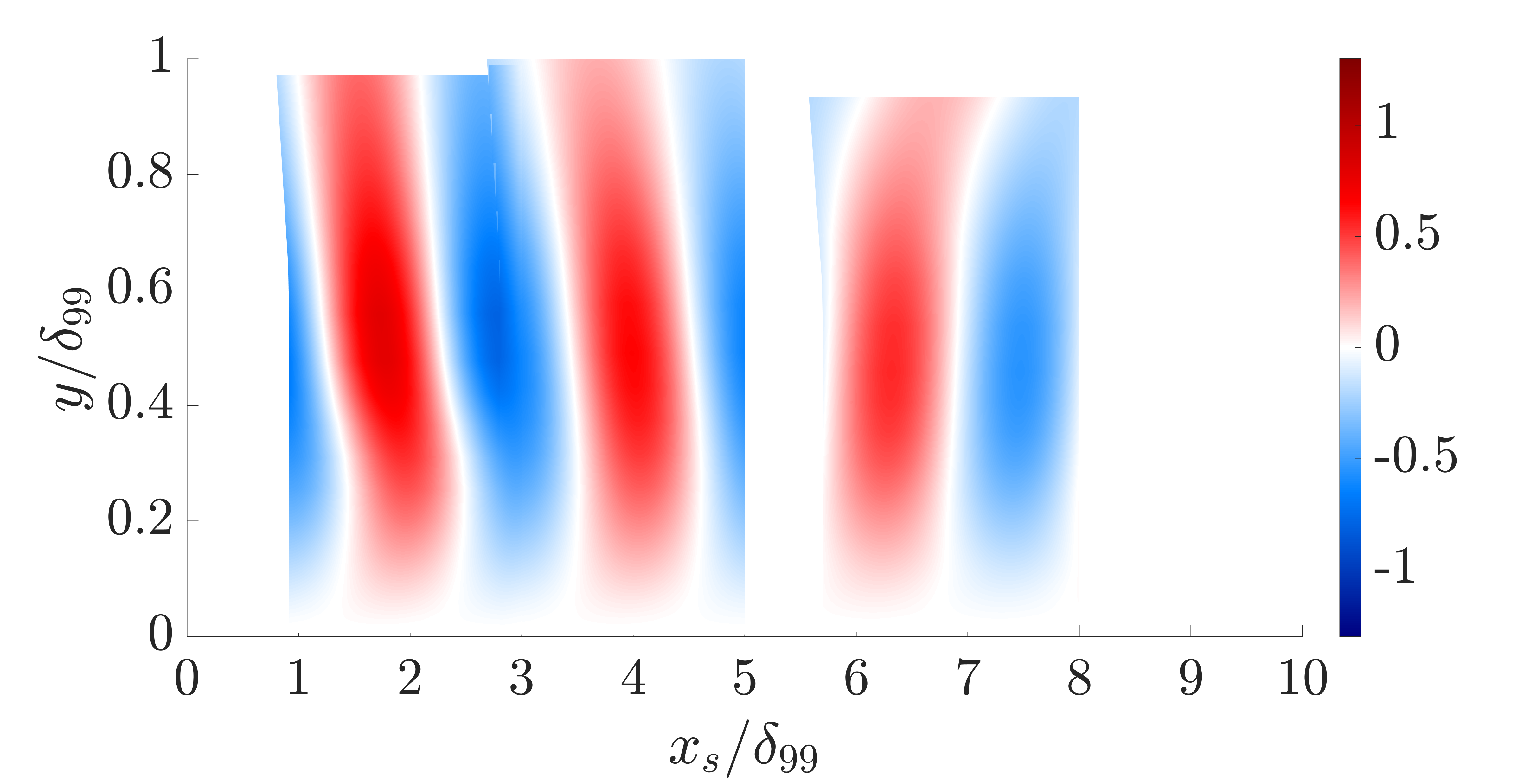}
   (c) $u_s^+$ \hspace{0.35\textwidth} (d)
$v_s^+$

    \includegraphics[width=0.49\textwidth]{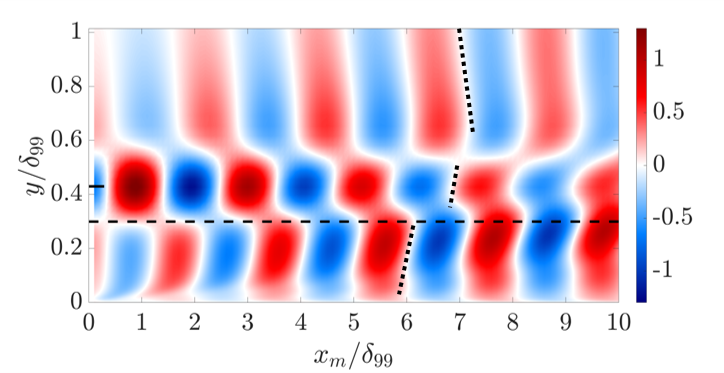}
    \includegraphics[width=0.49\textwidth]{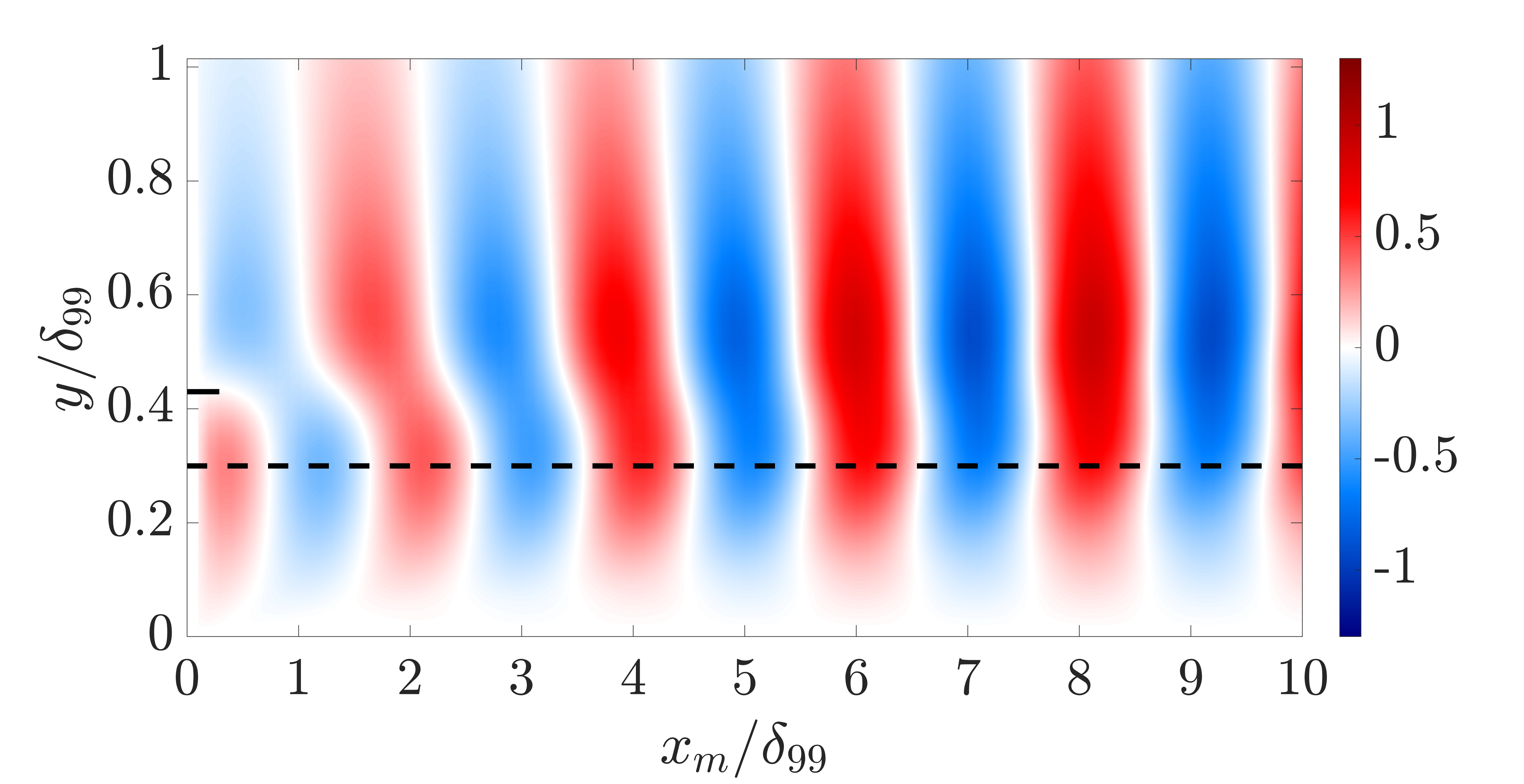}
    \caption{{\color{black}Downstream evolution of (a) $u_s^+$ and (b) $v_s^+$ from experiments,} (c) $u_s^+$ from the model in \eqref{eq:spatial_u_s}, and (d) $v_s^+$ from the model in \eqref{eq:spatial_v_s} with $f_p=80$~Hz, $y_p/\delta_{99}=0.3$. The dotted line ($\cdots$) in panel (c) indicates the downstream distance after three periods. In panels (c)-(d), the long horizontal line ($\dashed$) represents  $y_p/\delta_{99}$ and the short horizontal line ($\mline\mline$) indicates $y_f=y_p+0.13\delta_{99}$.}
    \label{fig:spatial_external_forcing_f_80Hz_y_p_0.3}
\end{figure}

Figure \ref{fig:spatial_external_forcing_f_80Hz_y_p_0.3} presents (a) $u_s^+(x_s,y)$ and (b) $v_s^+(x_s,y)$ as a function of pseudo-streamwise location $x_s$ and wall-normal height $y$ obtained from experimental measurements. Panels (c) and (d) respectively show the corresponding  computed values from equations \eqref{eq:spatial_u_s} and \eqref{eq:spatial_v_s}. Note that the gap of experimental data near $x_s/\delta_{99}\in [5,6]$ in figures \ref{fig:spatial_external_forcing_f_80Hz_y_p_0.3} (a) and (b) is because we limit the construction of pseudo-spatial evolution to one period (see e.g., $u_s^+(x_s,y)$ near $x_s/\delta_{99}=8$). 
\rev{We do not extrapolate measurement data to $x_s/\delta_{99}\in [0,0.9]$ as we expect the phase speed there to vary significantly and the underlying assumption of equation \eqref{eq:x_m_x_s} to be violated.} Here, we note that the values computed using spatial input-output analysis show good qualitative agreement with the experimental measurements at $x_m/\delta_{99}\in [1,5]$ in terms of both their amplitude and shape over wall-normal distance. Figure \ref{fig:spatial_external_forcing_f_80Hz_y_p_0.3}(c) at $x_m/\delta_{99}\approx8$ also shows a phase shift between the central and top regions compared with $x_m/\delta_{99}\in [1,5]$, which is consistent with the variation of  the experimentally obtained $u_s^+$ over pseudo-streamwise location $x_s$ shown in figure \ref{fig:spatial_external_forcing_f_80Hz_y_p_0.3}(a). This phase shift observations for both experimental and model results are consistent with phase-locked velocity $\tilde{u}^+$ at different downstream locations shown in figure \ref{fig:spatial_external_forcing_comparison_exp_modal}. \rev{This} phase shift over the downstream extent can be understood in terms of the difference in the phase speeds over the wall-normal extent. This difference is illustrated through tracking the evolution of a structure in each of the three regions over three periods; this distance is indicated by a dotted line ($\cdots$) in figure \ref{fig:spatial_external_forcing_f_80Hz_y_p_0.3} (c). The differences in the locations of these lines clearly illustrate the effect of the wall-normal dependent phase speed. More specifically, the flow structures in the top region and the central region are traveling slightly faster than those in the bottom region. The shape of phase-locked velocity observed in figure \ref{fig:spatial_external_forcing_comparison_exp_modal} is a  direct result of these differences. Changes in the phase speed of the large-scale structures as a function of wall-normal heights have been associated with the stretching and intensifying of the legs of hairpin vortices as they propagate downstream \citep{Adrian2007}. The results in panels (a) and (c) of figure \ref{fig:spatial_external_forcing_comparison_exp_modal}  highlight the benefit of an analysis method that enables analysis over a wide range of streamwise wavenumbers.

\begin{figure}
    \centering

   (a) \hspace{0.45\textwidth} (b)
   
    \includegraphics[width=0.49\textwidth]{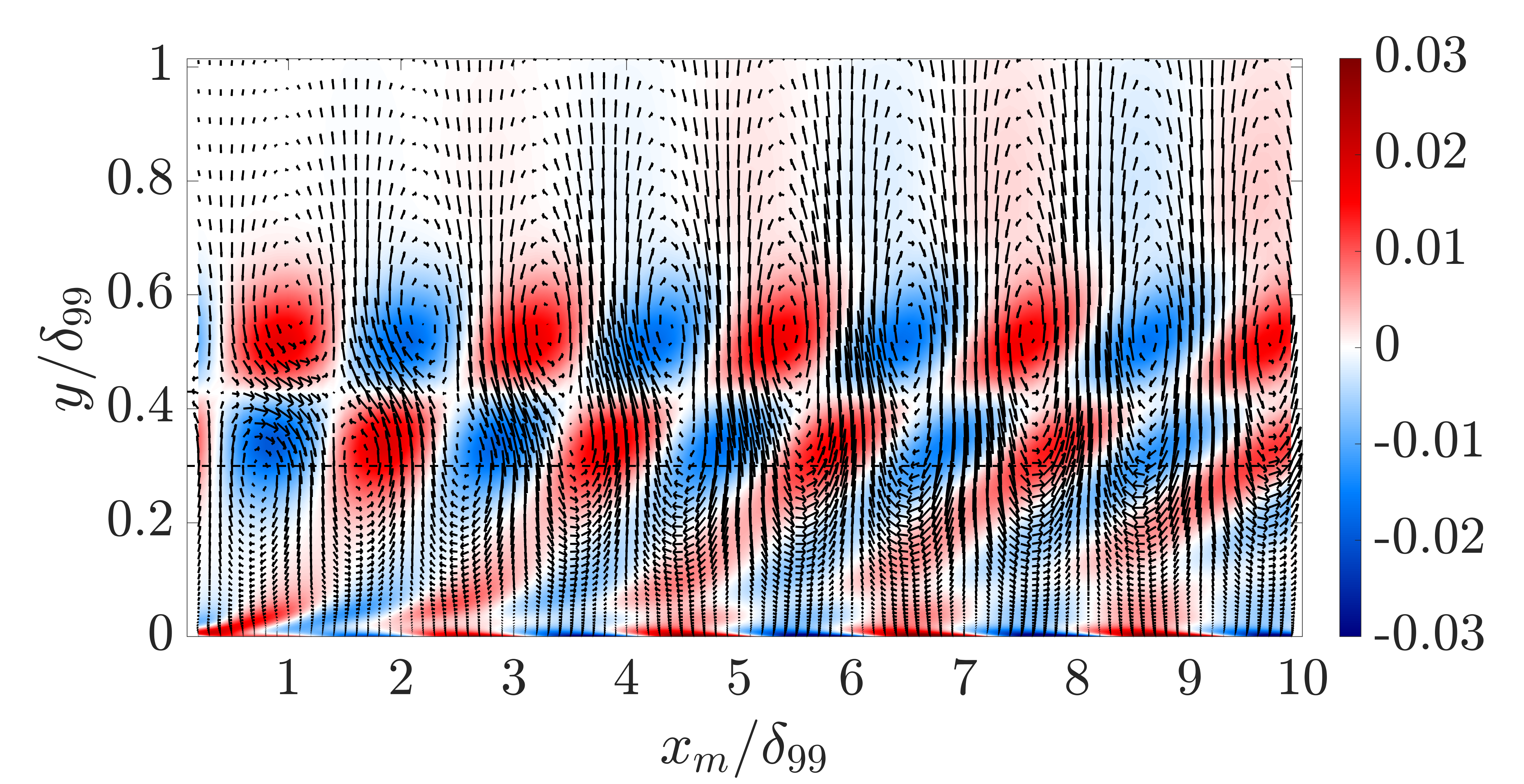}
    \includegraphics[width=0.49\textwidth]{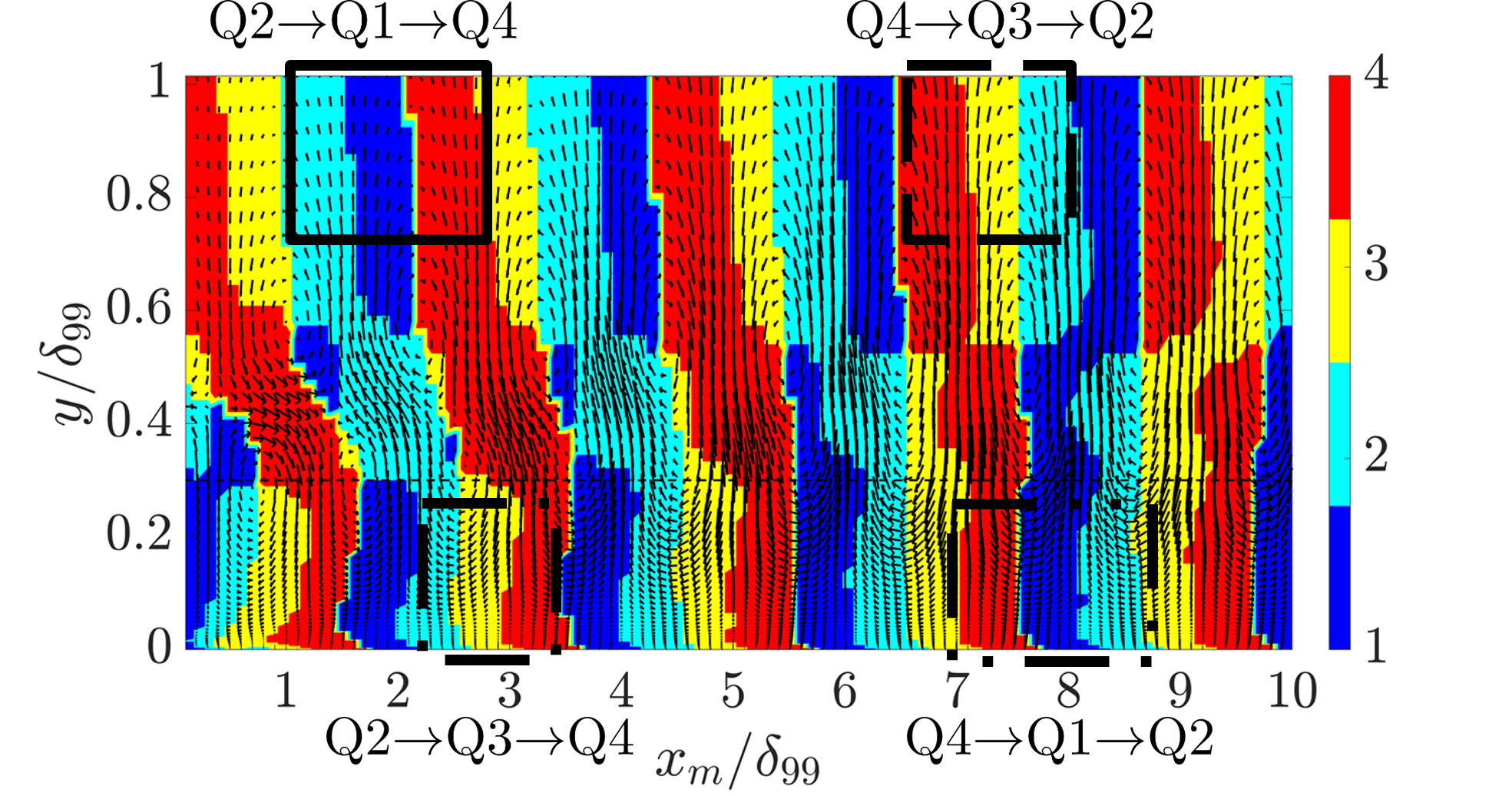}    \caption{The contour in panel (a) is spanwise vorticity $\omega_{z,s}^+$ and the contour colors in panel (b) indicate  quadrant numbers. The velocity vector field $(u_s^+, \:\: v_s^+)$ is superimposed on contours in panels (a) and (b). In panel (b), the blue color indicates quadrant Q1, cyan indicates  quadrant Q2,  yellow corresponds to quadrant Q3, and the red color denotes quadrant Q4. The boxes with ($\mline\mline$), ({\color{black}$\dashed$}), ({\color{black}$\dashdot$}), and ({\color{black}$\dashdotdot$}) outlines examples of
   Q2$\rightarrow$Q1$\rightarrow$Q4,  Q4$\rightarrow$Q3$\rightarrow$Q, Q2$\rightarrow$Q3$\rightarrow$Q4, and  Q4$\rightarrow$Q1$\rightarrow$Q trajectories, respectively.}
      \label{fig:spatial_external_forcing_f_80Hz_y_p_0.3_vorticity_quad}
\end{figure}

The wall-normal velocity $v_s^+$ obtained from the experimental data and the model are respectively shown in figures \ref{fig:spatial_external_forcing_f_80Hz_y_p_0.3} (b) and \ref{fig:spatial_external_forcing_f_80Hz_y_p_0.3}(d), which also show good agreement. In contrast to the streamwise component, the $v_s^+$ obtained from both experimental measurements and spatial input-output analysis are nearly uniform across the wall-normal height. Such a nearly uniform wall-normal velocity is consistent with observations in \citep{jacobi2011dynamic,huynh2020characterization} based on planar PIV measurements. Figure \ref{fig:spatial_external_forcing_f_80Hz_y_p_0.3_vorticity_quad}(a) presents the spanwise vorticity $\omega_{z,s}^+$ as contours with velocity vectors $(u_s,v_s)$ superimposed. Here, we can see that this body force model generates counter-rotating spanwise vorticity near the inflow region. As the actuated large-scale structures propagate downstream, the bottom spanwise vorticity becomes more inclined towards the wall.

Combined information from streamwise and wall-normal velocity can be used to provide insight into the influence of large-scale structures on the Reynolds shear stress. In order to study these stresses, we combine the spatial input-output framework with quadrant analysis \citep{wallace1972wall,Wallace2016} \rev{to classify the shear stress distribution of the spatially evolving flow field. We then compare} the modal structures resulting from actuation with coherent motion in canonical wall-bounded turbulence.  The quadrants are defined in terms of the $u_s$ and $v_s$ phase-locked velocities obtained from equations \eqref{eq:spatial_u_s} and \eqref{eq:spatial_v_s}. We adopt the traditional definitions for each quadrant, more specifically the first quadrant Q1  corresponds to outward motions $(u_s>0,v_s>0)$, the second quadrant Q2 represents ejections $(u_s<0, v_s>0)$, the third quadrant Q3 corresponds to inward motion $(u_s<0, v_s<0)$ and the fourth quadrant Q4 behavior corresponds to sweeps $(u_s>0, v_s<0)$ \citep{Wallace2016}. Figure \ref{fig:spatial_external_forcing_f_80Hz_y_p_0.3_vorticity_quad}(b) plots regions of the flow field corresponding to each of these quadrants as a function of streamwise distance in different colors (contours) with the velocity vectors $(u_s,v_s)$ superimposed. Here, the Q4 and Q2 quadrant events are strongest in the central region, particularly close to the actuator $x_m/\delta_{99}\leq 5.5$,  while Q1, Q2, Q3, and Q4 quadrant events appear to be equally distributed throughout the top and bottom regions, as well as in the center region further downstream, i.e. at $x_m/\delta_{99}>7$.  Moving from left to right (along the downstream direction), quadrant events occur in the order Q4$\rightarrow$Q3$\rightarrow$Q2$\rightarrow$Q1 in the top region. However, the  quadrant event order changes to Q1$\rightarrow$Q2$\rightarrow$Q3$\rightarrow$Q4 in the bottom region. The ordering in both of these regions is consistent with the counter-rotating vorticity patterns appearing in the top and bottom regions in figure  \ref{fig:spatial_external_forcing_f_80Hz_y_p_0.3_vorticity_quad}(a). Quadrant trajectory patterns of Q2$\rightarrow$Q1$\rightarrow$Q4, Q2$\rightarrow$Q3$\rightarrow$Q4, Q4$\rightarrow$Q1$\rightarrow$Q2, and Q4$\rightarrow$Q3$\rightarrow$Q2 were shown to be the most prominent in the dynamics and transport of near-wall turbulence in a previous study that employed quadrant analysis to characterize 36 distinct evolution patterns for $(u,v)$ in turbulent pipe flow \citep{nagano1995coherent}. We outline instances of these four important quadrant trajectories using  boxes with different line types in figure \ref{fig:spatial_external_forcing_f_80Hz_y_p_0.3_vorticity_quad}(b), where the  ($\mline\mline$), ({\color{black}$\dashed$}), ({\color{black}$\dashdot$}) and ({\color{black}$\dashdotdot$}) boxes respectively encompass  Q2$\rightarrow$Q1$\rightarrow$Q4,  Q4$\rightarrow$Q3$\rightarrow$Q2,   Q2$\rightarrow$Q3$\rightarrow$Q4, and  Q4$\rightarrow$Q1$\rightarrow$Q2 trajectories. These quadrant trajectories indicate that the interactions of synthetic large scales with the boundary layer are consistent with the dynamics of a canonical TBL, suggesting the promise of this type of flow interrogation. The ability to perform this detailed study of the downstream evolution and interactions of synthetically generated large-scale structures in actuated TBLs  highlights the benefits of spatial input-output analysis.

\subsection{Effect of actuation frequency and actuator height}
\label{subsec:spatial_varying_freq_height}

\begin{figure}
    \centering
      (a) $u_s^+$ at $f_p=20$~Hz \hspace{0.35\textwidth} (b) $v_s^+$ at $f_p=20$~Hz
      
    \includegraphics[width=0.49\textwidth]{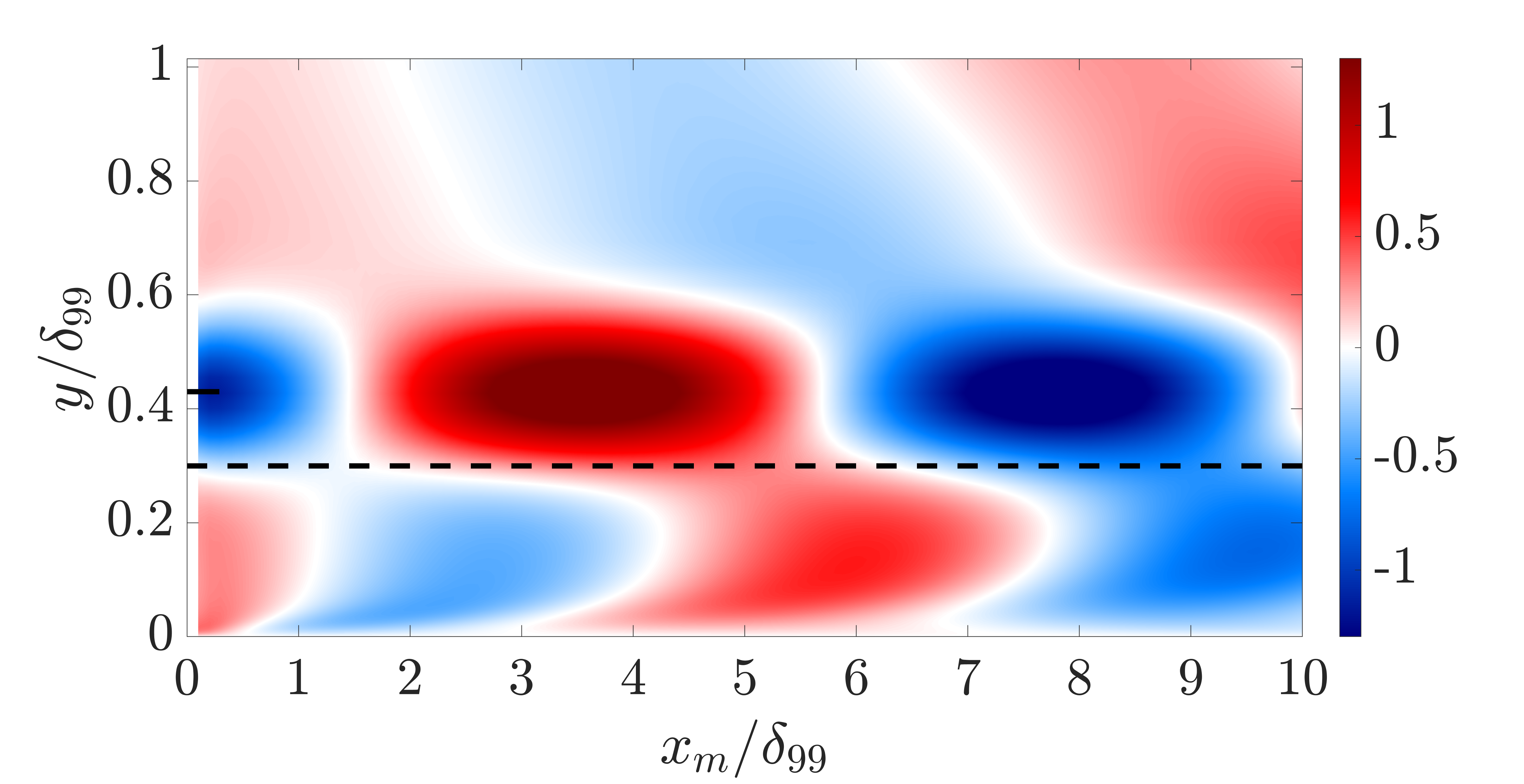}
    \includegraphics[width=0.49\textwidth]{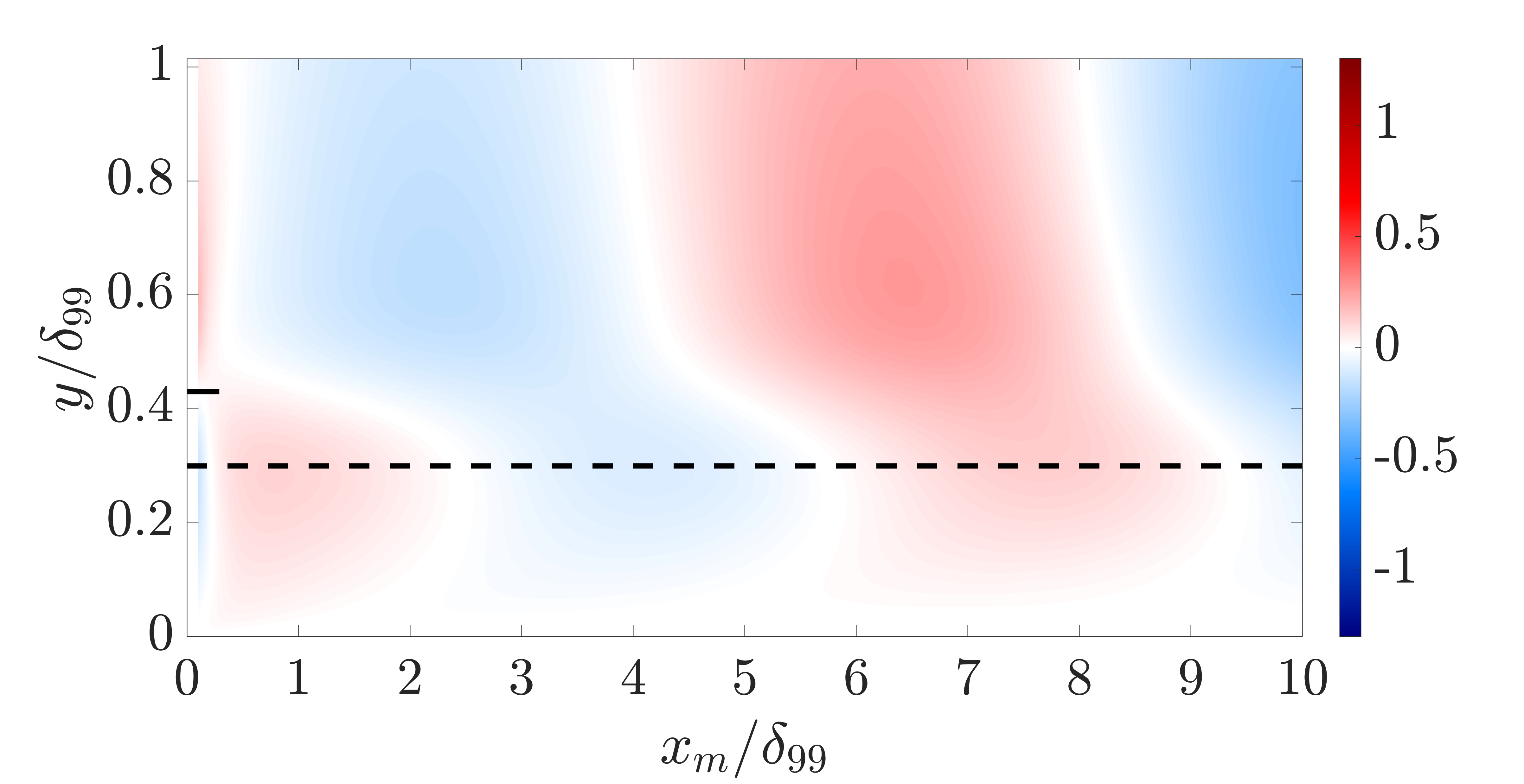}
    
   (c)  $u_s^+$ at $f_p=80$~Hz \hspace{0.35\textwidth} (d) $v_s^+$ at $f_p=80$~Hz
   \includegraphics[width=0.49\textwidth]{f80_yp03_u_dot.png}
    \includegraphics[width=0.49\textwidth]{f80_yp03_v.png}
    
    (e) $u_s^+$ at $f_p=200$~Hz \hspace{0.35\textwidth} (f)
    $v_s^+$ at $f_p=200$~Hz
   \includegraphics[width=0.49\textwidth]{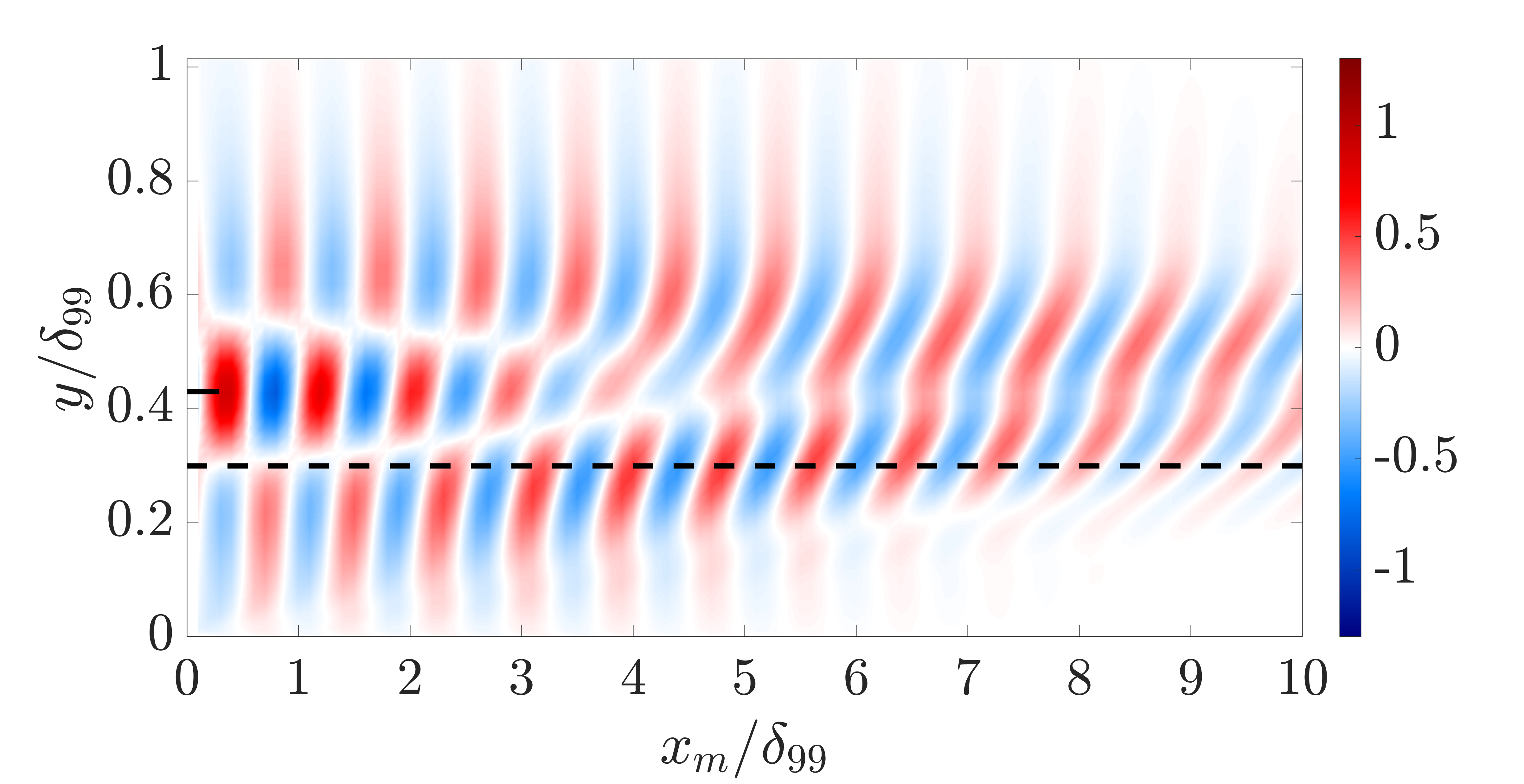}
    \includegraphics[width=0.49\textwidth]{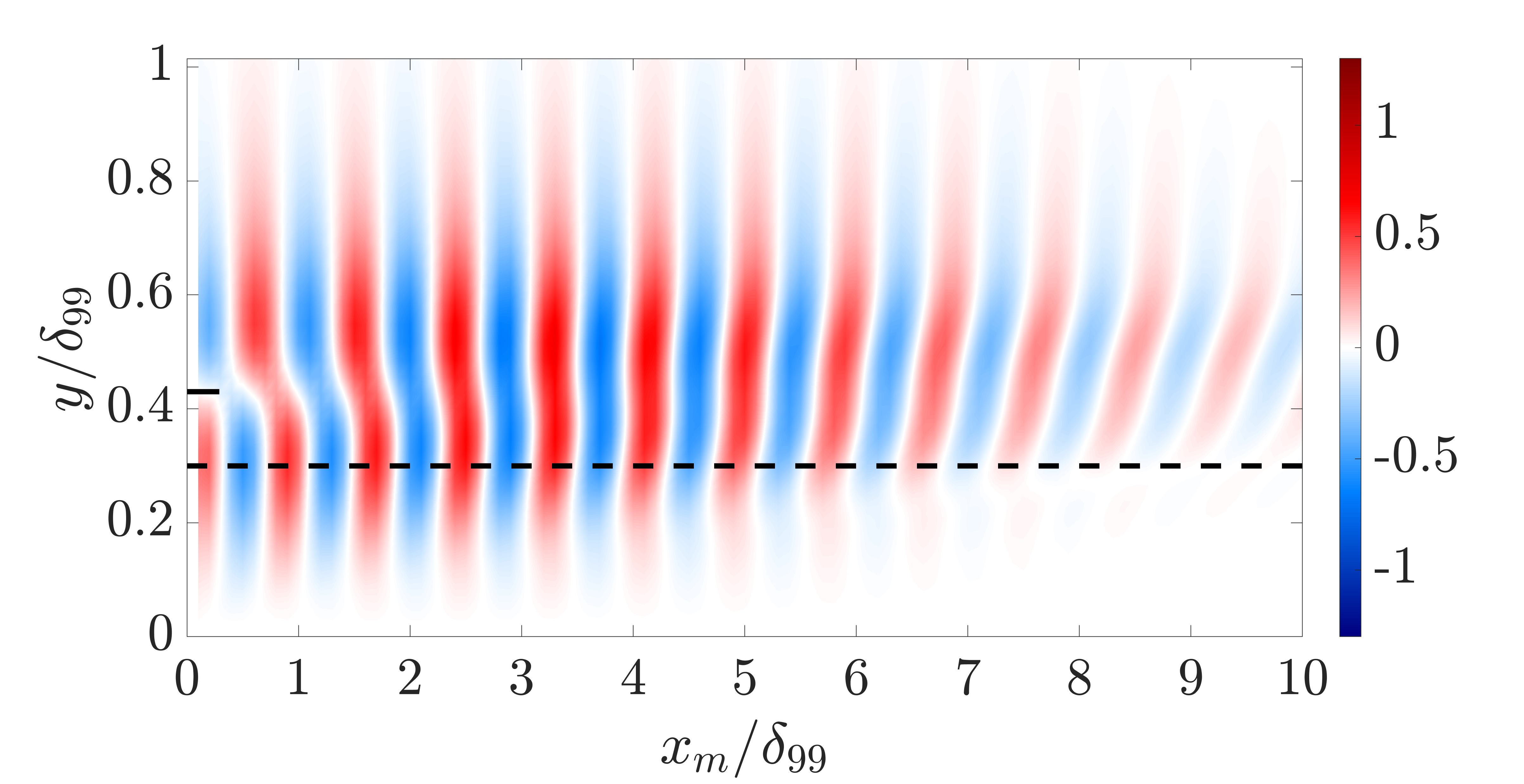}
   
    \caption{Downstream evolution of $u_s^+$ and $v_s^+$ at $f_p=20$~Hz, 80~Hz and 200~Hz. The plate height is $y_p/\delta_{99}=0.3$ for all cases. In all panels, the long horizontal line ($\dashed$) marks  $y_p/\delta_{99}$ and short horizontal line ($\mline\mline$) indicates $y_f=y_p+0.13\delta_{99}$. }
    \label{fig:spatial_external_forcing_frequency_velocity}
\end{figure}

\begin{figure}
    \centering
      (a) $\omega_{z,s}^+$ at $f_p=20$~Hz \hspace{0.23\textwidth} (b) Quadrant numbers at $f_p=20$~Hz
     \includegraphics[width=0.49\textwidth]{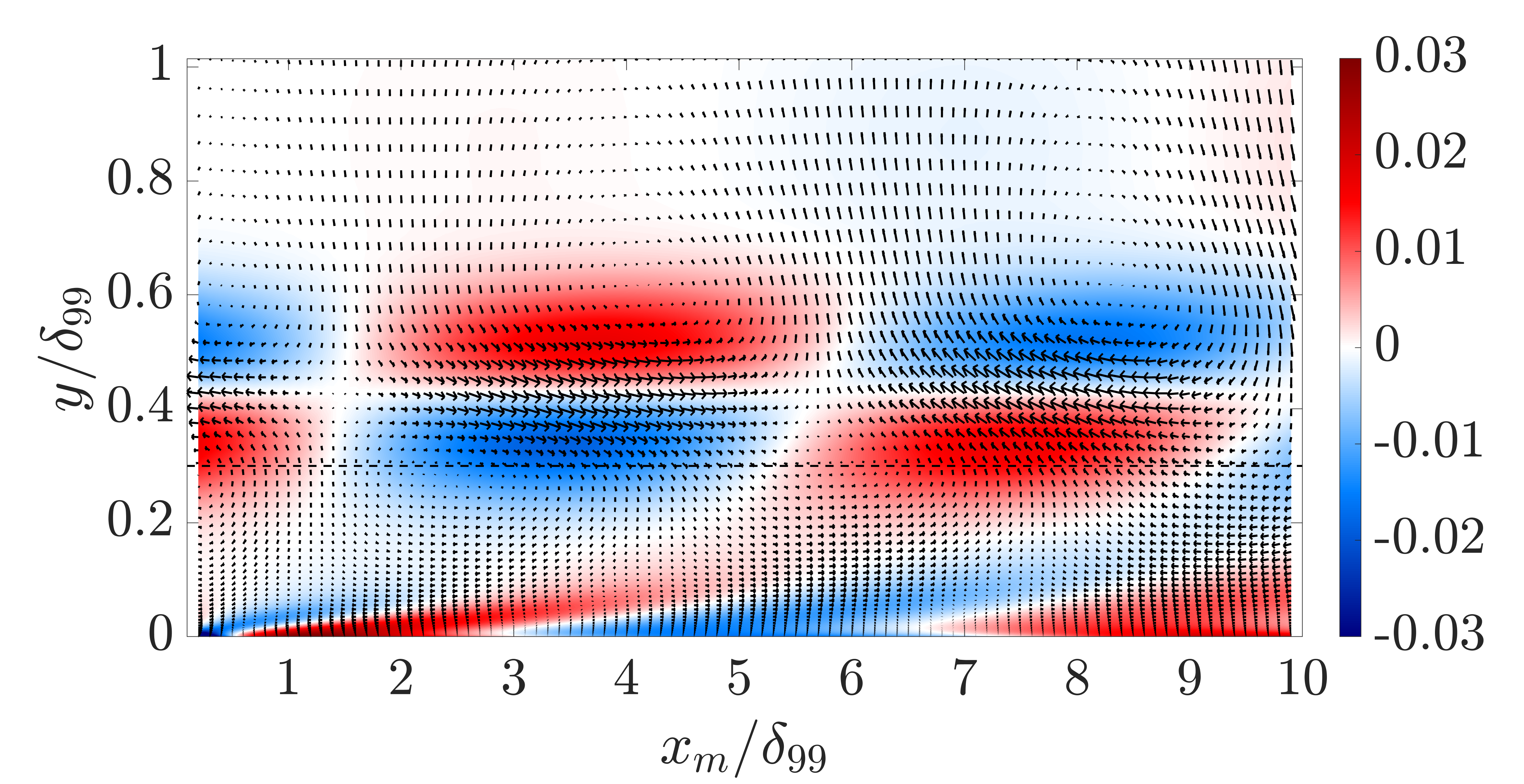}
    \includegraphics[width=0.49\textwidth]{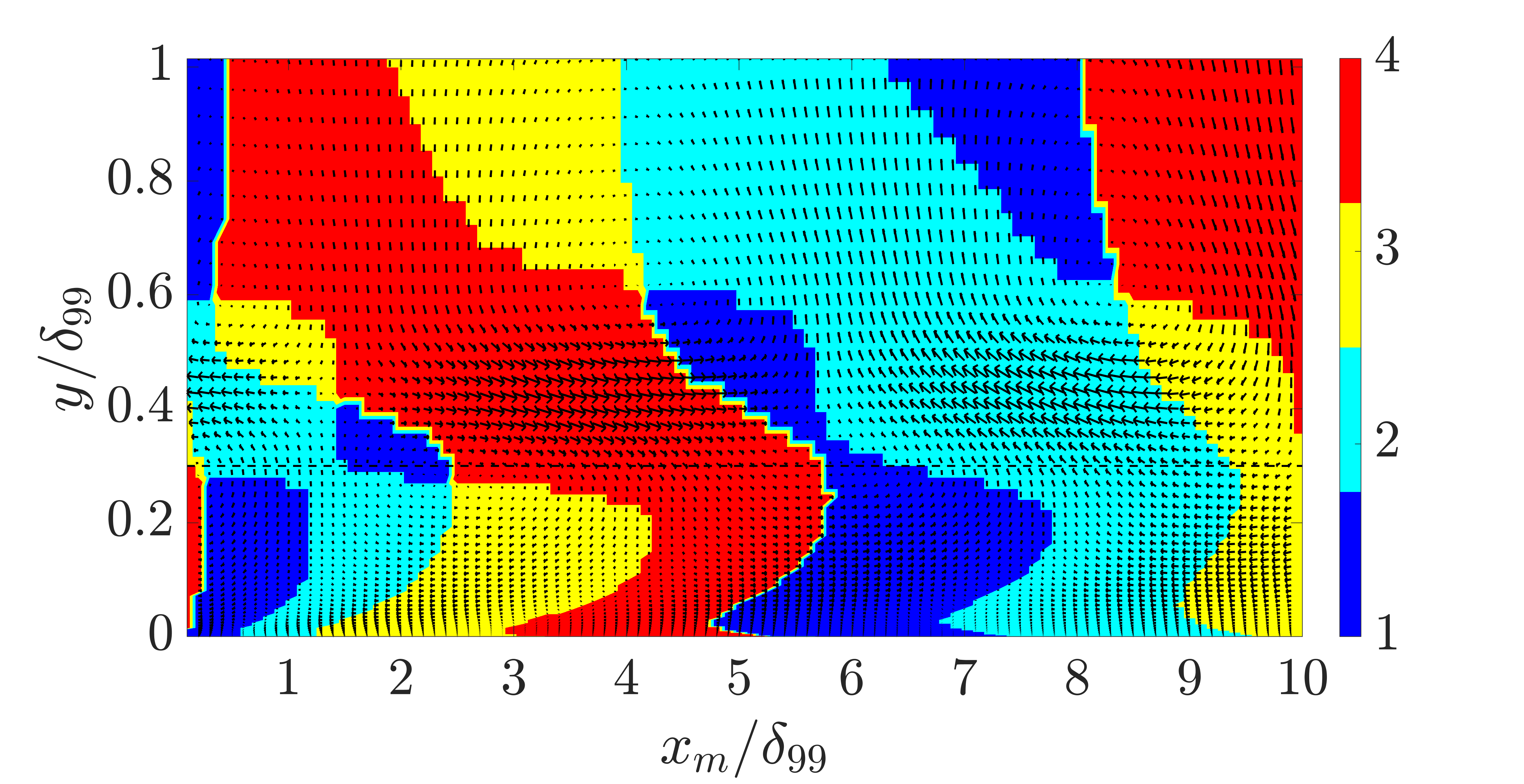}

   (c) $\omega_{z,s}^+$ at $f_p=80$~Hz \hspace{0.23\textwidth} (d) Quadrant numbers at $f_p=80$~Hz
    \includegraphics[width=0.49\textwidth]{f80_yp03_omega.png}
    \includegraphics[width=0.49\textwidth]{f80_yp03_quad_outline.png}  
    
    (e) $\omega_{z,s}^+$ at $f_p=200$~Hz \hspace{0.23\textwidth} (f)
    Quadrant numbers at $f_p=200$~Hz
    \includegraphics[width=0.49\textwidth]{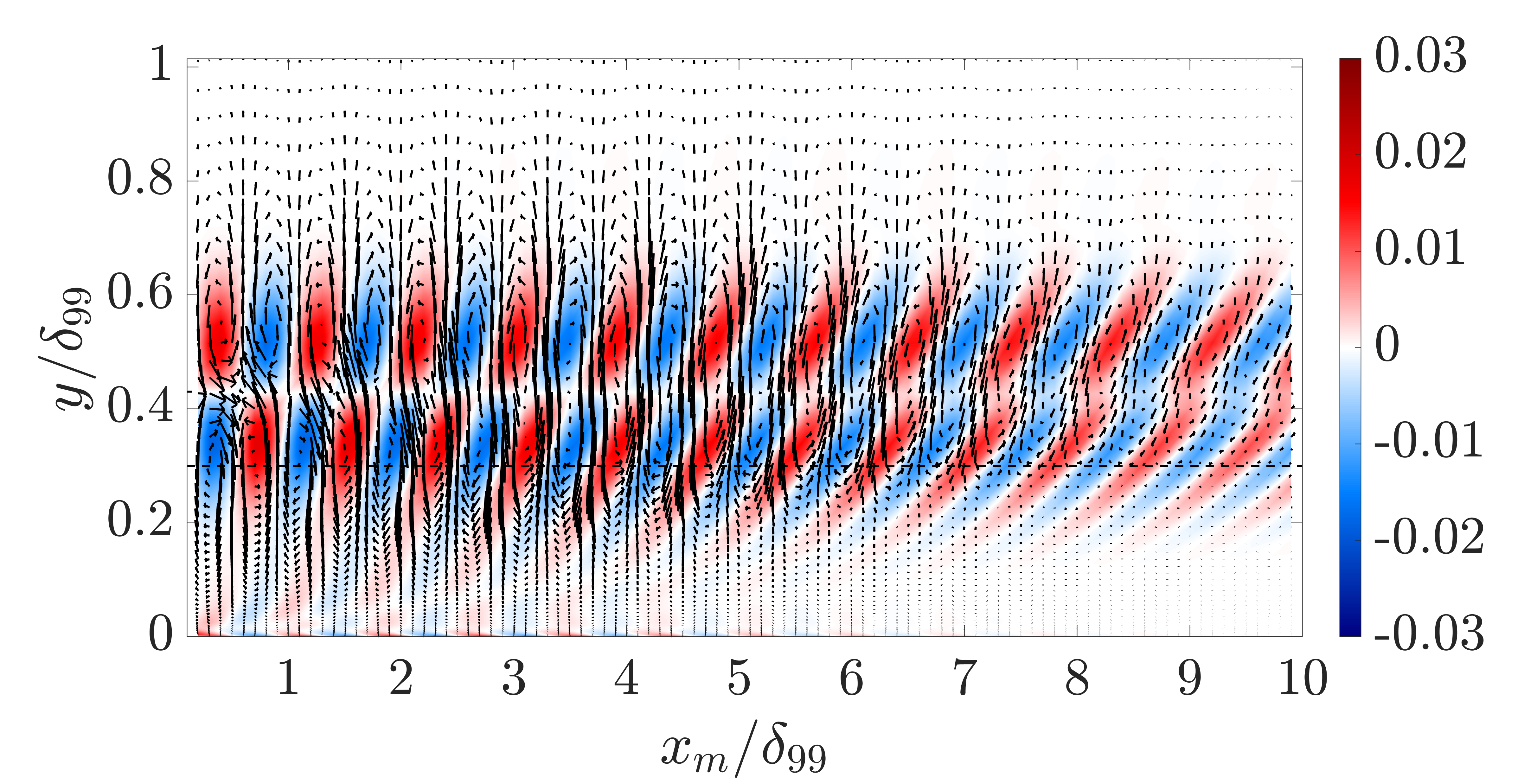}
    \includegraphics[width=0.49\textwidth]{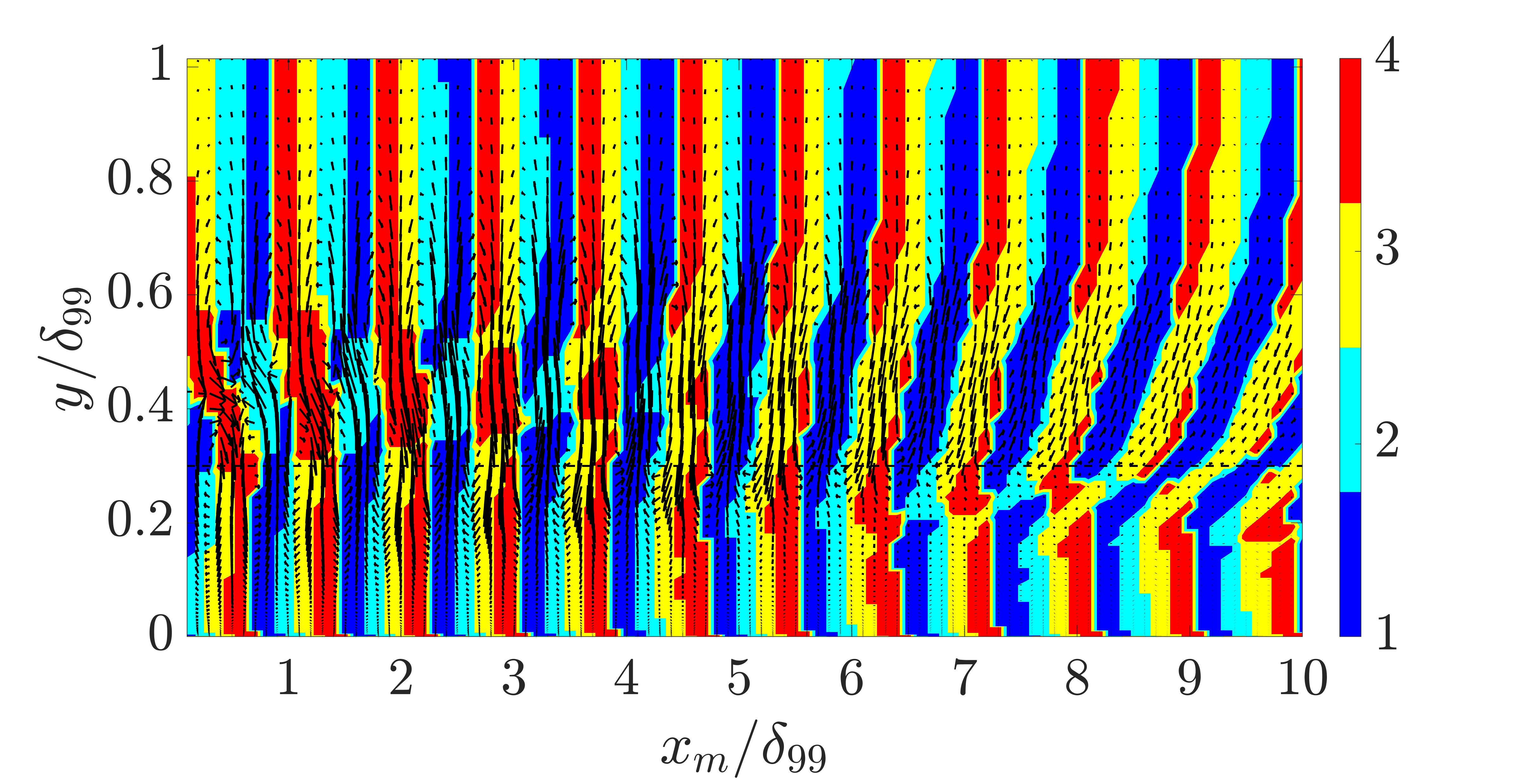}
    
    \caption{Downstream evolution of spanwise vorticity $\omega_{z,s}^+$, and quadrant numbers at $f_p=20$~Hz, 80~Hz and 200~Hz. All results are associated with $y_p/\delta_{99}=0.3$. The velocity vector field $(u_s^+, \:\: v_s^+)$ is superimposed on contours. In panels (b), (d), and (f), the blue color indicates quadrant Q1, cyan indicates quadrant Q2, yellow corresponds to quadrant Q3, and the red color denotes quadrant Q4. }
    \label{fig:spatial_external_forcing_frequency_vorticity_quad}
\end{figure}

\begin{figure}
    \centering
      (a) $u_s^+$ at $y_p/\delta_{99}=0.1$ \hspace{0.35\textwidth} (b) $v_s^+$ at $y_p/\delta_{99}=0.1$
      \includegraphics[width=0.49\textwidth]{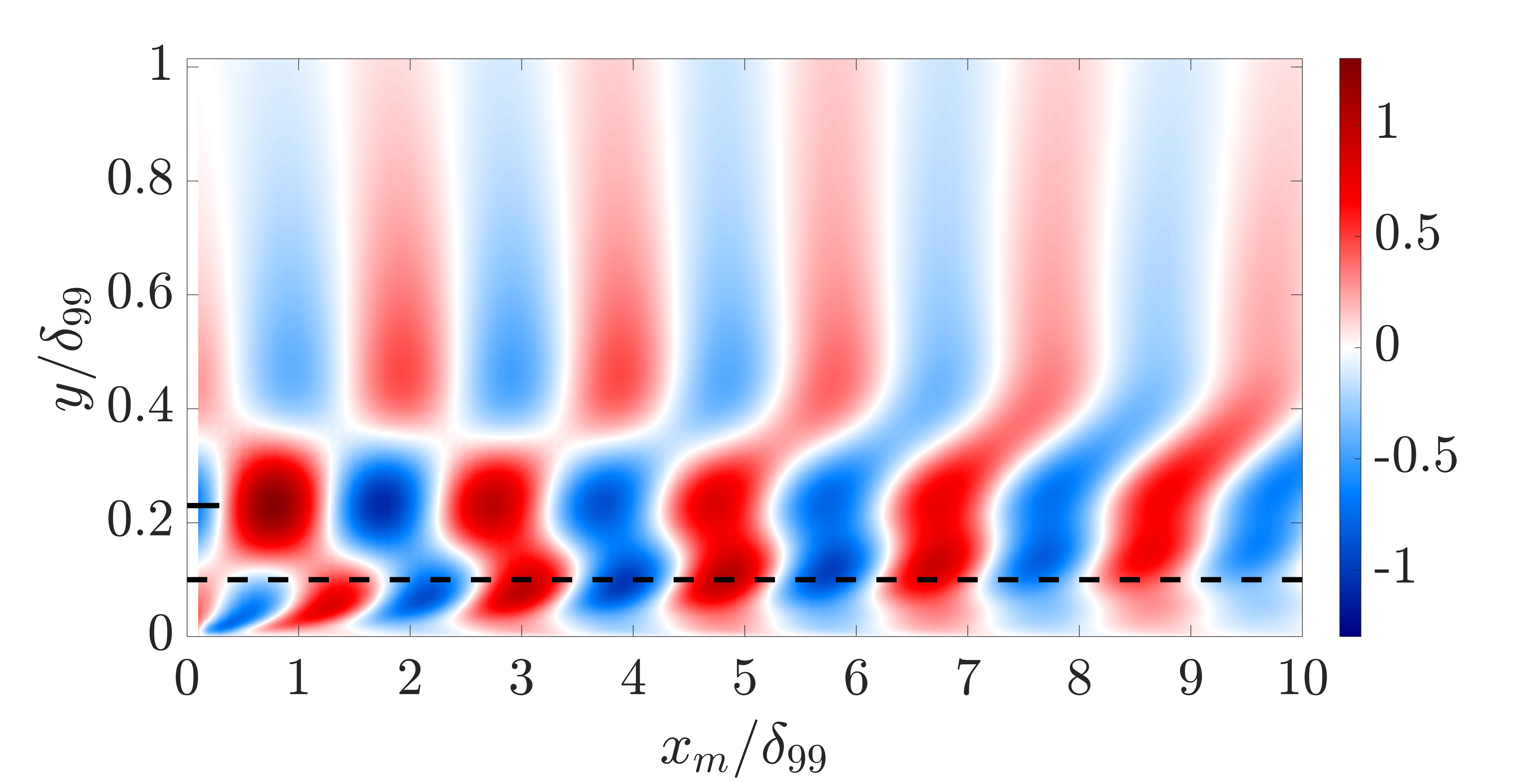}
    \includegraphics[width=0.49\textwidth]{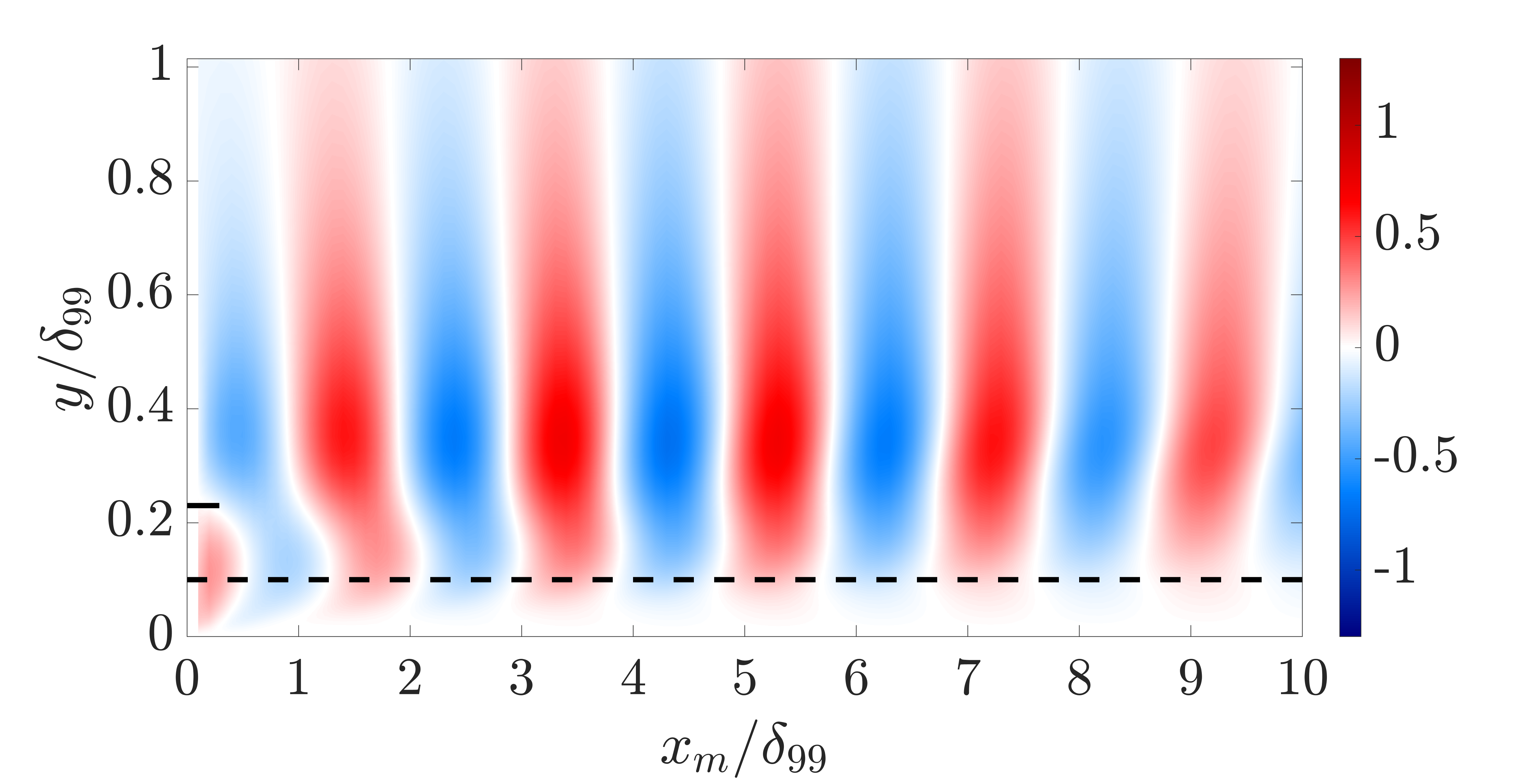}
    
   (c) $u_s^+$ at $y_p/\delta_{99}=0.3$ \hspace{0.35\textwidth} (d) $v_s^+$ at $y_p/\delta_{99}=0.3$
      \includegraphics[width=0.49\textwidth]{f80_yp03_u_dot.png}
    \includegraphics[width=0.49\textwidth]{f80_yp03_v.png}
    
       (e) $u_s^+$ at $y_p/\delta_{99}=0.5$ \hspace{0.35\textwidth} (f) $v_s$ at $y_p/\delta_{99}=0.5$

    \includegraphics[width=0.49\textwidth]{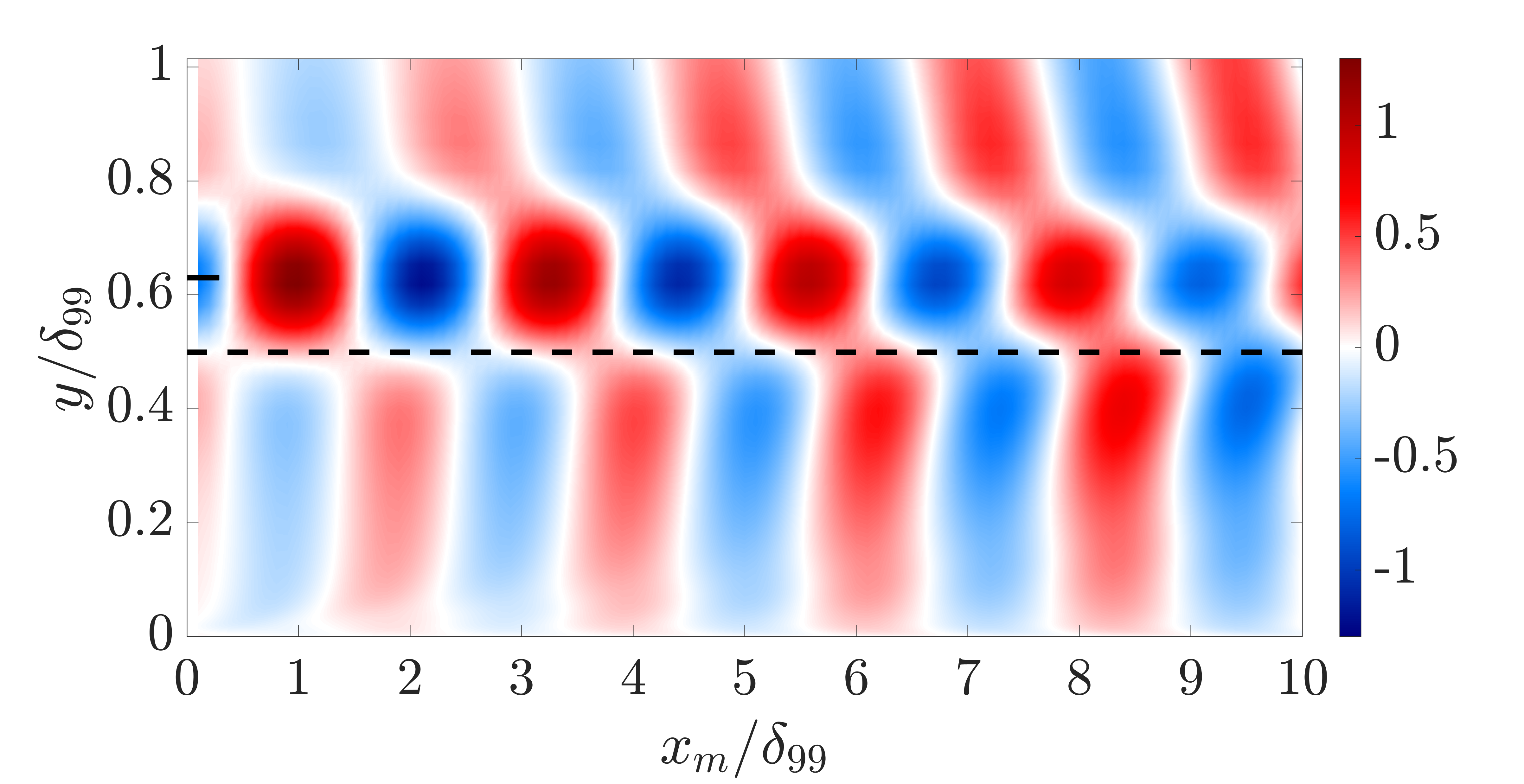}
    \includegraphics[width=0.49\textwidth]{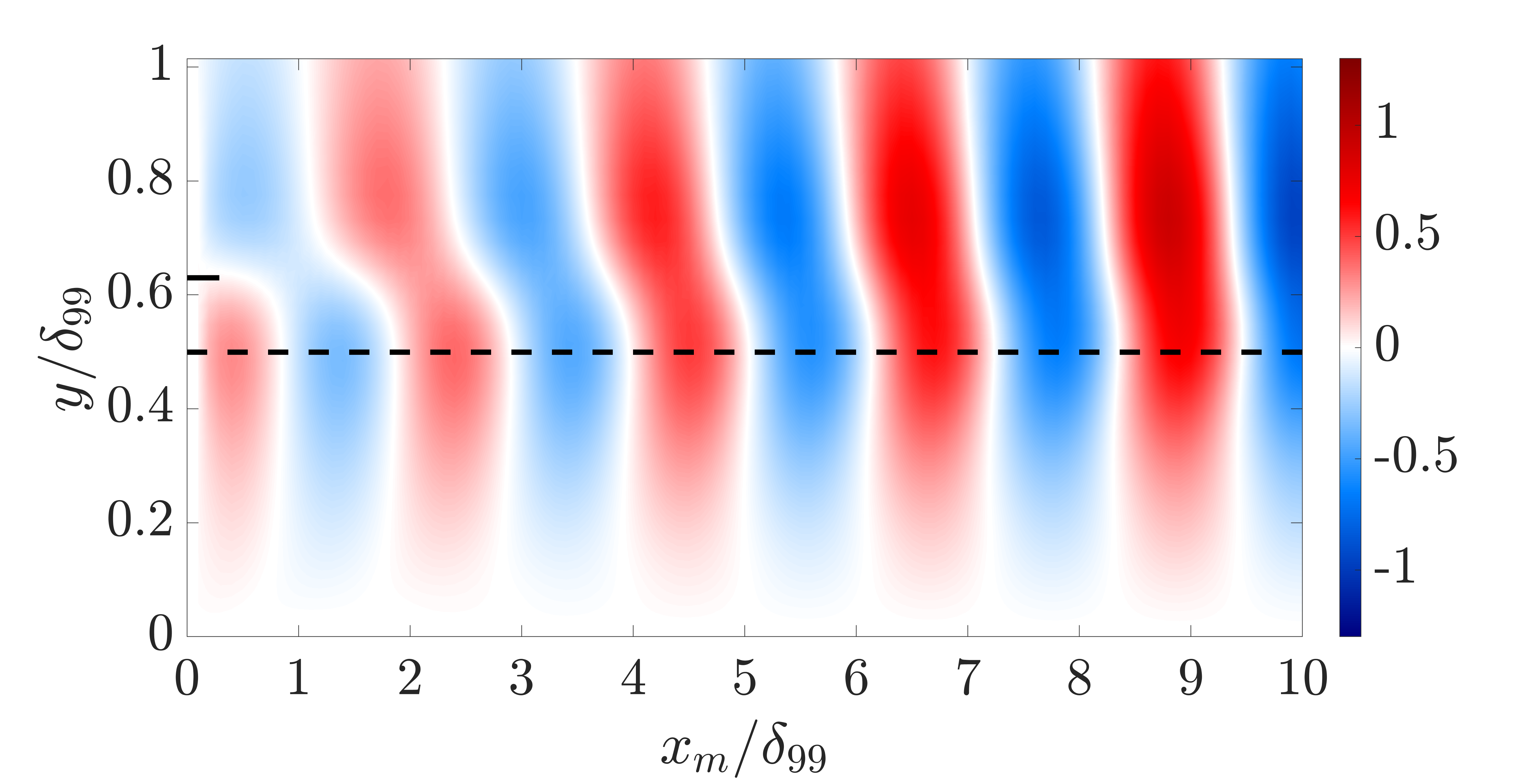}
   
    \caption{Downstream evolution of $u_s^+$ and $v_s^+$ at  $y_p/\delta_{99}=$0.1, 0.3 and 0.5. All results are associated with $f_p=80$~Hz. In all panels, the long horizontal line ($\dashed$) marks the location $y_p/\delta_{99}$ and the short horizontal line ($\mline\mline$) indicates $y_f=y_p+0.13\delta_{99}$.}
    \label{fig:spatial_external_forcing_height_velocity}
\end{figure}

\begin{figure}
    \centering
      (a)  $\omega_{z,s}^+$ at $y_p/\delta_{99}=0.1$ \hspace{0.23\textwidth} (b) Quadrant numbers at $y_p/\delta_{99}=0.1$
     \includegraphics[width=0.49\textwidth]{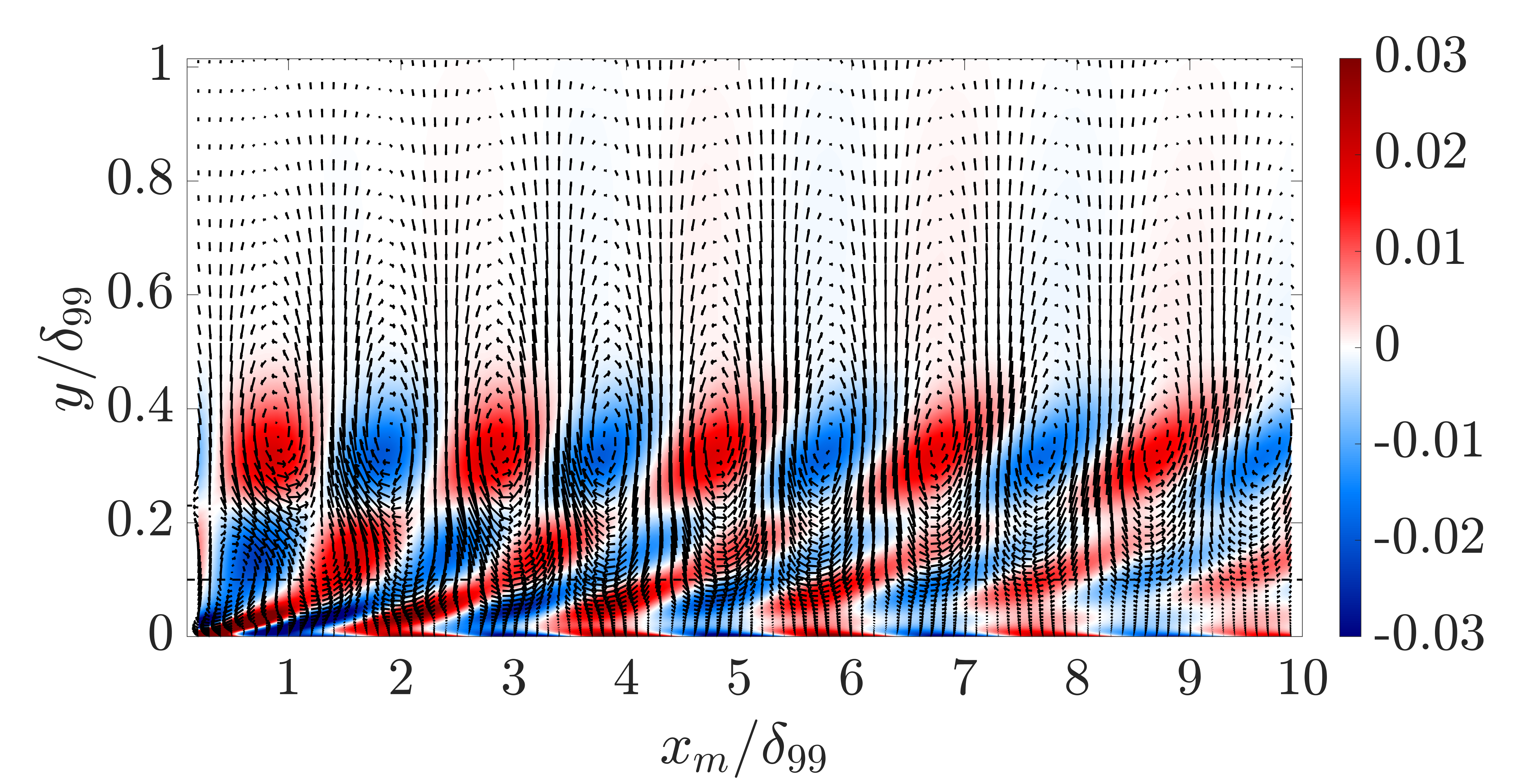}
    \includegraphics[width=0.49\textwidth]{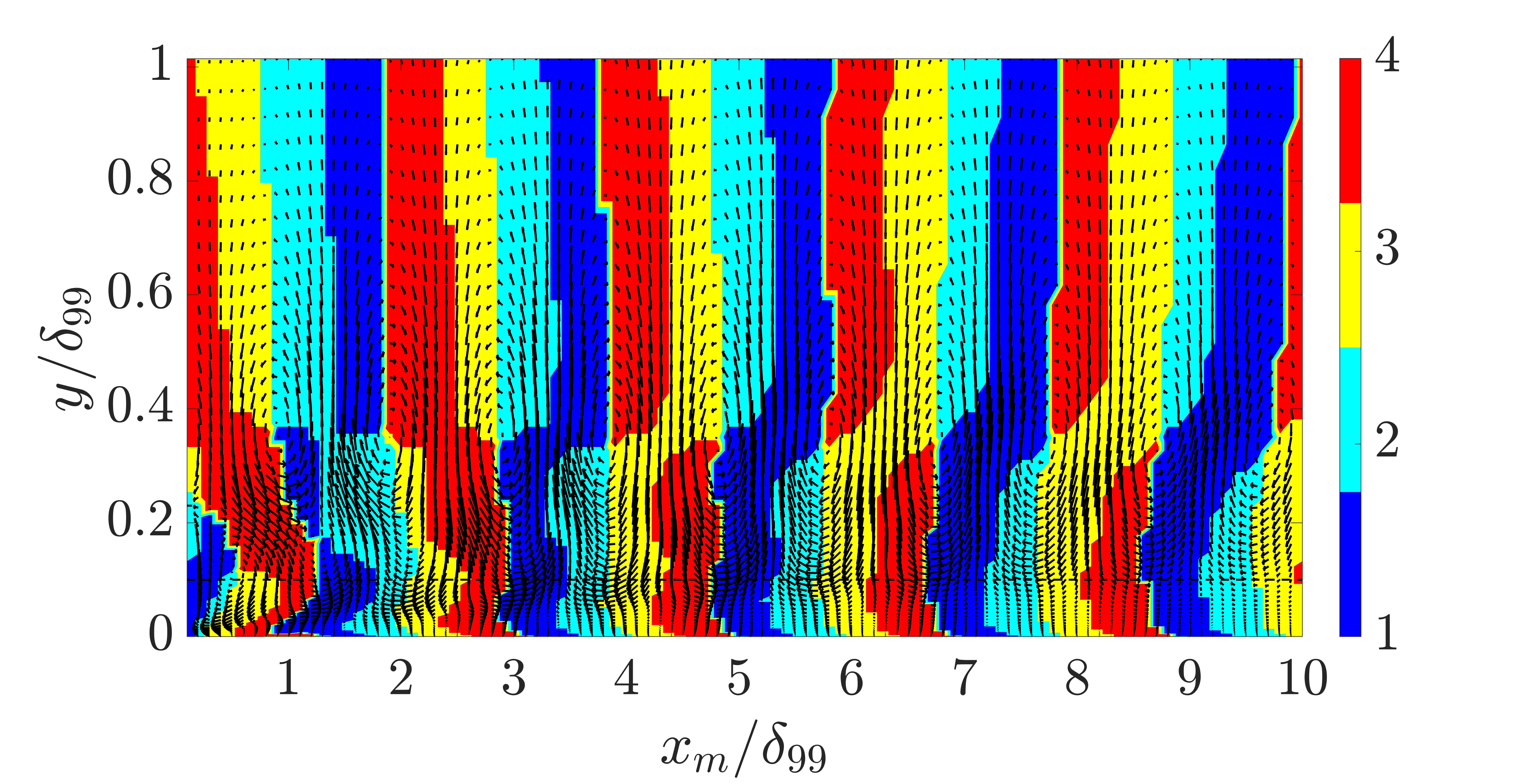}
    
   (c) $\omega_{z,s}^+$ at $y_p/\delta_{99}=0.3$ \hspace{0.23\textwidth} (d)
   Quadrant numbers at $y_p/\delta_{99}=0.3$
    \includegraphics[width=0.49\textwidth]{f80_yp03_omega.png}
    \includegraphics[width=0.49\textwidth]{f80_yp03_quad_outline.png}  
    
       (e) $\omega_{z,s}^+$  at $y_p/\delta_{99}=0.5$ \hspace{0.23\textwidth} (f) Quadrant numbers at $y_p/\delta_{99}=0.5$
       
   \includegraphics[width=0.49\textwidth]{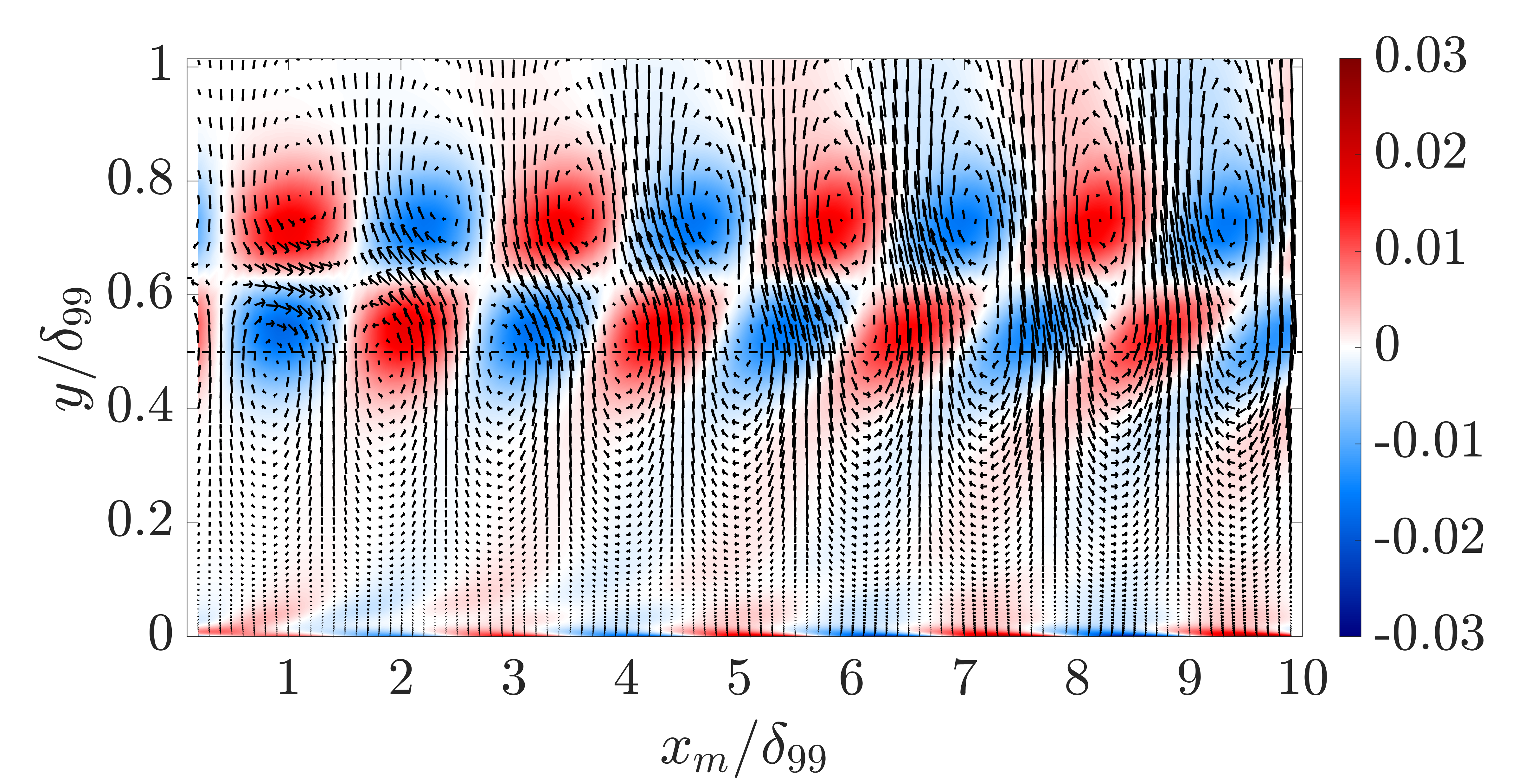}
    \includegraphics[width=0.49\textwidth]{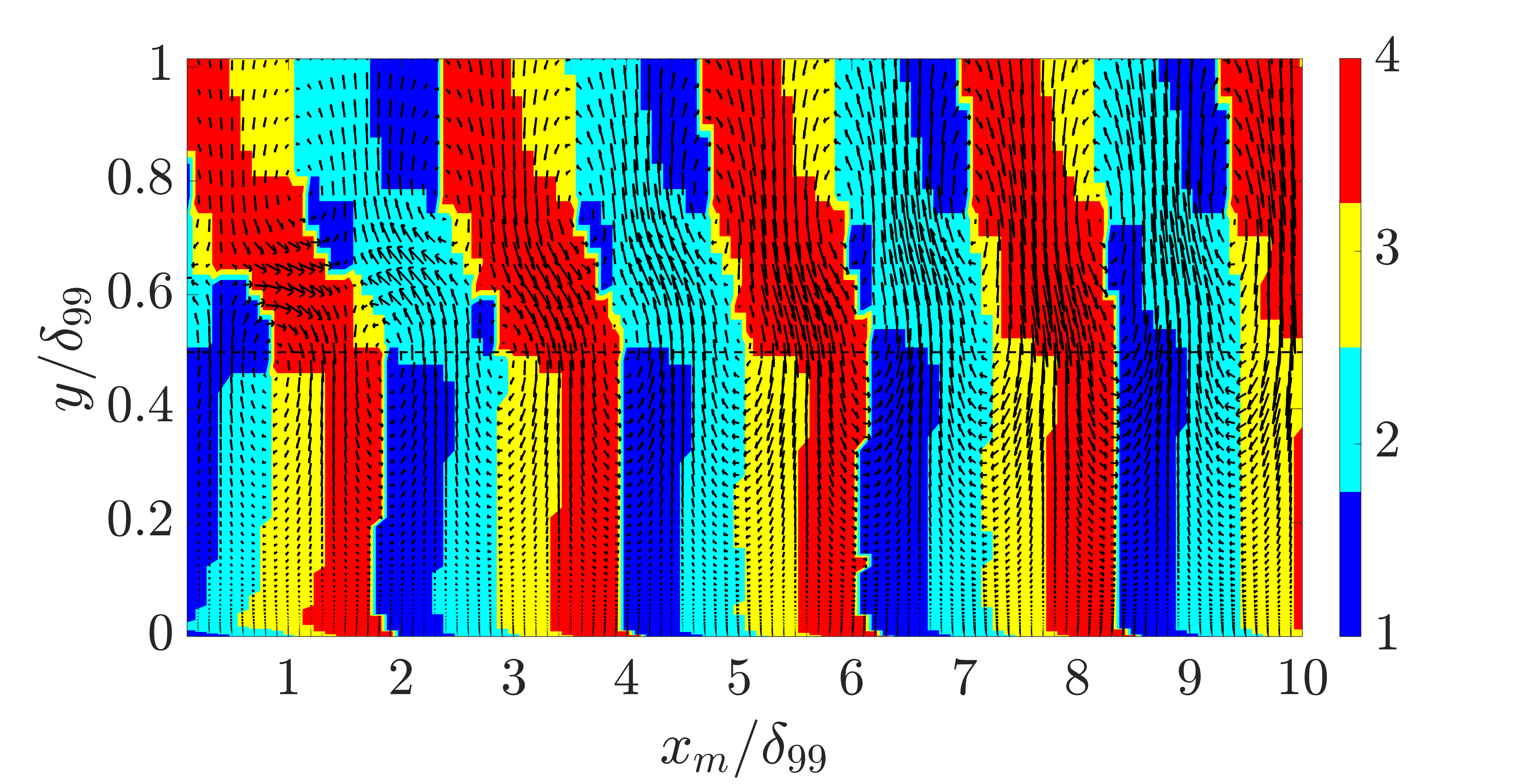}
       \caption{Downstream evolution of spanwise vorticity $\omega_{z,s}^+$ and quadrant numbers with the velocity vector field $(u_s^+, \:\: v_s^+)$ superimposed with $y_p/\delta_{99}=$0.1, 0.3 and 0.5. The actuation frequency is $f_p=80$~Hz for all panels. In panels (b), (d), and (f), Q1 events are indicated as blue contours, Q2 events are cyan, Q3 events are yellow color, and Q4 events are indicated by red contours.}

    \label{fig:spatial_external_forcing_height_vorticity_quad}
\end{figure}

In this subsection, we employ spatial input-output analysis to study the effect of changes in actuation frequency $f_p$ and actuator plate height $y_p$.  We first investigate the effect of actuation frequency, by introducing two additional frequencies, $f_p=20$~Hz and $f_p=200$~Hz, while keeping the plate height fixed at $y_p/\delta_{99}=0.3$. Then, we analyze the effect of varying the actuator height to values $y_p/\delta_{99}=0.1$ and $y_p/\delta_{99}=0.5$ for the fixed actuation frequency $f_p=80$~Hz.  

Figure \ref{fig:spatial_external_forcing_frequency_velocity}(a)-(b) show modal velocity components $u_s^+$, and $v_s^+$ obtained for an actuation frequency of $f_p=20$~Hz ($0.0996{U}_\infty/\delta_{99}$ and $f_p^+=0.0034$). These results indicate that the streamwise wavelength of the actuated structures is longer than those in figure \ref{fig:spatial_external_forcing_f_80Hz_y_p_0.3}(c) and (d), where the   actuation frequency is $f=80$~Hz, which is consistent with the lower frequency of the actuation. Note, the two modal velocity components for $f=80$~Hz, from figure \ref{fig:spatial_external_forcing_f_80Hz_y_p_0.3}(c) and (d) are replotted here in panels (c)-(d) of figure \ref{fig:spatial_external_forcing_frequency_velocity} for ease of comparison. Figure \ref{fig:spatial_external_forcing_frequency_velocity}(e)-(f) plots the same quantities as in figure \ref{fig:spatial_external_forcing_frequency_velocity}(a)-(b) for actuation at $f_p=200$~Hz ($0.996{U}_\infty/\delta_{99}$ and $f_p^+=0.034$). Here, the large-scale structures show a much smaller streamwise wavelength and decay much faster with downstream distance; see e.g., the streamwise velocity of the central region shown in \ref{fig:spatial_external_forcing_frequency_velocity}(e). Similar variations with temporal frequencies were observed by \citet{huynh2020characterization}, who found a linear correlation between temporal frequency and streamwise wavenumber. \rev{We also observe that the flow structures close to the wall at high frequency $f_p=200$~Hz are vanishing at downstream location $x/\delta_{99}\in [7,10]$ in figure \ref{fig:spatial_external_forcing_frequency_velocity}(e)-(f), which suggests that flow structures close to the wall due to off-wall actuation persist for a shorter downstream distance when their streamwise wavelength is smaller. Instead, the flow structures associated with lower frequency in figure \ref{fig:spatial_external_forcing_frequency_velocity}(a)-(d) display longer streamwise wavelength and extend their footprint towards the wall. This behavior is consistent with the observation that large-scale structures associated with large streamwise wavelengths have a footprint that extends further towards the wall \citep{mathis2009large,saxton2017coherent}.} Furthermore, the streamwise velocity at low-frequency $f_p=20$~Hz in figure \ref{fig:spatial_external_forcing_frequency_velocity}(a) is stronger than that seen in the structures generated by higher frequency actuation, $f_p=200$~Hz  in figure \ref{fig:spatial_external_forcing_frequency_velocity}(e). Instead, the amplitude of wall-normal velocity for the lower frequency $f_p=20$~Hz actuation in figure \ref{fig:spatial_external_forcing_frequency_velocity}(b) is  smaller than that due to the higher frequency $f_p=200$~Hz actuation in figure \ref{fig:spatial_external_forcing_frequency_velocity}(f). \rev{This phenomenon can be qualitatively understood from the two-dimensional continuity equation, which suggests the vertical velocity amplitude is proportional to the streamwise wavenumber. As shown in figure \ref{fig:spatial_external_forcing_frequency_velocity}, the flow structures associated with $f_p=20$~Hz possess the largest wavelength (or smallest wavenumber) leading to the smallest amplitude among these three different frequencies. However, for the high frequency results ($f_p=200$~Hz) in figure \ref{fig:spatial_external_forcing_frequency_velocity}(f), the spatial transient growth is weaker and the wall-normal velocity amplitude shows faster downstream decay compared with the case when $f_p=80$~Hz. The faster decay leads to vertical velocity amplitudes smaller than that in the data for $f_p=80$~Hz in figure \ref{fig:spatial_external_forcing_frequency_velocity}(d) at downstream location $x_m/\delta_{99}\approx 8$. This $f_p=80$~Hz case displaying the largest wall-normal velocity amplitude also coincides with the frequency leading to the largest modulation coefficient between phase-locked velocity and residual turbulence \citep[figure 3.10]{lozier2021turbulent}.
}

Figure \ref{fig:spatial_external_forcing_frequency_vorticity_quad} shows the spanwise vorticity computed from equation \eqref{eq:spatial_omega_s} and the results of a quadrant trajectory analysis for the  $f_p=20$~Hz, $80$~Hz, and $200$~Hz cases. 
For all of these frequencies, the spanwise vorticity in figures \ref{fig:spatial_external_forcing_frequency_vorticity_quad}(a), (c), (e) suggest that the actuated large-scale structures are more inclined towards the wall as they propagate downstream due to their height-dependent phase speed. Comparing the quadrant analysis at actuation frequency $f_p=20$~Hz, $80$~Hz, and $200$~Hz in figures \ref{fig:spatial_external_forcing_frequency_vorticity_quad}(b), (d), and (f), it is clear that the ejection (Q2) and sweep (Q4) events occupy a larger extent of the $(x_m,y)$ plane at the lowest frequency. This prevalence of Q2 and Q4 events are also observed at $f_p=80$~Hz (replotted here as figure \ref{fig:spatial_external_forcing_frequency_vorticity_quad}(d)) but are restricted to downstream regions close to the actuator location $x_m/\delta_{99}\lesssim 5$. \rev{Table \ref{tab:quadrant_ratio} displays the relative prevalence of each quadrant event, computed as $Qi/\sum_{j=1}^4 Qj$, (i=1,2,3,4), over the spatial extent $x_m/\delta_{99}\in [0,10]$ and $y/\delta_{99}\in [0,1]$. We compute this ratio for each forcing injection height $y_p$ and forcing frequency $f_p$ considered here. The prevalence of Q2 and Q4 events of $f_p=20$~Hz and $f_p=80$~Hz with $y_p/\delta_{99}=0.3$ can be also reflected in table \ref{tab:quadrant_ratio} with prevalence ratios greater than  25\%. } 
\begin{table}
    \centering
    \caption{Ratio (\%) of each quadrant events over all four quadrant events $Qi/ \sum_{j=1}^4 Qj$, (i=1,2,3,4) occupying the region $x_m/\delta_{99}\in [0,10]$ and $y/\delta_{99}\in [0,1]$ for different actuator heights and frequencies.}
    \begin{tabular}{cccccc}
    \hline
      $y_p/\delta_{99}$ & $f_p$(Hz) & $Q1/ \sum_{j=1}^4 Qj$ & $Q2/ \sum_{j=1}^4 Qj$ & $Q3/ \sum_{j=1}^4 Qj$ & $Q4/ \sum_{j=1}^4 Qj$ \\
        \hline
        0.3 & 20 & 16.2 & 34.2 & 18.0 & 31.6\\ 
        0.3 & 80 & 20.5 & 27.9& 21.9 & 29.7  \\
         0.3& 200 &28.9 & 22.1 & 28.2 & 20.8 \\
          0.1& 80 &  25.8 & 24.4& 25.3 & 24.4\\
           0.5& 80 & 18.8& 31.0& 19.9& 30.4 \\
           \hline
    \end{tabular}
    \label{tab:quadrant_ratio}
\end{table}

Close to the actuator $x_m/\delta_{99}\lesssim 2$, the quadrant order at high actuation frequency $f_p=200$~Hz in figure \ref{fig:spatial_external_forcing_frequency_vorticity_quad}(f) looks similar to the previous analysis at $f_p=80$~Hz in figure \ref{fig:spatial_external_forcing_frequency_vorticity_quad}(d), where the behavior is separated into different vertical bands with alternating Q1 \rev{or} Q2 quadrant and Q3 \rev{or} Q4 quadrant (Q1/Q2-Q3/Q4) events. However, farther downstream, the events in figure \ref{fig:spatial_external_forcing_frequency_vorticity_quad}(f) at $f_p=200$~Hz corresponding to quadrant Q1 and Q3 behavior are more prevalent and stronger than the quadrant Q2 and Q4 events. \rev{This larger prevalence of Q1 and Q3 events for $f_p=200$~Hz is quantified in table \ref{tab:quadrant_ratio}, which indicates that both of these types of events occur with prevalence ratios greater than  25\%.} This can be related to the observation that Q2 and Q4 are associated with a larger time scale (smaller frequency) than Q1 and Q3 events in fully developed turbulent channel flow  \citep{wallace1972wall,Wallace2016}. \rev{An increase in Q1 and Q3 quadrant events are shown to be associated with a negative contribution to Reynolds shear stress; see e.g., \citep{Wallace2016}. A reduction in Q2 and Q4 events has also been observed in turbulent channel flow with active or passive drag reduction \citep{choi1993direct,choi1994active}. This observation suggests further analyzing the potential to achieve drag reduction by high frequency actuation, which we leave  as a topic of future work.} {\color{black}The observation that quadrant events occur in the order Q4$\rightarrow$Q3$\rightarrow$Q2$\rightarrow$Q1 at the top region quadrant event order changes to Q1$\rightarrow$Q2$\rightarrow$Q3$\rightarrow$Q4 at the bottom region is also consistent with important quadrant events characterized in turbulent pipe flow \citep{nagano1995coherent}.}

Finally, we study the effect of different plate heights given a fixed actuation frequency of $f_p=80$~Hz. Figure \ref{fig:spatial_external_forcing_height_velocity} shows $u_s^+$ and $v_s^+$ computed for the cases with actuator heights $y_p/\delta_{99}=0.1$, $y_p/\delta_{99}=0.3$, and $y_p/\delta_{99}=0.5$. Comparing the plots of $u_s^+$ in figure \ref{fig:spatial_external_forcing_height_velocity} (a), (c), and (e), it is clear that the characteristic streamwise wavelength is longer when the actuator height is higher. This phenomenon results from a larger phase speed associated with the central region due to a larger local mean velocity at a higher plate height. Here, the effect of differences in phase speed between the central region and bottom region is more visible than in the results for $y_p/\delta_{99}=0.1$ with $y_p/\delta_{99}=0.3$. This larger difference is due to the larger mean velocity gradient in the near-wall region. The wall-normal velocity $v_s^+$ generated through actuation at different actuator heights $y_p$ in figures \ref{fig:spatial_external_forcing_height_velocity}(b), (d), and (f), shows nearly uniform behavior across the wall-normal height for downstream positions $x_m/\delta_{99}\in[0,10]$ for all cases. 

Figure \ref{fig:spatial_external_forcing_height_vorticity_quad} presents $\omega_{z,s}^+$, and quadrant trajectories associated with actuation at heights $y_p/\delta_{99}=0.1$, $y_p/\delta_{99}=0.3$, and $y_p/\delta_{99}=0.5$. The spanwise vorticity $\omega_{z,s}$ in figure \ref{fig:spatial_external_forcing_height_vorticity_quad}(a), (c), and (e) indicates similar patterns for all of these actuator heights. The quadrant analysis results in figure \ref{fig:spatial_external_forcing_height_vorticity_quad}(b), (d), and (f) for these different actuator heights are separated into different vertical bands with alternating Q1 \rev{or} Q2 quadrant and Q3 \rev{or} Q4 quadrant (Q1/Q2-Q3/Q4) activity. The quadrant order remains the same as the case with plate height $y_p/\delta_{99}=0.3$, i.e.,  Q4$\rightarrow$Q3$\rightarrow$Q2$\rightarrow$Q1 at the top region, and Q1$\rightarrow$Q2$\rightarrow$Q3$\rightarrow$Q4 at the bottom region. 
This suggests that quadrant trajectory orders observed in canonical wall-bounded turbulent flows are robust to the height where the large-scale structures are introduced.

\section{Conclusions and future work}
\label{sec:spatial_spatial_conclusion}

In this work, we use the one-way spatial integration method of \citet{towne2015one} to develop a spatial input--output analysis approach that does not require specification of a single streamwise wavenumber. This approach naturally has the advantage of naturally producing a wall-normal dependent phase speed allowing the computation of a local convective velocity of the actuated large-scale structures.
We focus on the particular problem of a low Reynolds number turbulent boundary layer where a synthetic large-scale structure is introduced through a spanwise-uniform DBD plasma actuator, whose effect is modeled as a streamwise body force associated with a dominant temporal frequency.

We first demonstrate the proposed spatial input-output based analysis produces phase-locked velocities with large-scale structures reminiscent of those obtained by experimental measurements employing hot-wire anemometry and a phase-locked analysis. We further predict the decreasing inclination angle of the associated large-scale structures as they propagate downstream, which illustrates the benefit of an analysis method that emits a wall-normal dependent  phase speed.  We use the quadrant analysis in  \citep{wallace1972wall,Wallace2016} to classify the shear stress distribution of the spatially evolving flow field. The results indicate that ordering the field based on these quadrants produces a trajectory order (Q4$\rightarrow$Q3$\rightarrow$Q2$\rightarrow$Q1 in the top region and Q1$\rightarrow$Q2$\rightarrow$Q3$\rightarrow$Q4 in the bottom region) similar to that observed in turbulent pipe flow \citep{nagano1995coherent}. This ordering (spatial progression of quadrant behaviors) is found to be independent of actuator height, while actuator height instead determines a phase speed for flow structures that is close to the local mean velocity at that height. The analysis also captures the relationship between changes in the actuation frequency and the greater prevalence of  different shear stress patterns, particularly the association of greater Q2 and Q4 activity with a larger time scale \citep{wallace1972wall}.   These observations further support the fact that the synthetic large-scale structures interact with the TBL in a manner consistent with naturally occurring large-scale structures. It therefore suggests that progress in analyzing the dynamics of large-scale structures can be made by studying the effect of introducing external perturbations in a controlled manner.

The results demonstrate that the proposed spatial input-output analysis can provide insights into the large-scale flow structures induced by temporally periodic and spatially localized perturbations in wall-bounded turbulent flows.  The method can be naturally extended to study flow structures with spanwise variation by setting $k_z\neq 0$ and spanwise velocity by modifying the output operator. This method may be further extended to analyze flow structures and \rev{potential drag reduction} induced by more comprehensive actuators by modifying the forcing function and associated input operator. This change would allow the analysis of different types of flow perturbations than the set-up considered here.

\section*{Acknowledgment}
The authors gratefully acknowledge support from the Office of Naval Research (ONR) through grant number N00014-18-1-2534. C.L., I.G., and D.F.G. also acknowledge support from the US National Science Foundation (NSF) through grant number CBET 1652244. C.L. appreciates the support from the Chinese Scholarship Council.


\appendix
\section{Asymptotic consistent turbulent boundary layer profile}

\label{sec:spatial_appendix_TBL_mean}

Here, we describe the asymptotic consistent turbulent boundary layer profile developed by \citet{monkewitz2007self}, which is also previously used; e.g., in \citet{cossu2009optimal}. 
The mean profile is provided by:
\begin{align}
    U=u_\tau[U_i^+(y^+)-U^+_{log}(y^+)+U_e^+(Re_{\delta_*})-U_w^+(\eta)].
\end{align}
$u_\tau$ is the wall friction velocity, $y^+=yu_\tau/\nu$ is the wall-normal location in the inner units, and $U_e^+=U_e/u_\tau$ is the free stream velocity $U_e$ scaled with $u_\tau$. $Re_{\delta_*}=U_e\delta_*/\nu$ is the Reynolds number scaled on the displacement thickness length scale, and $\eta=y/\Delta$ is the wall-normal coordinate scaled with the Rotta-Clauser outer length scale $\Delta=\delta_* U_e^+$. The inner and the outer coordinates are related by $y^+=Re_{\delta_*}\eta$. Then we have the explicit formula for these mean velocity from \citet{monkewitz2007self}:
\begin{subequations}
\begin{align}
    U_i^+(y^+)=&0.68285472\; \ln(y^{+2}+ 4.7673096y^+ + 9545.9963)\nonumber\\
    &+1.2408249\; \arctan(0.010238083y^+ + 0.024404056)\nonumber \\
    &+1.2384572\;\ln(y^+ + 95.232690)-11.930683\nonumber\\
    &-0.50435126\;\ln(y^{+2}- 7.8796955y^+ + 78.389178)\nonumber\\
    &+ 4.7413546\;\arctan(0.12612158y^+ - 0.49689982)\nonumber\\
    &- 2.7768771 \;\ln(y^{+2}+ 16.209175y^+ + 933.16587)\nonumber\\
    &+ 0.37625729\; \arctan(0.033952353y^+ + 0.27516982)\nonumber\\
    &+ 6.5624567\; \ln(y^+ + 13.670520)+ 6.1128254,\\
    U^+_{log}(y^+)=&\frac{1}{\kappa}\ln(y^+)+B,\\
    U_e^+(Re_{\delta_*})=&\frac{1}{\kappa}\ln(Re_{\delta_*})+C,\\
    U_w^+(\eta)=&[\frac{1}{\kappa}E_1(\eta)+w_0]\frac{1}{2}[1-\tanh(\frac{w_{-1}}{\eta}+w_2\eta^2+w_8\eta^8)],
\end{align}
\end{subequations}
where $\kappa=0.384$, $B=4.17$, $C=3.3$, $w_0=0.6332$, $w_{-1}=-0.096$, $w_2=28.5$, $w_8=33000$, and $E_1(\eta)=\int^{\infty}_\eta \frac{e^{-t}}{t}dt$. These analytical expressions are validated to be the same as the mean profile at $Re_\tau=690$ obtained from direct numerical simulations \citep{simens2009high,jimenez2010turbulent} (\url{https://torroja.dmt.upm.es/turbdata/blayers/low_re/profiles/}).

\bibliography{main_spatial_clean}
\bibliographystyle{new-aiaa_no_link}
\end{document}